\documentclass{article}
\usepackage{graphicx}

\makeatletter    
     
\def\eqalign#1{\null\,\vcenter{\openup\jot\m@th \let\\=\crcr 
  \ialign{\strut\hfil$\displaystyle{##}$&$\displaystyle{{}##}$\hfil
      \crcr#1\crcr}}\,}

\def\meqalign#1{\null\,\vcenter{\openup\jot\m@th \let\\=\crcr 
  \ialign{\strut\hfil$\displaystyle{##}$&&$\displaystyle{{}##}$\hfil
      \crcr#1\crcr}}\,}
\makeatother   
\def\journalfont{\it}   
\def\jou#1{{\journalfont #1\ }}
\def\joudef#1#2{\def #1{\jou{\ignorespaces #2}}}

\joudef{\AAA}{  Astron.\ Astrophys.}
\joudef{\AIP}{  Adv.\ Phys.}
\joudef{\AM}{   Ann.\ Math.}
\joudef{\AP}{   Ann.\ Phys.\ (N.Y.)}
\joudef{\AOP}{   Ann.\ Phys.\ (N.Y.)}
\joudef{\APJ}{  Astrophys.\ J.}
\joudef{\CJP}{  Can.\ J.\ Phys.}
\joudef{\CMP}{  Commun.\ Math.\ Phys.}
\joudef{\CQG}{  Class.\ Quantum Grav.}
\joudef{\GRG}{  Gen.\ Rel.\ Grav.}
\joudef{\IJMP}{ Int.\  J.\ Mod.\ Phys.}
\joudef{\IJTP}{ Int.\  J.\ Theor.\ Phys.}
\joudef{\JKPS}{  J.\ Korean.\ Phys.\ Soc.}
\joudef{\JMP}{  J.\ Math.\ Phys.}
\joudef{\JPAMG}{ J.\ Phys.\ A: Math.\ Gen.}
\joudef{\MNRAS}{ Mon.\ Not.\ R.\ Ast.\ Soc.}
\joudef{\NAT}{  Nature}
\joudef{\NCIM}{ Nuovo Cim.}
\joudef{\NUCP}{ Nuc.\ Phys.}
\joudef{\NCB}{  Il Nuovo Cimento ``B}
\joudef{\PL}{   Phys.\ Lett.}
\joudef{\PR}{   Phys.\ Rev.}
\joudef{\PREP}{ Phys.\ Rep.}
\joudef{\PRL}{  Phys.\ Rev.\ Lett.}
\joudef{\PTP}{  Prog.\ Theor.\ Phys.}
\joudef\RMP{  Rev.\ Mod.\ Phys.}
\joudef\SPJ{  Sov.\ Phys.\ JETP}

\joudef{\AIHP}{  Ann.\ Inst.\ H. Poincar\'e}
\joudef{\AM}{    Ann.\ Math.}
\joudef{\JETP}{  Sov.\ Phys.\ J.E.T.P.}
\joudef{\JP}{    J.\ Phys.}

\def\vol#1{{\bf #1}}   
\def\xc{\hbox{\rlap{\hskip 1.5pt\raise .75pt\hbox{--}}$\Xscr$}}
\def\half{{\textstyle {1\over2}}}

  \def\Cscr{{\cal C}} \def\Dscr{{\cal D}}
\def\Escr{{\cal E}}  \def\Gscr{{\cal G}} \def\Hscr{{\cal H}}
   
\def\Mscr{{\cal M}}   
   \def\Tscr{{\cal T}}
   \def\Xscr{{\cal X}}

\def\pmb#1{\setbox0=\hbox{$#1$}%
  \kern-.025em\copy0\kern-\wd0
  \kern.05em\copy0\kern-\wd0
  \kern-.025em\raise.0433em\box0}
\def\pmbs#1{\setbox0=\hbox{$\scriptstyle #1$}%
  \kern-.0175em\copy0\kern-\wd0
  \kern.035em\copy0\kern-\wd0
  \kern-.0175em\raise.0303em\box0}
\def\bfbeta{\pmb{\beta}}    \def\bfsbeta{\pmbs{\beta}}
  \def\bfdelta{\pmb{\delta}}
\def\bfkappa{\pmb{\kappa}}  \def\bfskappa{\pmbs{\kappa}}
\def\bfpi{\pmb{\pi}}
  \def\bfsalpha{\pmbs{\alpha}}
\def\bfrho{\pmb{\rho}}  \def\bfsigma{\pmb{\sigma}} 

\def\bff#1{\pmb{#1}}
\def\bfs#1{\pmbs{#1}}

\def\Gammait{{\mit\Gamma}}

\def\oversymbol#1#2{\vbox{\ialign{##\crcr \hfil$#1$\hfil\crcr
   \noalign{\kern1pt\nointerlineskip}%
   \hbox{$\hfil\displaystyle#2\hfil$}\crcr}}}
\def\fraction#1#2{{\textstyle{#1\over#2}}}  

\def\sterling{\hbox{\it\char'44}}
\def\ketl{\langle}        \def\ketr{\rangle}
\def\rarrow{\to}  \def\relv{\bigm|} 

\def\diag{{\rm diag}}

\def\s{\tilde}
\def\A{\hat}
\def\ctr#1{\hfil#1\hfil} \def\lft#1{#1\hfil} \def\rt#1{\hfil#1}
\def\inr{\in}
\def\bfbbeta{{\bfbeta}}
\def\bfe{{\bf e}}
\def\dl{\dot}
\def\odl#1{\oversymbol{\circ}{#1}}
\def\du{\dot}
\def\b{\vec}
\def\wedget{\wedge}
\def\sst{}
\def\s#1{\tilde #1}
\def\dott#1{\dot{#1}{}}
\def\tb{{\bar t}}

\def\boxit#1{\vbox{\hrule height .8pt\hbox{\vrule width .8pt\hskip 3pt
  \vbox{\vskip 3pt #1 \vskip 3pt}\hskip 3pt\vrule width .8pt}
  \hrule height .8pt}}

\def\headerplainsetup{\typeout{************ using PLAIN TeX page dimensions.}%
      \topmargin=0pt                     
      \advance\voffset by -\headheight   
      \advance\voffset by -\headsep      
      \oddsidemargin=0pt \evensidemargin=0pt       
      \textheight=8.9truein \textwidth=5truein}  

\begin{document}

\renewcommand{\textfraction}{.15}

{\small\noindent
Reformatted with corrections from 
{\it Proc.\ Int.\ Sch.\ Phys.\ ``E. Fermi" Course\\ LXXXVI (1982)
on ``Gamov Cosmology"\/} (R. Ruffini, F. Melchiorri, Eds.),\\
North Holland, Amsterdam, 1987, 61--147.}
\vskip 24pt

\begin{flushleft}
{\Large\bf SPATIALLY HOMOGENEOUS DYNAMICS:\\[5pt] 
A UNIFIED PICTURE\footnote{This work was supported by National Science Foundation grant
No. PHY--80--07351 and typeset with \TeX.}
}\\[20pt]
Robert T. Jantzen\footnote{Present address:
Department of Mathematical Sciences, Villanova University, Villanova, PA 19085 \qquad[ {\tt http://www.homepage.villanova.edu/robert.jantzen} ]}
\\
Harvard-Smithsonian Center for Astrophysics\\
60 Garden St., Cambridge, MA 02l38
\end{flushleft}

\begin{abstract}
The Einstein equations for a perfect fluid spatially homogeneous spacetime
are studied in a unified manner by retaining the generality of certain
parameters whose discrete values correspond to the various Bianchi types of
spatial homogeneity.  A parameter dependent decomposition of the metric
variables adapted to the symmetry breaking effects of the nonabelian Bianchi
types on the ``free dynamics" leads to a reduction of the equations of motion
for those variables to a 2-dimensional time dependent Hamiltonian system
containing various time dependent potentials which are explicitly described
and diagrammed.  These potentials are extremely useful in deducing the gross
features of the evolution of the metric variables.
\end{abstract}

\section{Introduction}

Although interest in spatially homogeneous cosmological models peaked in the
early seventies, it was not until the late seventies that a more
unified picture of the dynamics of these models developed, following the
proper recognition of the role played in the problem by the gauge freedom
of general relativity [1--3].   
Since a number of books [4,5]
and review articles [6--9]
exist which discuss spatially homogeneous cosmology at
various levels and from various points of view, it seems appropriate here
to emphasize certain aspects of the subject which are not well covered in
the literature.  Remarkably there is still something new left to say on this
topic nearly a decade after its most active period of research.  The present
discussion will seek to generalize many concepts which have already appeared
in the context of particular symmetry types or special initial data and
fit them together into a single unified picture of spatially homogeneous
dynamics.  It should be emphasized that although spatially homogeneous
cosmological models are usually studied for a (nearly) discrete set of
parameter values corresponding to the (nearly) discrete set of Bianchi
types [10--14],
both the metric and field equations depend analytically on
a 4-dimensional space of essential parameters (of which at most three may
be simultaneously nonzero).  By varying these parameters continuously, one
may deform each of the various symmetry types into each other and thus relate
properties of one Bianchi type to those of another, the more specialized
symmetry types occurring as singular limits of more general types.

Lagrangian or Hamiltonian techniques enable one to associate a 
finite dimensional classical mechanical system with the ordinary differential
equations equivalent to the spatially homogeneous Einstein equations and thus
offer a convenient means of visualizing the dynamics and of understanding
its qualitative features.  These techniques, which provide the framework of
this exposition, were pioneered by Misner [15--18]  
and followed through by
Ryan [19--23],
leading to an alternative but equivalent description [24] 
of the qualitative results obtained by 
Lifshitz, Khalatnikov and Belinsky [25--29] 
for the evolution of certain spatially homogeneous cosmological
models near the initial singularity using piecewise analytic approximations.
This latter work was later confirmed and extended by Bogoyavlensky, Novikov
and Peresetsky using powerful techniques from the qualitative theory of
differential equations [30--35]. 
Qualitative studies of various perfect
fluid models including the regime away from the initial singularity should
also be noted [36--40].

The present discussion has as its foundation previous papers of 
the author [1--3, 41--43].  
No attempt will be made to review the large body of important
work which preceded them.  References [4--9] 
adequately serve this function.
In particular the bibliographies of the Ryan-Shepley book [5] 
and the recent review by MacCallum [9] 
provide an exhaustive list of relevant research papers.

It turns out that the key to our problem is a simple adage of mathematical
physics:  whenever a symmetric matrix is encountered, diagonalize it.  First
one diagonalizes the symmetric tensor density associated with the structure
constant tensor whose components completely determine locally the type
of spatial homogeneity.  The three diagonal components plus an additional
parameter characterizing the trace of the structure constant tensor are the
four parameters referred to above.  Next one diagonalizes the component matrix
of the spatial metric, maintaining the values of these parameters.  This
leads to a decomposition of the gravitational configuration space variables
(namely the component matrix of the spatial metric) into two sets of three
variables, one set parametrizing the diagonal values of the metric component
matrix which are readily interpreted in terms of the action of the
3-dimensional scale group (independent rescaling of the unit of length along
orthogonal directions) and another set specifying the diagonalizing matrix.
This decomposition incredibly simplifies the field equations since the
diagonalizing variables correspond to pure gauge directions, reflecting the
effect on the metric variables of spatial diffeomorphisms which are
compatible with the spatial homogeneity.  Scalar functions such as the
spatial scalar curvature which are gauge invariant depend at most on some
of the diagonal variables, for example.  Finally one diagonalizes the DeWitt
metric [44] 
on the configuration space $\Mscr$ of spatial metric 
component matrices by choosing an orthogonal basis of the Lie algebras of the
scale group and of the 3-dimensional group $\A G$ used in the metric
diagonalization; the DeWitt metric is important since the ``free dynamics"
is equivalent to geodesic motion for this metric.  The advantages of diagonal
matrices over general symmetric matrices are obvious; all matrix operations
(multiplication, determinant, inverse) become trivial and functions of these
matrices have a much simpler dependence on the individual components.
Diagonalizing a quadratic form kinetic energy function also greatly
simplifies the equations of motion.

A familiar example from classical mechanics which proves useful as an
analogy is the problem of the motion of a rigid body [45,46].  
Such an
analogy was in fact first introduced for general spacetimes by Fischer
and Marsden [47] 
in their original discussion of the role of the lapse and
shift in the three-plus-one formulation of the Einstein equations.  It is
even more appropriate for the spatially homogeneous spacetimes where
the correspondence is nearly complete.  The usual synchronous gauge
spatial frame is analogous to the space-fixed axes in the rigid body problem.
The ``diagonal gauge" spatial frame which diagonalizes both the spatial
metric and the symmetric tensor density associated with the structure
constant tensor corresponds to the body-fixed axes which diagonalize the
moment of inertia tensor.  The special orthogonal group $SO(3,R)$ generalizes
to the relevant 3-dimensional diagonalizing matrix group $\A G$.  Since the
components of the structure constant tensor must remain fixed under its
action, this group is a matrix representation of a subgroup of the special
automorphism group of the given Bianchi type Lie algebra (the subgroup is
unimodular since its Lie algebra is required to be offdiagonal).  The
concept of angular velocity also has an analogue which is closely related to
the shift vector field whose associated time dependent spatial
diffeomorphism drags the synchronous spatial frame into the diagonal gauge
spatial frame by inducing the time dependent frame transformation which
orthogonalizes the spatial frame.  However, since the action of $\A G$ on the
metric configuration space represents an orbital motion, the situation is
more involved.

At this point it is helpful to keep in mind the problem of the nonrelativistic
motion of a particle in a spherically symmetric potential (the central force
problem).  Here the symmetry group $SO(3,R)$ is a subgroup of the group of
motions of the Euclidean metric on the configuration space $R^{3}$.  By
introducing spherical coordinates one separates the configuration space
variables into angular variables describing the orbits of $SO(3,R)$, namely
2-spheres except for the fixed point at the origin where the orbit dimension
degenerates, and radial variables which describe the directions orthogonal to the orbits.  The components of orbital angular momentum arise from evaluating the moment function [40] 
for the action of $SO(3,R)$ on $R^{3}$ in the
standard basis of its Lie algebra and may be interpreted as the inner
products of the standard basis of rotational Killing vector fields with the
velocity of the system.  Since $SO(3,R)$ is a symmetry group of the dynamics,
angular momentum is conserved and the problem is then reduced to 
1-dimensional
radial motion in a new potential, the angular momentum contributing
an effective potential (the centrifugal potential) to the original radial
potential.  When the latter potential is absent, the case of the motion
of a free particle, it is of course simplest to consider only radial orbits
which lead to the simplest representation of the free (straight line) motion,
namely geodesics of the Euclidean metric.  This restriction to radial orbits
is possible because of the additional translational symmetry which allows one
to transform the angular momentum to zero.

In spatially homogeneous dynamics the Euclidean metric on $R^{3}$ is replaced
by the Lorentzian DeWitt metric on the 6-dimensional space $\Mscr$ of spatial
metric component matrices.  The decomposition of the Euclidean space
variables goes over roughly into the 3-dimensional space of ``diagonal"
variables and the 3-dimensional space of ``offdiagonal" variables as described
above.  The offdiagonal variables describe the orbits of the action on
$\Mscr$ of the matrix group $\A G$ and the diagonal variables describe the
orthogonal directions.  The moment function for the action of $\A G$ on 
$\Mscr$
is the analogue of the orbital angular momentum.  Its components in a certain 
basis of the matrix Lie algebra $\A g$ of $\A G$ will be seen below to 
correspond to the
space-fixed components of the spin angular momentum in the rigid body
analogy, thus neatly intertwining these two classical analogies.  The free
motion (geodesics of the DeWitt metric) is most easily represented as purely
diagonal, using the larger isometry group $SL(3,R)$ of the DeWitt metric to
transform away the angular momentum associated with any particular subgroup 
$\A G$.
The overall scale of the metric matrix represented by its determinant
(product of its diagonal values) corresponds to the single timelike direction
and the free motion is subject to an additional energy constraint requiring
the geodesic to be null.  Apart from the freedom to rescale and translate
the affine parameter of these diagonal null geodesics, there is a 1-parameter
family of them, parametrized by the angle of revolution of the 2-dimensional
null cone in the space of diagonal metric matrices; these are the well known
Kasner solutions [48]. 
However, a geodesic which has zero angular momentum
for a particular choice of the group $\A G$ will have nonzero angular momentum for almost all other choices of this group.

In the central force problem the separation of radial and angular variables
is clean.  In our problem the symmetry group of the free dynamics is $SL(3,R)$
and only the corresponding division of variables into a conformal metric (unit
determinant) and a scale variable (the metric determinant which parametrizes
the orbits of $SL(3,R)$) is clean.  The division of variables into two 
orthogonal sets of three
diagonal variables and three offdiagonal variables is instead highly ambiguous.
There is essentially a 2-parameter family of subgroups $\A G$ of $SL(3,R)$ with
3-dimensional offdiagonal matrix Lie algebras whose orbits are almost
everywhere transversal to the diagonal submanifold of $\Mscr$.  The orbits of
any of these subgroups may be used to perform the decomposition, which is
automatically orthogonal with respect to the DeWitt metric.  The
addition to the free system of the spatial scalar curvature as a potential
breaks down the $SL(3,R)$ symmetry uniquely to one of these subgroups in the
general case, although some degeneracy remains in some of the more specialized
symmetry types (Bianchi types II and V; Bianchi type I is the free system
and the symmetry remains unbroken).  This breakdown of $SL(3,R)$ symmetry to a
particular subgroup $\A G$ depends continuously on the parameters which 
specify
the Lie algebra of the spatial homogeneity group, just as the symmetry 
breaking
scalar curvature potential itself depends continuously on these parameters.
When the subgroup $\A G$ has a compact subgroup, the diagonal/offdiagonal
decomposition develops a singularity where the orbit dimension degenerates 
from
its generic value three, similar to the coordinate singularity of spherical
coordinates at the origin of $R^{3}$.  These points of the configuration
space turn out to be associated with additional continuous symmetry of the
spatial metric; they are protected by angular momentum barriers in the same
way as is the origin of $R^{3}$ in the central force problem.  

The concept of angular momentum links the rigid body and central force
analogies.  The time dependent diagonalizing matrix is a curve in the matrix
group $\A G$, representing the orbital motion of the system.  (In fact the 
matrix
group $\A G$ directly parametrizes the points of each orbit.)  The tangent 
vector
of this curve, namely the velocity associated with the offdiagonal variables,
represents the orbital velocity of the configuration space point.  Given a
basis of the matrix Lie algebra $\A g$ of $\A G$, one determines a 
corresponding basis
of the tangent space at the identity and two global frames on $\A G$, one left
invariant and one right invariant, which reduce to this basis at the identity.
The left and right invariant frame (contravariant) components of the velocity
tangent vector correspond respectively to the space-fixed and body-fixed
components of the angular velocity in the rigid body analogy.  These components
are related to each other not by $\A G$ itself as in the previously defined 
space-fixed and body-fixed components but by the adjoint representation of 
$\A G$.
The association of left and right with space and body assumes that the
diagonal gauge frame is related to the synchronous gauge frame by a passive
transformation, corresponding to a right action of $\A G$ on $\Mscr$ and since
$\A G$ is identified with its orbits this becomes right translation of $\A G$ 
into itself.

By the local
identification of $\A G$ with its orbits in $\Mscr$ (the ``offdiagonal
variables"), one may use the DeWitt metric on the orbit to lower the
indices of the velocity tangent vector, leading to what are analogous to
the space-fixed (left invariant frame) components and body-fixed (right
invariant frame) components of the angular momentum.  Thus the frame
components of the DeWitt metric along the orbit act as the components of the
moment of inertia tensor in the rigid body analogy.  The ``space-fixed
components of the angular momentum" are the inner products of the left
invariant frame vectors with the velocity.  These vector fields generate
the right translations and hence the right action of $\A G$ on $\Mscr$ and are
therefore Killing vector fields of the DeWitt metric.  The space-fixed
components of the angular momentum are thus the components of the moment
function for the action of $\A G$ on $\Mscr$ and therefore the components of 
the
orbital angular momentum, which are conserved for the free motion.  Note,
however, that the body-fixed components are related to these constant
components by the adjoint transformation and so are in general time dependent. 
Furthermore, the right invariant frame components of the DeWitt metric must
be independent of the orbital variables since right translation is an isometry.
By properly choosing the basis of $\A g$, these components may in fact be
diagonalized.

One advantage of the present problem over the analogous classical problems
is that one is free to reparametrize the time variable by introducing
a nontrivial spatially homogeneous lapse function.  Lapse functions which
depend only on the metric component matrix correspond to conformally
rescaling the DeWitt metric and the scalar curvature potential.  However,
in general the lapse function may also depend on time derivatives of the
metric components leading to a much larger freedom in the Hamiltonian
system (a freedom not permitted in the Lagrangian approach).  Often
special choices other than the usual cosmic proper time are suggested by the
dynamics which help to simplify its description.  Two such choices are
the Misner $\Omega$-time [15] 
related to the logarithm of the metric
determinant and his supertime [17], 
where the lapse is simply related to
the metric determinant.  The latter choice of time is also crucial to the
Belinsky-Lifshitz-Khalatnikov analysis of the dynamics near the initial
singularity.  For the free dynamics, they are both affine parameters for the
geodesic motion. 

Of course all of these remarks will become clearer once explicit notation
and formulas are introduced.  The result of this formal manipulation
is a reduction of the Einstein equations to a 2-dimensional Hamiltonian system
with time dependent potentials associated with the spatial curvature,
with the centrifugal forces arising from the ``motion" of the diagonalizing
spatial frame, and with the energy-momentum of the source of the gravitational
field, here assumed to be a perfect fluid.  This system must be supplemented
by the equations of motion of the source of course.  Explicit diagrams of the
various time dependent potentials are extremely useful in deducing the gross
features of the evolution of the metric variables; near the initial singularity
they may be used to construct ``diagrammatic solutions" of the field equations,
as done by Ryan for the Bianchi type IX case [20]. 

The main body of the paper is divided into three sections.  In the first of
these the parametrized spatially homogeneous spacetime and field equations
are introduced.  In the second the parametrized decomposition of the metric
variables is introduced and used to reduce the Einstein equations to a 
2-dimensional time dependent Hamiltonian system.  The potentials of this 
system
are then described in detail.  In the third section their qualitative effect 
on the dynamics is discussed.


\section{The $\Cscr_D$-parametrized Spatially Homogeneous 
Spacetime}

Before even beginning to discuss spatially homogeneous spacetimes, it is
worthwhile introducing some useful facts concerning the smooth action
of a Lie group $G$ on a manifold $M$ as a transformation 
group [49--52].
A left action is just a homomorphism $f: G\rarrow \Dscr (M)$ from the Lie group into the group
of diffeomorphisms of $M$, i.e. for $a\in G$ the corresponding transformation
$f_{a}$ satisfies under composition $f_{a_{1}} \circ f_{a_{2}} =
 f_{a_{1}a_{2}}$. By defining $f^{-1}_{a} \equiv f_{a^{-1}}$ one obtains an 
antihomomorphism $f^{-1}$ satisfying $f^{-1}_{a_{1}} \circ f^{-1}_{a_{2}} = f^{
-1}_{a_{2}a_{1}}$ which is the defining relation for a right action.  When 
$f$ is an isomorphism
$(G \simeq f_{G} \equiv \{ f_{a}\relv a \in G \})$
so that only the identity $a_{0}\in G$ acts as the identity transformation
on $M$, $f$ is called an effective action.  For example, any Lie group acts
effectively on itself on the left by left translation $L_{a}(a_{1}) = aa_{1}$
and on the right by right translation $R_{a}(a_{1})  = a_{1}a$ with
$G \simeq L_{G} \simeq R_{G}$.  (Note that $R^{-1}:G \rarrow \Dscr (G)$ is an
isomorphism and a left action.)  Introducing the redundant but useful notation
$a \cdot x \equiv f_{a}(x)$ for the transformation $f_{a}$ acting on
$x \in M$, denote the orbit of $x$ by $G \cdot x = \{ a \cdot x
\relv a \in G \} $, namely all points which can be reached from 
$x$ under the action of the group.

Not all of the transformations are effective in moving the point $x$.  Let
$G_{x} = \{ a \in G \relv a \cdot x = x \}$ be the
isotropy subgroup at $x$ of the action of $G$ on $M$, namely the subgroup of 
$G$
 which leaves $x$ fixed.  Intuitively one expects that at least locally the
orbit of $x$ is in a one-to-one correspondence with the smallest subset of
transformations which can move the point $x$ arbitrarily on its orbit.  This
notion is described by introducing the space $G/G_{x} = \{ aG_{x}
\relv a \in G \}$ of left cosets of the subgroup $G_{x}$ in $G$,
where each left coset $aG_{x} = \{ ab \relv b \in G_{x}
\} $ is an orbit of the right translation action of $G_{x}$ on $G$.
All elements of a given coset map $x$ to the same point of $M$ and elements
of different cosets necessarily map $x$ to different points as one may easily
check.  One can therefore extend the domain of the map $f$ from $G$ to 
$G/G_{x}$
when acting on $x$, i.e. $F_{x}(aG_{x}) = f_{a}(x)$ defines a map $F_{x}:
G/G_{x}\rarrow G \cdot x$ which turns out to be a diffeomorphism of the left
coset space onto the orbit [49].  
In particular the dimension of the
orbit is the difference in dimension of $G$ and $G_{x}$.  Note further that
fixing the point in question to be $x_{0}$, so a general point of the orbit
may be represented by $x = f_{a}(x_{0}) = F_{x_{0}}(aG_{x_{0}})$, then the
left action of $G$ on $M$ $\quad x \rarrow f_{a_{1}(x)} = f_{a_{1}a}(x_{0}) =
F_{x_{0}}(a_{1}aG_{x_{0}})$ corresponds to the left translation $aG_{x_{0}}
\rarrow a_{1}aG_{x_{0}}$ on the coset space.

In mathematics any orbit of a transformation group (therefore diffeomorphic
to $G/H$ for some subgroup $H$ of $G$) is called a homogeneous space, since all
points of the space are equivalent under the transformation group.  Here in
the context of general relativity, a narrower notion of homogeneous space
is required which incorporates not only the equivalence of the points of the
space but of the geometry as well.  A (pseudo-) Riemannian space $(M,
\hbox{\sl g})$
is called homogeneous if it is the orbit of an isometry group (invariance
group of the metric \hbox{\sl g}), in which case the action is said to be
transitive.  A simply transitive action is one in which the isotropy group
at every point of the single orbit is trivial (contains only the identity
$a_{0}$ of $G$); in this case the orbit and the group are diffeomorphic and
the left action of $G$ on $M$ corresponds to left translation on $G$.  Using 
the
diffeomorphism $F_{x_{0}}:G \rarrow M$ for an arbitrary point $x_{0}$ of M
to pull back the metric from $M$ to $G$, one therefore obtains a left invariant
metric on $G$.  Thus a homogeneous (pseudo-) Riemannian space with a simply
transitive isometry group is equivalent to a left invariant (pseudo-)
Riemannian manifold involving that group.  (For a right action one simply
replaces left by right everywhere in the above discussion; the choice of
left or right actions is a matter of convention.)

However, not all transformation groups act transitively and the interesting
question about a given action is how the orbits fit together to fill up
the entire manifold.  One may introduce an integer-valued function
$d_{G}(x) \equiv \hbox{dim}(G\cdot x)$ on $M$ whose value gives the dimension 
of the
orbit to which each point belongs; all those orbits of a given dimension form
a subspace called a stratum and the partitioning of $M$ into the various strata
is called a stratification [53].  
Note that it is easy to verify that if
$x_{2} = a_{21}\cdot x_{1}$ and $a\in G_{x_{1}}$, then $a^{-1}_{21}a
a_{21}\in G_{x_{2}}$, i.e. $G_{x_{2}} = a^{-1}_{21}G_{x_{1}}a_{21}$,
so the isotropy subgroups at different points of a given orbit are all
conjugate (and therefore isomorphic) subgroups of $G$.  Since $d_{G}(x) = 
\hbox{dim}\  G - \hbox{dim}\  G_{x}$, a decrease in the orbit dimension 
corresponds to an increase in the isotropy subgroup dimension.

Often the action of a Lie group $G$ on a manifold $M$ describes a symmetry, all
points of a given orbit being equivalent in some sense which depends on the
context, and one is interested in how things change in the directions
``orthogonal" or ``oblique" (``transversal") to the orbits.  It is therefore
natural to introduce the orbit space $M/G = \{ G\cdot x\relv
x\in M\}$; however, due to the varying dimension of the
orbits, this is not a manifold.  For nice enough actions one can usually 
choose a subspace of $M$ (a submanifold with or without boundary or a 
collection
of such subspaces which intersects each orbit only once or finitely many times)
such that its intersection with the ``generic" stratum of maximum dimension
orbits is a submanifold whose tangent space is complementary (``transversal")
to the orbit tangent space at each intersection point and hence this
submanifold is a local slice for the action on the generic stratum [53].
This ``slice" is very helpful in studying objects which are invariant under
the group and seem more complicated when studied on the entire space $M$.

For example, consider the rotations about the $z$-axis of $R^{3}$. The group
is $SO(2,R)\sim S^{1}$ acting as an isometry subgroup of the Euclidean 
metric
on $R^{3}$, the orbits are circles centered on the $z$-axis and lying in the 
planes of constant $z$, and the half plane $y=0,x>0$ directly parametrizes
the orbit space which is a manifold with boundary.  On the other hand the full
plane $y=0$ is a manifold intersecting the generic orbits (circles of nonzero 
radius) twice but having the advantage that the projection of all geodesics
of the Euclidean metric onto this manifold are smooth curves, while those
which intersect the $z$-axis suffer reflection at the boundary when projected 
onto the half plane. It is convenient to use the term ``slice" to refer to
either the plane or the half plane.

The spatially homogeneous spacetimes or ``Bianchi cosmologies" which are
studied here have a 3-dimensional isometry group $G$ acting simply 
transitively
on a 1-parameter family of spacelike hypersurfaces (the orbits) which
provides a natural slicing of the spacetime.  Each orbit is a homogeneous
Riemannian manifold and therefore isometric to a copy of $G$ equipped with a
left invariant Riemannian metric, namely the pullback of the induced spatial
metric on the orbit.  Rather than maintaining the distinction between $G$
and each orbit, it is simpler to identify the spacetime manifold $M$ with the
product manifold $R\times G$, where $R$ is the real line with natural
coordinate $t$ which
parametrizes the family of copies $G_{t} = \{ (t,x)\relv x\in G
\}$ of $G$ in $R\times G$, on each of which $G$ acts by left
translation.
However, this still leaves open the question of how these left invariant
Riemannian manifolds fit together into a spacetime and how the left
translations on each copy of $G$ in $R\times G$ fit together into a global
action of $G$ on $M$.

One needs to describe a threading of this natural slicing by a congruence
of curves in the spacetime (which is nowhere tangent to an orbit) which
will be identified with the $t$-lines in $R\times G$, the $t$-coordinate on
$R\times G$
corresponding to a given time function for the slicing.  Each such
identification leads to a different global reference system based on the
same natural slicing of the spacetime.  In order for the induced metric
on each copy of $G$ to be a left invariant metric, the class of threading
congruences must be compatible with the action of $G$ on the spacetime.  One
threading congruence and slicing parametrization is picked out uniquely by
the symmetry, namely the invariant congruence of geodesics orthogonal to the
orbits, the proper time along these geodesics measured from some initial
orbit serving to parametrize the family of orbits.  Identifying this
parametrized congruence with the $t$-lines of $R\times G$ establishes the so
called 
``synchronous reference system" [48]
adapted to the spatial homogeneity,
with the action of $G$ on $M = R\times G$ being $t$-independent left 
translation
on each copy of $G$.  Any other threading of the slicing may then be viewed as
a
$t$-dependent diffeomorphism of the family of copies $G_{t}$ of $G$ in $M$
relative
to the synchronous threading [47].  
Dragging along the $t$-dependent left
invariant spatial metric by this diffeomorphism will lead to the $t$-dependent
spatial metric in the reference system adapted to the new congruence.  This
will again be a $t$-dependent left invariant metric only if one restricts the
spatial diffeomorphism freedom, i.e. restricts the allowed class of threadings,
to be compatible with the group structure of $G$.  The compatibility condition
is that such diffeomorphisms map the space of left invariant tensor fields on
$G$ into themselves.  These consist of the left and right translations and the
automorphisms of $G$, the latter diffeomorphisms being those which preserve 
the group multiplication and form a finite dimensional Lie group $Aut(G) = 
\{ h\in\Dscr(G)\relv h(a_{1})\, h(a_{2}) = h(a_{1}a_{2})
\}$
called the automorphism group.  The translations and automorphisms together
form a semidirect product Lie group [51]
$L_{G}\times _{s}Aut(G) =
 R_{G}\times _{s}
Aut(G) \equiv\Dscr(g)$.

Before discussing in detail the structure of a spatially homogeneous spacetime,
it is worth understanding first the homogeneous Riemannian manifolds from which
they are constructed.  These are left invariant Rieman\-nian 3-man\-i\-folds
$(G,\hbox{\sl g})$, where {\sl g} is a left invariant Riemannian metric on the
3-dimensional Lie group $G$.  On each Lie group there is a natural 
identification
of the tensor algebra at any given point, say the identity $a_{0}\in G$,
with the algebra of either left or right invariant tensor fields.  Given a
tangent tensor at the identity, one can left (right) translate that tensor all
over the group using the differential of the unique left (right) translation
which maps the identity to each point of the group, thus obtaining a left
(right) invariant tensor field which coincides with the original tensor
at the identity.  In particular, given a basis $\A e = \{ \A e_{a}
\}$
 of the tangent space at the identity and its dual basis $\{ 
{\A{\omega}}
\null^{a}\}$ of covectors (satisfying $\A\omega^{a}(\A e_{b}) = 
\delta^{a}
{}_{b}$), one obtains a global left (right) invariant frame $e = \{
e_{a}\}$ ($e=\{ \s e_a\}$) and its dual frame 
$\{\omega^{a}\}$ ($\{{\s {\omega}}\null^a\}$) 
of left (right)invariant 1-forms on the group.  The
components of a given left (right) invariant tensor field in this frame
are just the components (namely constants) of the original tensor at the
identity with respect to the given basis of the tangent space there. For
example, a left invariant Riemannian metric may be expressed in the form
$$\hbox{\sl g} = g_{ab}\ \omega^{a}\otimes\omega^{b} ,\qquad g_{ab}=
\hbox{\sl g}(e_{a},e_{b})
\eqno (2.1)$$
where the constant matrix $\hbox{\bf g}=(g_{ab})$ is symmetric and 
positive-definite.
This relation in fact establishes a diffeomorphism (for each left invariant
frame $e$) from the space of left invariant metrics on $G$ onto the space 
$\Mscr$
of symmetric positive-definite matrices of the given dimension.  For dimension
three, $\Mscr$ is a 6-dimensional submanifold of the space $gl(3,R)$
of $3\times 3$ real matrices whose natural basis will be designated by 
$\{\hbox{\bf e}^{b}{}_{ a}\}$, in terms of which a matrix 
may be 
represented as $\hbox{\bf A} = (A^{a}{}_{b}) = A^{a}{}_{b}
\hbox{\bf e}^{b}{}_{a}$.

Let $g$ and $\s g$ denote the spaces of respectively left and right
invariant vector fields on $G$, each isomorphic as a vector space to the
tangent space at the identity $TG_{a_{0}}$ and having corresponding bases
$e$ and $\s e$ arising from some basis $\A e$ of $TG_{a_{0}}$.  These vector 
spaces turn
out to be closed under the Lie bracket operation and are therefore Lie
subalgebras of the infinite dimensional Lie algebra $\xc(G)$ of smooth
vector fields on $G$.  As a Lie subalgebra of $\xc(G)$, each generates a
finite dimensional subgroup of the group $\Dscr(G)$ of diffeomorphisms of $G$
into itself; $g\ (\s g )$ generates the action of $G$ on itself by right
(left) translation, with image diffeomorphism subgroup $R_{G}(L_{G})$.  The Lie
algebra $g$ of left invariant vector fields on $G$ is referred to as the
Lie algebra of the Lie group $G$.

A Lie group $G$ is completely determined locally by the structure of its Lie
algebra $g$.  Given a basis $e$ of $g$, this structural information
is contained in the collection of (constant) components of the structure
constant tensor defined by
$$[e_{a},e_{b}] = C^{c}{}_{ ab}e_{c}\qquad \hbox{or}
\qquad C^{c}{}_{ ab} = \omega^{c}
([e_{a},e_{b}])\ .\eqno(2.2)$$
Since $e$ is also a global frame on $G$ with dual frame $\{ 
\omega^{a}\} $, a standard formula gives the dual relation
$$d\omega^{a} = -\half C^{a}{}_{bc}\omega^{b}\wedge\,\omega^{c}.\eqno(2.3)$$
Similar formulas hold for $\s e$ and $\{{\s {\omega}}\null^{a}
\}$ except for a change in sign of the structure constant tensor 
components, while $[e_{a},{\s e}_{b}] =
0$ since the diffeomorphism subgroups $R_{G}$ and $L_{G}$ they generate 
commute
with each other due to the associativity of the group multiplication.  The
structure constant tensor components are not arbitrary but must be
antisymmetric in the lower indices and satisfy a quadratic identity imposed by
the cyclic Jacobi identity.  Let 
$$\Cscr = \{ C^{a}{}_{bc}\relv C^{a}_{(bc)} = 0 = C^{d}_{[ab}C^{e}{}_{
c]d}\} \eqno(2.4)$$
be the space of possible real structure constant tensor components, a
6-di\-men\-sion\-al space for 3-dimensional Lie algebras.

Of course one may always choose another basis $\overline{e}_{a} = 
A^{-1b}_{\hskip 13pt a}
e_{b}$
of $g$ leading to new structure constant tensor components
$$\overline{C}^{a}{}_{bc} = A^{a}{}_{d}C^{d}{}_{fg}A^{-1f}_{\hskip 13pt b}
A^{-1g}_{\hskip 13pt c}
\equiv j_{\hbox{\bf A}}(C^{a}{}_{bc})\eqno(2.5)$$
which describes the same Lie algebra structure.  In fact when the structure
constant tensor components of two different Lie algebras of the same dimension
are related in this way, the Lie algebras are called isomorphic and represent
the same abstract Lie algebra.  (A simple change of basis leads to bases of
the two Lie algebras with identical structure constant components.)  When
$\overline{C}^{a}{}_{bc} = C^{a}{}_{bc}$ so that the components of the 
structure constant
tensor are invariant under the linear transformation, then  {\bf A}  is the 
matrix of an automorphism of the Lie algebra into itself.  In other words the
isotropy group of the above left action $j$ of the general linear group on 
$\Cscr$
at $C^{a}{}_{bc}$ is just the matrix representation of the automorphism group
$Aut(g)$ of the Lie algebra with respect to the basis $e$; denote this 
matrix group by $Aut_{e}(g)$.  The orbits of the action of the general 
linear group on $\Cscr$ correspond to the isomorphism classes of structure 
constant tensors.
These isomorphism classes are designated by their Roman numeral Bianchi type
following the original classification scheme of Bianchi [10].

In three dimensions the structure constant tensor is easily decomposed into
its irreducible parts under the action of the general linear group $GL(3,R)$,
greatly simplifying matters.  One may dualize the antisymmetric pair of indices
leading to an equivalent second rank contravariant tensor density whose
antisymmetric part may be represented as the dual of a covector, leading to the
following decomposition due to Behr [13,14]
$$\eqalign{
        C^{ab} &=\half C^{a}{}_{cd}\epsilon^{bcd}=C^{(ab)}+C^{[ab]} =
        n^{ab}+\epsilon^{abc}a_{c}\cr
        C^{a}{}_{bc} &= C^{ad}\epsilon_{dbc}=\epsilon_{bcd}n^{ad}+a_{f}\delta
        ^{fa}_{bc},\quad a_{f}=\half C^{a}{}_{fa}\cr
        0&=a_{f}n^{fa}=a_{f}C^{fa}=a_{f}C^{f}{}_{ab}\ .\cr } \eqno(2.6)$$
The Jacobi identity requires that the covector be annihilated by the
symmetric tensor density.  When this covector is nonzero, one may introduce
a scalar $h$ by the following formula [38]
$$a_{a}a_{b}=\half h\,\epsilon_{acd}\epsilon_{bfg}n^{cf}n^{dg}\ .
\eqno2.7)$$
These objects transform under the left action (2.5) of $GL(3,R)$ on $\Cscr$ in 
the following way
$$\eqalign{
     \overline{n}^{ab} &= (\det \hbox{\bf A})^{-1}A^{a}{}_{f}A^{b}{}_{g}
        n^{fg}\equiv         j_{\hbox{\bf A}}(n^{ab})\cr
     \overline{a}_{b}&=a_{c}A^{-1c}_{\hskip 13pt b}\equiv 
        j_{\hbox{\bf A}}(a_{b}),
     \qquad \overline{h}=h\equiv j_{\hbox{\bf A}}(h).\cr}\eqno(2.8)$$

One may always diagonalize the symmetric component matrix $\hbox{\bf n}
=(n^{ab})$ by
an orthogonal transformation with matrix $\hbox{\bf O}\in O(3,R)$ . 
(The eigenvalues of {\bf n} change sign if $\det \hbox{\bf O} = -1$.)  The
Jacobi
identity guarantees that the covector may be chosen to lie along the dual of
one of the eigenvectors of {\bf n}, thus reducing the components of the 
structure
constant tensor to the following ``standard diagonal form"
$$\eqalign{
    \hbox{\bf n}&= \hbox{diag}(n^{(1)},n^{(2)},n^{(3)})\ ,\qquad a_{f}
    =a\delta^{3}{}_{f}
    \quad (a\geq 0)\cr
    a n^{(3)} &=0\ ,\qquad a^{2}=h\,n^{(1)}n^{(2)}\ .\cr}\eqno(2.9)$$
Denote the corresponding subspace of $\Cscr$ by $\Cscr_{D}$; this subspace
turns out to
contain all the interesting information.(It is in fact a ``slice" for the 
action
of the orthogonal group on $\Cscr$.)  If $e$ is the basis of a Lie algebra 
whose
structure constant tensor components are in standard diagonal form, then the
Lie brackets of the basis vectors are given by
$$[e_{2},e_{3}] = n^{(1)}e_{1}-ae_{2},\qquad [e_{3},e_{1}]=n^{(2)}e_{2}+ae_{1},
\qquad [e_{1},e_{2}]=n^{(3)}e_{3}\ .\eqno(2.10)$$

\begin{table} 
$$
\vbox{\tabskip 0pt
        \def \|{\vrule height 11pt depth 5pt}
\hbox to 11cm{\ }     \hrule
\halign to 11cm{#\tabskip 0pt plus 100pt &#& #
&$\ctr{#}$&#&$\ctr{#}$&#&$\ctr{#}$&#&$\ctr{#}$&#&$\ctr{#}$&# 
& #& #
&$\ctr{#}$&#&$\ctr{#}$&#&$\ctr{#}$&#&$\ctr{#}$&#&$\ctr{#}$&
#\tabskip 0pt \cr
\|&\hfill Class A\hfill&&
&&&&&&&&&\|&
   \hfill Class B\hfill&&
&&&&&&&&&\| \cr
  \noalign{\hrule}
\|&\hfill  Type\hfill&\|&
n^{(1)}&\|&n^{(2)}&\|&n^{(3)}&\|&a&\|&h&\|&
   \hfill Type\hfill&\|&
n^{(1)}&\|&n^{(2)}&\|&n^{(3)}&\|&a&\|&h&\| \cr
  \noalign{\hrule}
\|&\hfill I                &\|&0&\|& 0&\|& 0&\|& 0&\|&-
&\|&\hfill V                &\|&0&\|& 0&\|& 0&\|& 1&\|&- &\|\cr
\|&\hfill II               &\|&0&\|& 0&\|& 1&\|& 0&\|&-
&\|&\hfill IV               &\|&1&\|& 0&\|& 0&\|& 1&\|&- &\|\cr
\|& &\|& &\|&  &\|&  &\|&  &\|&
&\|&\hfill III$\equiv$VI$_{-1}$
                           &\|&1&\|&-1&\|& 0&\|& 1&\|&-1  &\|\cr
\|&\hfill VI$_0$           &\|&1&\|&-1&\|& 0&\|& 0&\|&0
&\|&\hfill VI$_{h\neq 0,-1}$
                           &\|&1&\|&-1&\|& 0&\|& a&\|&-a^2&\|\cr
\|&\hfill VII$_0$            &\|&1&\|& 1&\|& 0&\|& 0&\|&0
&\|&\hfill VII$_{h\neq 0}$
                           &\|&1&\|& 1&\|& 0&\|& a&\|&a^2 &\|\cr
\|&\hfill VIII             &\|&1&\|& 1&\|&-1&\|& 0&\|& 0
&\|&                        &\|& &\|&  &\|&  &\|&  &\|&    &\|\cr
\|&\hfill IX               &\|&1&\|& 1&\|& 1&\|& 0&\|&  0 
&\|&                        &\|& &\|&  &\|&  &\|&  &\|&    &\|\cr
\noalign{\hrule}   }}
$$
\caption{
Canonical values of the 
standard diagonal form structure constant tensor components for each Bianchi
type.
}
\end{table}

Standard diagonal form is preserved by all diagonal matrix transformations
(provided the third diagonal component is positive when $a>0$) and certain
permutations.  Such transformations may be used to further reduce these
components to canonical values for each orbit.  Only the absolute value of the
signature of {\bf n}, the constant $h$ when well defined and the vanishing or
nonvanishing of the covector with component row vector $(a_{f})$ are invariant
under the general linear group.  By normalizing the nonzero diagonal values
of {\bf n} to absolute value unity, permuting these diagonal values if 
necessary
and changing their overall sign using reflection matrices of negative
determinant, while normalizing $a$ to unity when nonzero and $h$ is undefined,
one may arrive at the particular choice of canonical values of the structure
constant tensor components listed in Table 1 for each isomorphism class or
Bianchi type.  The apparently odd choice for Bianchi type II reflects a
prejudice which tries to associate the third basis vector with a preferred
basis vector of the Lie algebra (2.10) when possible.  (A conflict arises for
Type IV which does not allow a choice corresponding to the type II choice.)
Figure 1 represents the space $\Cscr_{D}$ as a 3-plane (class A submanifold) 
and
an orthogonal half 3-plane (class B submanifold) in $R^{4}$ and indicates the
canonical points of $\Cscr$ corresponding to Table l.  Although $\Cscr_{D}$ is 
 the
union of two manifolds, it is clearly not a manifold itself.  It is convenient
to think of $\Cscr_{D}$ as stratified by values of the integer pair $(
\hbox{rank {\bf
n}},d_{_{\Cscr_D}})$, where $d_{_{\Cscr_D}}$ is the reduced orbit dimension,
namely the dimension of the intersection of an orbit with $\Cscr_{D}$; the
strata are then labeled by the Roman numerals (excluding III and omitting
subscripts on VI and VII) of the Bianchi types.  Only types VI$_{h\leq 0}$ 
and
VII$_{h\geq 0}$ represent strata consisting of a family of orbits; the
remaining strata are themselves orbits.

\begin{figure}[t!] 
\begin{center}
\includegraphics[width=.9\textwidth]{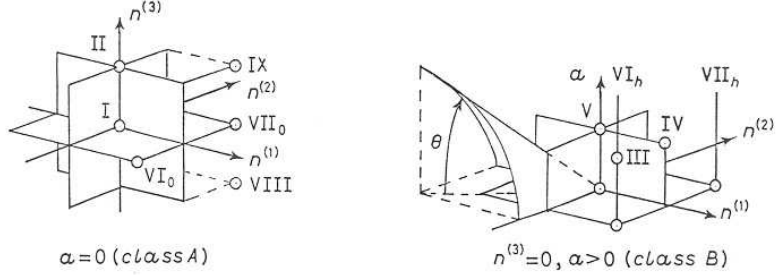}
\end{center}
\caption{The parameter space 
$\Cscr_D$ as
a 3-plane (class A) and an orthogonal half 3-plane (class B) in $R^4$ with
coordinates $(n^{(1)}, n^{(2)}, n^{(3)}, a)$, showing the canonical
representatives of each Bianchi type.  Of the 8 open octants in the class A
case, 2 and 6 respectively represent type IX and VIII, while half of the 12
open faces bounding these octants represent type VII$_0$ and the other half
type VI$_0$; the 6 coordinate open half lines represent type II and the origin
type I.  Similarly in the class B case, half of the 4 open octants are 
associated with each of the 1-parameter family of Bianchi types VI$_h$ and
VII$_h$, a single isomorphism class corresponding to a constant value surface
of the function $h=a^2(n^{(1)}n^{(2)})^{-1}$.  A typical such surface is 
illustrated in one octant, the angle $\theta$ given by 
$\tan\theta= |h/2|^{1/2}$; 
those in the remaining octants are obtained by rotation through multiples of 
$\pi/2$, $h$ alternating in sign for a given magnitude $|h|$.  The 4 vertical
open faces bounding these octants all represent type IV and the positive 
$a$-axis type V, with the $a=0$ plane giving the class A limit of each type.}  
\end{figure}

The space $\Cscr_{D}$ is very useful in describing the notion of Lie algebra
contraction [50].
Consider the effect on $\Cscr_{D}$ of an arbitrary 
positive-definite diagonal matrix transformation, i.e. an element of the 
3-dimensional abelian ``scale group" $Diag(3,R)^{+}$ (the identity component 
of
the diagonal subgroup  $Diag(3,R)$ of $GL(3,R)$ whose Lie algebra $diag(3,R)$
consists of the diagonal elements of $gl(3,R)$ ) which represents
 independent 
scalings of the standard basis vectors of $R^3$ or of the basis vectors of 
any 3-dimensional vector space.  Such a matrix may be represented in the form
$$\eqalign{
        e^{\bfbbeta}& =\hbox{diag}(e^{\beta^{1}},e^{\beta^{2}},e^{\beta^{3}})
       \quad     \in \ Diag(3,R)^{+} \cr
   \bfbeta&=\hbox{diag}(\beta^{1},\beta^{2},\beta^{3})\quad \in\ diag(3,R)
       \ , \cr }\eqno(2.11)$$
and its left action on $\Cscr_{D}$ via (2.5) is
$$\overline{n}^{(a)}= j_{\hbox{$e^{\bfbbeta}$}}(n^{(a)})=
      e^{2\beta^{a}-(\beta^{1}+\beta^{2} +\beta^{3})}n^{(a)}
\quad,\qquad\overline{a}= j_{\hbox{$e^{\bfbbeta}$}}(a)=e^{-\beta^{3}}a\ . 
\eqno(2.12)$$
The barred components represent another point in the same orbit as long as
the scale transformation is nonsingular.  However, if one takes a singular
limit a point on the boundary of an orbit can be reached resulting in a change
of Bianchi type.  This is called Lie algebra contraction [50].

When a given stratum consists of a family of orbits as is the case for
types VI$_{h\leq 0}$ and VII$_{h\geq 0}$, in order to arrive at a point of 
the
boundary of a given stratum not at the boundary of the starting orbit, one
must allow motion transversal to the orbits; such a motion is called a Lie
algebra deformation.  For these two Bianchi types, changing the parameter $h$
represents a Lie algebra deformation.   
For example, a type IV or V point of $\Cscr_{D}$ can
be reached from a type VI$_{h\neq 0}$ or type VII$_{h\neq 0}$ 
point only
by a deformation.

\begin{table} 
\setlength{\unitlength}{0.8cm}
\begin{picture}(12,7)(-1,0)

\put (-1,7){\line(1,0){15}}
\put (0,7.5){\makebox(0,0){Dim(${\cal O} \cap {\cal C}_D)$}}
\put (3,7.5){\makebox(0,0){Class A}}
\put (6.5,7.5){\makebox(0,0){Class B}}
\put (11,7.5){\makebox(0,0){Canonical $SAut_e(g) \subset SL(3,R)$
}}

\put (11,6.5){\makebox(0,0){$S0(3,R)$ \quad $SL(2,1)$
}}
\put (11,4){\makebox(0,0){$T_2 \times_s Ad^0_3$
}}
\put (11,1.5){\makebox(0,0){$T_2^T \times_s SL(2,3)$ \quad $T_2 \times_s SL(2,3)$
}}
\put (11,0){\makebox(0,0){$SL(3,R)$
}}

\put (0,6.5){\makebox(0,0){3}}
\put (0,4){\makebox(0,0){2}}
\put (0,1.5){\makebox(0,0){1}}
\put (0,0){\makebox(0,0){0}}

\put (3,6.5){\makebox(0,0){IX $\bullet\,\bullet\,\bullet$ VIII 
}}
\put (6.5,5){\makebox(0,0){VII$_{h>0}$ $\bullet\,\bullet\,\bullet$ VI$_{h<0}$ 
}}
\put (3,3){\makebox(0,0){VII$_0$ $\bullet\,\bullet\,\bullet$ VI$_0$ 
}}
\put (6.5,3){\makebox(0,0){IV 
}}
\put (3,1.5){\makebox(0,0){II 
}}
\put (6.5,1.5){\makebox(0,0){V 
}}
\put (3,0){\makebox(0,0){\ I 
}}

\put(2,6){\vector(0,-11){2.5}}
\put(4,6){\vector(0,-11){2.5}}
\put(5.5,4.5){\vector(-3,-1){3}}
\put(7.5,4.5){\vector(-3,-1){3}}
\put(3.5,6){\vector(-1,-2){1.25}}
\put(5.5,4.5){\vector(1,-1){1}}
\put(7.5,4.5){\vector(-1,-1){1}}
\put(1.85,2.75){\vector(1,-1){1}}
\put(4,2.75){\vector(-1,-1){1}}
\put(6.5,2.75){\vector(0,-1){1}}
\put(3,1.25){\vector(0,-1){1}}
\put(6.5,2.75){\vector(-3,-1){3}}
\put(6.5,1.25){\vector(-3,-1){3}}
\end{picture}
\medskip
\caption{
The ``reduced" specialization diagram describing the possible
Lie algebra contractions and deformations of the Bianchi types and their
orbit dimensions restricted to the standard diagonal form subspace $\Cscr_D$
of the space $\Cscr$ of possible structure constant tensors. Not shown are
direct paths between Bianchi types which may be connected through other
Bianchi types by the indirect paths shown.  The final column indicates 
the identity component of the canonical special automorphism matrix group
of each type, with notation explained in appendix B.
}
\end{table}

Apart from trivial permutations, each of the canonical points of $\Cscr_{D}$ 
may undergo such Lie algebra contractions and/or deformations to arrive at
canonical points lying in the same stratum or in lower-dimensional strata
at the boundary of the given stratum.  The various possibilities are
illustrated in Table 2 following MacCallum [39]. 
Each class B
Bianchi type has a corresponding class A limit obtained by the contraction
(V, IV) or deformation (VI$_{h}$, VII$_{h}$) $\quad a\rarrow 0$ shown in 
Table 2.
(Unfortunately the type II components arising from this contraction of the
canonical type IV components differ from the canonical type II components
by a permutation.)  By extending the scale group to the complex domain, one
may perform a rotation in the complex plane which directly connects the
canonical components of certain Bianchi types.  For example, the scaling
$\hbox{diag}(e^{i\theta},e^{i\theta},1)$ with $\theta\in[0,{\pi \over 2}]$
is a path connecting the canonical components of types VIII and IX, which
represent inequivalent real forms of the same complex Lie algebra.  The types
VII$_{h\neq 0}$ and VII$_{h\neq 0}$ also require analytic 
continuation of the
parameter $a$ as well. Such pairs are connected by horizontal dotted lines
in Table 2.

If the real scaling matrix $e^{\bfsbeta}$ is highly anisotropic, i.e. ``nearly
singular", its action on a canonical point of $\Cscr_{D}$ may simulate a Lie
algebra contraction on the space of functions on $\Cscr_{D}$:  the value
of a function of the structure constant tensor components at the image
point will approximately equal its value at the nearest point of the boundary. 
Thus under very anisotropic scalings, functions of a given Bianchi type
structure constant tensor approach those of a contracted type.  This is
a very useful way of viewing the behavior of highly anisotropic Bianchi
cosmologies, as will be described below.
 
    Equations (2.10) represent a $\Cscr_D$-parametrized Lie algebra   
$g_{_{\Cscr_D}}$ .  Using a trick involving the linear adjoint group,
one may realize this Lie algebra as the Lie algebra of left invariant vector
fields on a $\Cscr_D$-parametrized simply connected Lie group
$G_{_{\Cscr_D}}$, where the parametrization arises by introducing canonical
coordinates of the second kind with respect to the basis $e_{_{\Cscr_D}}$
of   $g_{_{\Cscr_D}}$ whose brackets are given by (2.10).  
These coordinates have range $R^3$ and are global for all Bianchi types
but type IX where the simply connected group manifold is instead $S^3$
and these coordinates form a local patch centered at the identity.  These
results are summarized in appendix A and illustrate the elegant consequences
of the first diagonalization referred to in the introduction.

     The result of this long digression is the $\Cscr_D$-parametrized simply
connected Lie group $G_{_{\Cscr_D}}$ which enables one to simultaneously 
describe
all 3-dimensional (simply connected) Lie groups.  One may next introduce
the $\Cscr_D$-parametrized left invariant Riemannian 3-manifold 
($G_{_{\Cscr_D}},
\hbox{\sl g}_{_{\Cscr_D}}$) or ``homogeneous Riemannian 3-space" with metric
$$\hbox{\sl g}_{_{\Cscr_D}}=g_{ab}\ \omega^a_{_{\Cscr_D}}\otimes\omega^b_
{_{\Cscr_D}}\ 
   ,\eqno(2.13)$$
where $e_{_{\Cscr_D}}$ and $\{\omega^a_{_{\Cscr_D}}\}$ are the
explicit $\Cscr_D$-parametrized fields given by formulas (A.9) and the
component matrix
$\hbox{\bf g}=(g_{ab})=g_{ab}\hbox{\bf e}^b{}_{a}$ with determinant 
$g=\det\hbox{\bf g}$
lies in the 6-dimensional space $\Mscr\subset GL(3,R)$ of component matrices of 
positive-definite inner products on $R^3$.  This space, through (2.13),
parametrizes the space $\Mscr_L(G_{_{\Cscr_D}})$ of left invariant metrics on 
the
Lie group $G_{_{\Cscr_D}}$. Since it is a bit awkward, the subscript $\Cscr_D$
will
usually be omitted in what follows.  Before moving on to the
$\Cscr_D$-parametrized spatially homogeneous spacetime, it pays to examine the 
curvature of the $\Cscr_D$-parametrized homogeneous Riemannian 3-space and the 
isometry classes
of the space of such Riemannian manifolds.  The latter question leads to the
second diagonalization mentioned in the introduction.

\noindent\hskip 14.97pt
     The isometry classes of $\Mscr_{L}(G)$ are its intersections with the
orbits of the diffeomorphism group $\Dscr(G)$ on the space of all smooth
Riemannian metrics on $G$.  Consider instead the largest subgroup of 
$\Dscr(G)$
which acts on the space $\Mscr_{L}(G)$, i.e. which maps all left invariant
metrics into left invariant metrics under the dragging along action.  This
is possible only if it maps the Lie algebra $g$ into itself and hence
the space
of all left invariant tensor fields into itself under dragging along.  The
orbits of this group on $\Mscr_{L}(G)$ should correspond to the isometry
classes of left invariant metrics; it is assumed that they do.

     The ``symmetry compatible subgroup" of $\Dscr(G)$ having this property,
already designated by $\Dscr(g)$ earlier in this section, is the
semidirect
product Lie group of translations and automorphisms of $G^{\ \ (1)}$
$$\Dscr(g)=L_{G}\times_s Aut(G)=R_G\times_s Aut(G)\ .\eqno(2.14)$$
The equality of the two semidirect products is connected with the adjoint
group $\ AD_{G}\ \subset$ $ Aut(G)$ of $G$, also called the group of inner 
automorphisms 
of $G$.  In addition to the effective left action of any Lie group $G$ on
itself
by left translation $L$ and inverse right translation $R^{-1}$ which are
commuting actions due to the associativity of the group multiplication, $G$
may act on itself on the left by inner automorphism:  $AD_{a}=L_{a}\circ
R^{-1}_{a}=R^{-1}_{a}\circ L_{a}$.  The image group $AD_{G}$ is a homomorphic
subgroup of the automorphism group of $G$ but is not necessarily isomorphic 
to $G$.  
They are isomorphic and the adjoint action effective when $AD_{a_0}$ is the
only transformation acting as the identity, i.e. when the center $C(G)=
\{ a\in G\relv AD_{a}=Id\}$ of $G$ is trivial (contains only
the identity $a_0$).  Returning to (2.l4), the fact that $AD^{-1}_{a}\circ
\alpha\in Aut(G)$ if $\alpha\in Aut(G)$ together with the identity
$L_{a}\circ \alpha=R_{a}\circ (AD^{-1}_{a}\circ \alpha)$ explains the
equality of the two semidirect products.

     This discussion may be repeated at the Lie algebra level for the Lie
algebra $\xc(g)$ which generates $\Dscr(g)$, a semidirect sum 
Lie subalgebra
of the Lie algebra $\xc(G)$ of smooth vector fields on $G$
$$\xc(g)=\s g \oplus_{s}\, aut(G)
   = g\oplus_s aut(G)\ .\eqno(2.15)$$
Here $aut(G)$ generates $Aut(G)$ and ${\rm ad}(G)=g - \s g 
\equiv\{ X^{a}(e_{a}- \s
e_{a})\relv(X^{a})\in R^{3}\}$ generates the adjoint group.

     When the group $\Dscr(g)$ acts on $g$ by dragging along, 
$L_{G}$ has no
action by definition, while $R_{G}$ and $AD_{G}$ have the same action,
inducing inner automorphisms of the Lie algebra:  $R^{-1}_{a}X=AD_{a}X
\equiv Ad(a)X\in g$ for $X\in g$.  The image subgroup $Ad(G)$ of
the general linear group of $g$ is called the linear adjoint group and 
when $G$ is connected coincides with the group $IAut(g)\subset 
Aut(g)$ of inner automorphisms of $g$  ; its matrix representation 
$Ad_{e}(g)$ with respect to a
basis $e$  of $g$ is exploited in appendix A.  Similarly by dragging 
along,
$Aut(G)$ induces the action on $g$  of the full group $Aut(g)$ of 
automorphisms of $g$ when $G$  is simply connected as is assumed here.  
Left invariant tensor fields undergo the transformation associated with the 
corresponding tensor representation of this group.
Similarly when $\xc(g)$ acts on $g$  by Lie derivation (define 
${\rm ad}(\xi)X=\sterling_{\xi}X$ for $\xi\in\xc(g)$ and $X\in g$ and 
let ${\rm ad}_{e}(\xi)$
be the matrix of ${\rm ad}(\xi)$ with respect to the basis $e$ of $g$), $\s g$
has no effect, while $g$ and ${\rm ad}(G)$ have the same action, 
inducing inner derivations of $g$ (the image Lie subalgebra $\hbox{ad}(g)
\subset aut(g)$), while $aut(g)$ induces the
action of the full Lie algebra of derivations $der(g)= 
aut(g)$ of $g$,
with matrix representation $aut_e(g)$. The Lie algebra $aut(g)$ generates 
the automorphisms of $g$ .

     Thus when $\Dscr(g)$ acts on the left invariant metric (2.13) by 
dragging along, the component matrix $\hbox{\bf g}$ undergoes the appropriate 
transformation law associated with the matrix automorphism group.  
If $\hbox{\bf A}\in Aut_{e}(g)$ is
an induced matrix automorphism of $g$ , this transformation law is
$$\hbox{\bf g}\in\Mscr\qquad\rarrow\qquad f_{\hbox{\bf A}}(\hbox{\bf g})=
\hbox{\bf A}^{-1T}\hbox{\bf g}\hbox{\bf A}^{-1}\ .\eqno(2.16)$$
The orbits of this action of $Aut_{e}(g)$ on $\Mscr$ through the 
correspondence
(2.13) represent the isometry classes of left invariant metrics.  However,
just as the space $\Cscr$ could be reduced to its essential structure by
diagonalization, here too the ``offdiagonal" metric matrix variables are
superfluous and all the essential information is carried by the diagonal
submanifold $\Mscr_{D}$ of $\Mscr$, assuming that $e$ is a basis of $g$
whose structure constant tensor components belong to $\Cscr_{D}$.  This 
submanifold $\Mscr_{D}$, like the space $\Cscr_{D}$, is also a ``slice" for 
the natural
action of the orthogonal group on the full space.  (The reduction of
$\Mscr\times\Cscr$ to $\Mscr_{D}\times\Cscr_{D}$ is a consequence of the well 
known fact that
one can always simultaneously diagonalize two real symmetric matrices by an
orthogonal transformation.)  In
fact $\Mscr_{D}$ is a ``slice" for the action (2.16) of any 3-dimensional
subgroup ${\A G}\subset GL(3,R)$  whose matrix Lie algebra $\A g$ has a basis 
$\{\bfkappa_{a}\}$ with the following property:  
for each 
cyclic permutation $(a,b,c)$ of $(1,2,3)$, the matrix $ \bfkappa_{a}$ 
belongs to $\hbox{span}\ \{ \hbox{\bf e}^b{}_{c},\hbox{\bf e}^c{}_{
\ b}\}$, where 
$\{ \hbox{\bf e}^{a}{}_{b}\}$ is the natural
basis of $gl(3,R)$ already introduced above.  In appendix B, the matrix
automorphism group is described for the $\Cscr_{D}$-parametrized Lie algebra 
$g$ .
It always contains such a subgroup $\A G$, which may be used to map a general
point of $\Mscr$ to the diagonal submanifold $\Mscr_{D}$.  It therefore
suffices to consider those automorphisms which map $\Mscr_{D}$ into itself
to determine the isometry classes.  Since $\Mscr_{D}$ is mapped into itself
by all permutations and diagonal transformations, it suffices to consider
elements of $Aut_{e}(g)$ of this type.

     $\Mscr_{D}$ consists of all diagonal matrices with positive entries and
clearly coincides with the scale group $Diag(3,R)^+$ as a submanifold of
$GL(3,R)$ .  They are best identified, however, in terms of the simply
transitive action of $Diag(3,R)^{+}$ on $\Mscr_{D}$ using the identity matrix
$\hbox{\bf 1}\in\Mscr_{D}$ as a reference point, as described at the beginning 
of this section.  The abelian group $Diag(3,R)^+$ is most naturally 
parametrized
by its Lie algebra $diag(3,R)$ which in turn parametrizes $\Mscr_{D}$
$$\vcenter{\halign{  $\rt{#}$&$\lft{#}$\qquad
         &$\lft{#}$\qquad  &$\lft{#}$\cr
&\bfbeta=\hbox{diag}(\beta^{1},\beta^{2},\beta^{3})\ ,
&e^{\bfsbeta} =\hbox{diag}(e^{\beta^1},
  e^{\beta^2},e^{\beta^3})\ ,
&\hbox{\bf g}^{\prime} =f^{-1}_{\hbox{$e$}^{\bfsbeta}}(\hbox{\bf 1}) 
=e^{2\bfsbeta}\cr
&\inr&\inr&\inr\cr 
&diag(3,R)\qquad\quad\supset &Diag(3,R)^+\qquad= &\Mscr_{D}\qquad\ .\cr
}}\eqno(2.17)$$
The prime on $\hbox{\bf g}^{\prime}$ serves as a reminder of its diagonality,
while the right action $f^{-1}$ is used to conform with convention.  Through
(2.17) the single matrix $\bfbeta$ simultaneously represents three different
diagonal matrices.  The  special scale group $SDiag(3,R)^+ =Diag(3,R)^+
\cap SL(3,R)$ with Lie algebra $sdiag(3,R) = \{\bfbeta\in diag
(3,R)\relv \hbox{Tr}\ \bfbeta =0\}$ is related in a similar way to
the unimodular submanifold $\overline{\Mscr}_D =\Mscr_D\cap SL(3,R)$;
the following notation proves convenient  
$$\eqalign{\bfbeta&= \half\ln\hbox{\bf g}^{\prime}
=\beta^0\hbox{\bf1}+{\A {\bfbeta}}\ ,
\qquad\beta^0={1\over6}\ln g^{\prime}
={1\over3}\hbox{Tr}\ \bfbeta\ ,\cr
\beta^a&=\half\ln g_{aa}^{\prime}=\beta^0+ \A {\beta}^a\ ,
\qquad\beta^{ab}\equiv\beta^a-\beta^b= \A {\beta}^a-\A {\beta}^b
=\half\ln(g^{\prime}_{aa}/g^{\prime}_{bb})\ .\cr}\eqno(2.18)$$
Misner [15] introduced a basis of $diag(3,R)$ and $sdiag(3,R)$ which
is
orthonormal with respect to the inner product $\left<\left<,\right>\right>
_{_{DW}}={1\over6}\left<,\right>_{_{DW}}$
on $gl(3,R)$, where the DeWitt inner product and trace inner product on
$gl(3,R)$ are defined by
$$\eqalign{\left<\hbox{\bf A},\hbox{\bf B}\right>_{_{DW}} &=\hbox{Tr {\bf AB}}-
\hbox{Tr {\bf A}}\ \hbox{Tr {\bf B}}\ ,\qquad\left<\hbox{\bf A},
\hbox{\bf B}\right>
= \hbox{Tr {\bf AB}}\ ,\qquad \hbox{\bf A,B}\in
gl(3,R)\cr
\left<\left<,\right>\right>&= {1\over6}\left<,\right>\ .\cr}\eqno(2.19)$$
This basis and the corresponding parametrization of $diag(3,R)$ and 
$sdiag(3,R)$ are given by
$$\eqalign{\bfbeta&= \beta^A\hbox{\bf e}_A =\beta^0\hbox{\bf e}_0 +\beta^+
\hbox{\bf e}_++ \beta^-\hbox{\bf e}_-\cr
\{\hbox{\bf e}_0, \hbox{\bf e}_+ ,\hbox{\bf e}_-\} &=
\{\hbox{\bf 1},\hbox{diag}(1,1,-2),\sqrt3\,\hbox{diag}(1,-1,0)
\}\cr
(\eta_{AB}) &=(\left<\left<\hbox{\bf e}_A,\hbox{\bf e}_B\right>\right>_{_{DW}})
=\hbox{diag}(-1,1,1) =(\eta^{AB})\cr
{\A {\beta}}\null^1 &=\beta^++\sqrt3\,\beta^-\ ,\quad
{\A {\beta}}\null^2  =\beta^+-\sqrt3\,\beta^-\ ,\quad
{\A {\beta}}\null^3  =-2\beta^+\cr
\beta^{23} &=3\beta^+ -\sqrt3\,\beta^-\ ,\quad
\beta^{31}  =-3\beta^+ -\sqrt3\,\beta^-\ ,\quad
\beta^{12}  =2\sqrt3\,\beta^-\ .\cr}\eqno(2.20)$$
These may be generalized by the definitions
$$ \eqalign{\beta^+_a &= -\half{\A {\beta}}\null^a\ ,\qquad\beta^-_a = 
     (4\sqrt3)^{-1}\epsilon_{abc}\beta^{bc}\ , \cr
  {\A {\bfbeta}} &= \beta^+_a \hbox{\bf e}_{a+} + \beta^-_a \hbox{\bf e}_{a-}
    \qquad (\hbox{no sum on $a$})\ ,\cr}\eqno(2.21)$$
with $\beta^{\pm} =\beta^{\pm}_{3}$ and the others obtained
by cyclic permutation of indices.  For each cyclic permutation $(a,b,c)$ of
$(1,2,3)$, the Taub submanifold $\Mscr_{T(a)} = \{ \hbox{\bf g}
^{\prime}\in \Mscr
_{D} \relv g^{\prime}_{bb} = g^{\prime}_{cc}\}$ and its unimodular 
submanifold $\overline{\Mscr}_{T(a)}\subset  \overline{\Mscr}_{D}$ may be 
equivalently defined by $\beta^{bc}=0$
or $\beta^{-}_{a}=0$.  They intersect at the isotropic submanifold
$\Mscr_{I}$ and $\overline{\Mscr}_I=\{ \hbox{\bf 1} \}$ 
respectively, for which
$\beta^{+} = \beta^{-} = 0$.  Since $\hbox{\bf g}^{\prime}= e^{2\beta^0}
e^{2\hat{\bfbbeta}} $, translation
along $\beta^0$ represents a conformal rescaling of the metric (2.13) under
which all curvatures scale by a factor $e^{q\beta^0}$ where q is an 
appropriate
dimension.  Thus the nontrivial information about curvature is associated
with the conformal submanifold $\overline{\Mscr}_{D}$ \quad (namely $
\beta^0=0$).

\begin{figure}[t!] 
\begin{center}
\includegraphics[width=.9\textwidth]{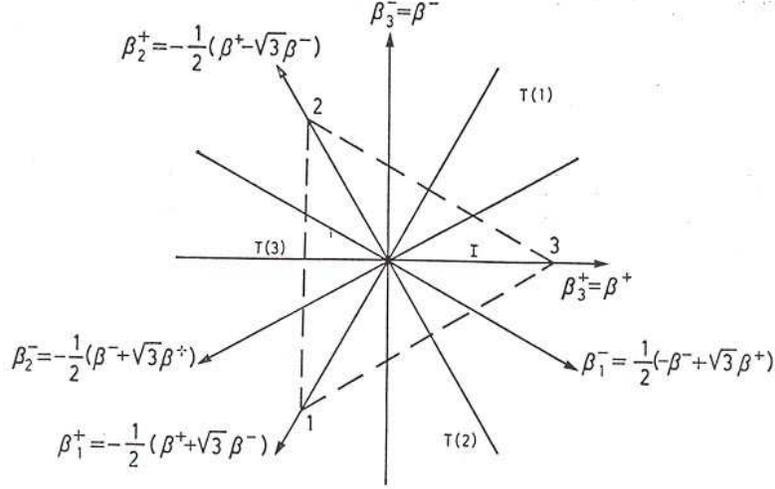}
\end{center}
\caption{The $\beta^+\beta^-$ plane:
$sdiag(3,R)\sim SDiag(3,R)^+\sim\overline{\Mscr}_D$.  Shown are the
three pairs of orthogonal axes related by rotations of angle $2\pi/3$.  The 
$\beta^+_a$-axis is the projection of the Taub submanifold $\Mscr_{T(a)}$ to
$\overline{\Mscr}_D$, while the origin represents the projection of the 
isotropic submanifold $\Mscr_I$.  Lines parallel to the sides of the triangle
are associated with constant values of the three diagonal components of the
conformal metric matrix while lines parallel to the bisectors of the vertex
angles are associated with constant values of their ratios.  Reflections and
permutations of the basis $e$ act on $\overline{\Mscr}_D$ as the symmetry
group of the equilateral triangle of this figure.}
\end{figure}

The $\beta^{+}\beta^{-}$ plane, i.e. $sdiag(3,R) \sim SDiag(3,R)
\sim\overline{\Mscr}_{D}$ is illustrated in Figure 2, indicating the Taub
and
isotropic submanifolds and each of the pairs of coordinate axes associated
with the three coordinate systems $\{\beta^{+}_{a} ,\beta^{-}_{a}
\}$.  Each pair of coordinate vectors, namely $\{
\hbox{\bf e}_{a+} ,
\hbox{\bf e}_{a-}\}$, is orthonormal with respect to the inner 
product $\langle\langle , \rangle\rangle$. 
Interpreting the $\beta^{+}\beta^{-}$ plane as $\overline{\Mscr}_{D}$, a 
cyclic
permutation of the basis $e$ of $g$ leads through (2.13) and (2.16) to 
a rotation
of the coordinate axes by $\pm 2\pi/3$, while a transposition of two basis
elements, say $e_{b}$ and $e_{c}$, leads to a reflection about the Taub
submanifold $\overline{\Mscr}_{T(a)}$, where (a,b,c) is a cyclic permutation 
of $(1,2,3)$ . 
The action of the discrete group of permutations on $\overline{\Mscr}_{D}$ 
thus coincides
with the symmetry group of the equilateral triangle shown in Figure 2, whose
sides are parallel to the constant value lines of the $\beta^{+}
_{a}$-coordinates and whose orthogonal bisectors are the Taub submanifolds.  
(Note that rotations about one of the frame vectors by $\pm\pi/2$ have the
same effect on the $\beta^+\beta^-$ plane as transpositions.)  The
translations of the $\beta^{+}\beta^{-}$ plane correspond to the action on
$\overline{\Mscr}_{D}$ of the special scale group.

The Lie algebra contractions of the space $\Cscr_{D}$ arising from singular
limits of the action of the scale group on this space are now easily 
described.  Let $\{\beta=s \hbox{\bf b}\relv s\in (-\infty, 0]
\}$ be a ray
from the origin of $diag(3,R)$ parametrized by $s$ and extend all
of the subscript and superscript notation of (2.17)-(2.21) to the constant
diagonal matrix {\bf b} ; then (2.12) becomes
$$ j^{-1}_{\hbox{$e^{\bfbbeta}$}} (n^{(a)},a) = (e^{s(b^0+4b^+_a)}n^{(a)},
    e^{s(b^0-2b^+_3)}a)\ .\eqno(2.22)$$
If $b^0\neq 0$, then one might as well set $b^0=1$.
In order that this have a finite limit as $s=\beta^{0}\rarrow -\infty$ leading
to a singular scale transformation $(\det e^{\bfsbeta}=e^{3\beta^{0}}\rarrow 0)$
which therefore induces a Lie algebra contraction, the following inequalities
must be satisfied when the corresponding structure constant tensor component
is nonvanishing
$$ n^{(a)} \neq 0:\quad b^+_a \geq -{1\over4}; \qquad a\neq 0: 
   \quad    b^+_3 \leq \half\ .\eqno(2.23)$$
These inequalities are illustrated in Figure 3.  The label of a given
dashed line indicates the structure constant tensor component which
remains fixed under the scaling (2.23) associated with the points of the
line, all points to the origin side of the line leading to a limit where
that structure constant tensor component goes to zero and all points to
the other side not leading to a finite limit if that component is nonzero.
Thus all points in the interior of the triangle 123 lead to the abelian limit
$(0,0,0,0)$, vertex 1:  $(0,n^{(2)},n^{(3)},0)$, vertex 2:  $(n^{(1)},0,
n^{(3)},0)$, vertex 3:  $(n^{(1)},n^{(2)},0,a)$, open side 23:  $(n^{(1)},
0,0,0)$, open side 31:  $(0,n^{(2)},0,0)$ and open side 12:  
$(0,0,n^{(3)},0)$.

\begin{figure}[t!] 
\begin{center}
\includegraphics[width=.9\textwidth]{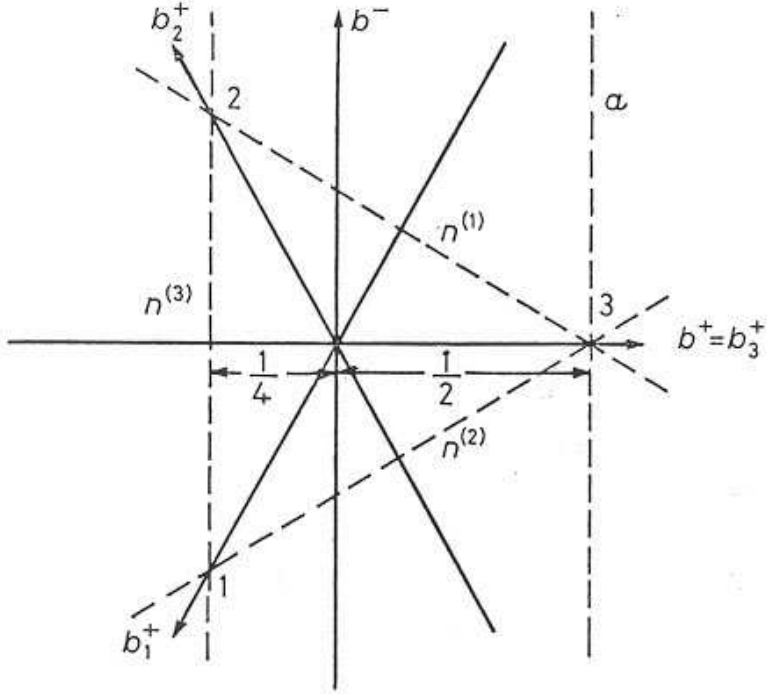}
\end{center}
\caption{The parameter space for the 
family of Lie algebra contractions of the space $\Cscr_D$.}
\end{figure}

On the other hand when $b^0=0$, at least one of the components of $e^{
\bfsbeta}$ must go to infinity as $s\rarrow -\infty$ so a finite limit
can result only if one of the components of {\bf n} is zero.  If $(a,b,c)$
is a cyclic permutation of (1,2,3) and $n^{(a)}=0$, then the limit $s\rarrow
-\infty$ of (2.22) will be finite only if $b^+_b\geq 0$ and $b^+_c\geq 0$,
which is the sector of the plane between the positive $b^+_c$ and $b^+_c$
axes, the limit being (0,0,0,0) between the axes, but with $j_{
\hbox{$e^{\bfbbeta}$}}(n^{(b)}, n^{(c)}, a) $ having the limit $(n^{(b)},0,0)$
on the $b^+_b$-axis and $(0,n^{(c)},0)$ on the $b^+_c$-axis. This class
of Lie algebra contractions might be called ``pure anisotropy" contractions.
  
Returning now to the question of isomorphism classes within the diagonal
submanifold $\Mscr_{D}$, namely the orbits of the action of the diagonal
and permutation automorphisms on $\Mscr_{D}$, one has four different cases
corresponding to the four categories of Table 2.  Modulo discrete
automorphisms, the diagonal automorphisms of the $\Cscr_D$-parametrized Lie
algebra are described for each of these categories in Appendix B.  For
the first category only permutation and reflection automorphisms act on
$\Mscr_{D}$ so $\Mscr_{D}$ itself locally parametrizes the space of
automorphism group orbits on $\Mscr$.  (For the canonical type IX case
the six sectors into which the three $\beta^{+}_{a}$ axes divide the
$\beta^{+}\beta^{-}$ plane are all isometric for a given value of $\beta^{0}$,
but in the canonical type VIII case only reflection about $\overline{\Mscr}
_{T(3)}$
connects isometric points.)  For the second category of Table 2,  there exists
a diagonal automorphism subgroup generated by the matrix $\hbox{\bf I}^{(3)}=
\hbox{diag}(1,1,0)$ when $n^{(3)}=0$, leading to translations along
$\beta^{+}=\beta^{+}_{3}$ in the $\beta^{+}\beta^{-}$ plane, so $\beta^{-}=
\beta^{-}_{3}$ together with $\beta^{0}$ locally parametrize the orbit space. 
The result for the other components $(n^{(3)}\neq 0)$ of $\Cscr_D$
belonging to this category may be obtained by cyclic permutation.  For the
third category automorphisms induce translations along both $\beta^{+}$ and
$\beta^{-}$ so all points of the $\beta^{+}\beta^{-}$ plane are equivalent
and $\beta^{0}$ alone parametrizes the orbit space, while in the abelian case
$\Mscr$ consists of a single orbit.

For the upper two categories of Table 2, the generic points of $\Mscr_{D}$
belong to an orbit on $\Mscr$ having three dimensions transversal to 
$\Mscr_{D}$ and at most a discrete isotropy group.  However, on certain
submanifolds of $\Mscr_{D}$ and for certain Bianchi types the full orbit
dimension decreases and only one or no directions remain transversal to
$\Mscr_{D}$.  At these points the isotropy group has dimension greater than
zero corresponding to additional symmetries of the metric (2.13).  This
occurs at the Taub submanifold $\Mscr_{T(a)}$ when $n^{(b)}=n^{(c)}$ ($(a,b,c)
$ is a cyclic permutation of $(1,2,3)$) corresponding to local rotational 
symmetry
(necessarily the index $a=3$ in the class B case) and at the isotropic 
submanifold
$\Mscr_{I}$ when $n^{(1)}=n^{(2)}=n^{(3)}$ corresponding to isotropy.  For
the lower two categories the isotropy group has generic dimension greater than
zero so there are always additional symmetries.  However, the Taub 
submanifolds
are still relevant to spacetime symmetries.  The choice of canonical 
components
for Bianchi type II was made so that $\Mscr_{T(3)}$ is associated with
additional spacetime symmetry for all canonical points of $\Cscr_{D}$.  This 
is discussed in greater detail elsewhere [43].
For the noncanonical
points of $\Cscr_D$ the submanifolds relevant to additional symmetry change 
as described below.

     The discussion of isometry classes and additional symmetries is not just
an interesting aside, but is important for appreciating the symmetries of the
scalar curvature $R$ of the metric (2.13).  This function on $\Mscr\times 
\Cscr_{D}$ is a scalar under a change of basis $e$ of $g$ and hence is
invariant under the action of the matrix automorphism group on $\Mscr$ alone,
having a constant value on each orbit.  Using standard formulas one may easily
evaluate the components of the connection of the metric (2.13), raising and
lowering all indices with the component matrices {\bf g} and {\bf g}$^{-1}$
\quad
$$ \nabla_{\hbox{$e_a$}}e_b = \Gammait^c{}_{ab} e_c\ , \quad \Gammait^c{}_{ab} 
= \half C^c{}_{ab} + C_{(a\hskip 5pt b)}^{\hskip 7pt c}\ .\eqno(2.24)$$
Introducing the unit alternating (pseudo-)tensor $\eta_{abc} = g^{1\over2}
\epsilon_{abc}$ and the two matrices $\hbox{\bf m}=g^{-{1\over2}}\hbox{\bf ng}$ and 
$A=a_{c}\eta
^{ca}_{\hskip 6pt b}\hbox{\bf e}^{b}{}_{a}$, the Ricci tensor and scalar curvature 
of this
connection are then found to have the following expressions
$$ \eqalign{\hbox{\bf R} &= R^a{}_{b} \hbox{\bf e}^b{}_{a} =
   2\hbox{\bf m}^2 -\hbox{\bf m}\hbox{Tr}\ \hbox{\bf m}
  -\hbox{\bf 1}(\hbox{Tr}\ \hbox{\bf m}^2 -\half
  \hbox{Tr}^2 \hbox{\bf m} + 2a_ca^c) + [\hbox{\bf m,A}] \cr
  R &= \hbox{Tr}\ \hbox{\bf R} =-(\hbox{Tr}\ \hbox{\bf m}^2
  -\half \hbox{Tr}^2\hbox{\bf m}) -6a_ca^c\ ,\cr}\eqno(2.25)$$
where the conventions of Misner, Thorne and Wheeler [18] 
are followed for curvature
tensor definitions.  Assuming as always that $C^{a}{}_{bc}\in \Cscr_{D}$, 
these
are $\Cscr_{D}$-parametrized functions on $\Mscr$.  Note that for 
$\hbox{\bf g}\in\Mscr_{D}$
the Ricci tensor component matrix is diagonal except for the last term which
contributes a 12 (and 21) component in the class B case.  In the class A
case $e$ is then an orthogonal frame of Ricci eigenvectors, while linear
combinations of $e_{1}$ and $e_{2}$ must be taken to obtain such a frame in
the class B case, leading to structure constant tensor components not
belonging to $\Cscr_{D}$.

     One may also introduce a potential function and several 1-forms on 
$\Mscr$ which may be interpreted as force fields
$$ \eqalign{ U_G &= -g^{1\over2}R\cr
  G&= -g^{1\over2}G^{ab}dg_{ab} = -g^{1\over2}(R^{ab} - \half Rg^{ab})
    dg_{ab}\cr
  Q&= Q^{ab}dg_{ab} = 2g^{1\over2}(a^cC^{(a \hskip 7pt b)}_{\hskip 8pt c}
     -2a^aa^b)dg_{ab} \cr
   &= 2g^{1\over2}[\eta^{(a}_{\hskip 6pt cd} m^{b)c}a^d
    -3(a^aa^b-{1\over3}g^{ab}a_ca^c)]dg_{ab}\cr
  G&=-dU_G + Q\ .\cr}\eqno(2.26)$$
The scalar curvature potential function $U_{G}$ serves as a potential for the
Einstein force field $G$ in the class A case where $Q$ vanishes, but in the
class B case the Einstein force field has a nonpotential component $Q$ which
generically satisfies $Q \neq 0 \neq dQ$.  As a $\Cscr_{D}
$-parametrized function on $\Mscr_{D}$, the scalar curvature potential is
given explicitly by
$$ \vcenter{\halign{$\rt{#}$&$\ctr{#}$&$\lft{#}$\cr
U_G &\null=\null& e^{\beta^0} (V^{\ast}+6a^2e^{4\beta^+})\cr
 V^{\ast}&\null=\null &\half\sum\nolimits_{a=1}^3\, (n^{(a)})^2e^{-8\beta^+
   _a}- [n^{(2)}n^{(3)}e^{4\beta^+_1} +n^{(3)}n^{(1)}e^{4\beta^+_2}
  +n^{(1)}n^{(2)}e^{4\beta^+_3}]\cr &\null=\null&
2e^{4\beta^+}[\half(n^{(1)}e^{2\sqrt3\beta^-}
  -n^{(2)}e^{-2\sqrt3\beta^-})]^2\cr
  &&\quad-2n^{(3)}e^{-2\beta^+}[\half(n^{(1)}
  e^{2\sqrt3\beta^-} +n^{(2)}e^{-2\sqrt3\beta^-})]
  +\half(n^{(3)})^2e^{-8\beta^+}\ .\cr}}\eqno(2.27)$$
The values of the curvature function $V^{\ast}$ (a scalar density of weight
$2\over3$ which may be expressed in the form $V^{\ast}=g^{-{1\over6}}
\Gscr^{-1}_{abcd}n^{ab}n^{cd}$ using (2.40))  
at the canonical points of $\Cscr_{D}$ are
$$\vcenter{ \halign{\rt{#}&$\lft{#}$&$\lft{#}$\cr
  IX/VIII:\quad& 2e^{4\beta^+}\hbox{sinh}^22\sqrt3\beta^- 
    \mp 2e^{-2\beta^+}\hbox{cosh}\ 2\sqrt3\beta^- +\half e^{-8\beta^+} &\cr
  VII:\quad& 2e^{4\beta^+}\hbox{sinh}^22\sqrt3\beta^-&\hskip -100pt 
  \hbox{VI:}\quad 2e^{4\beta^+}\hbox{cosh}^22\sqrt3\beta^-\cr
  II:\quad& \half e^{-8\beta^+}&\hskip -100pt \hbox{IV:}\quad\half
      e^{-8\beta^+_1}\cr
  I,V:\quad& 0\ .& \cr}}\eqno(2.28)$$

Suppose $Y$ is a function on $\Mscr\times \Cscr$ which is a scalar
under a change of basis $e$ of $g$ and therefore satisfies 
$Y(\hbox{\bf g},C^{a}{}_{bc})=
Y(f_{\hbox{\bf A}}(\hbox{\bf g}),j_{\hbox{\bf A}}(C^{a}{}_{bc}))$ or 
equivalently $Y(f_{\hbox{\bf A}}^{-1}(\hbox{\bf g}),
C^{a}{}_{bc}) =Y(\hbox{\bf g}, j_{\hbox{\bf A}}(C^a{}_{bc}))$.  
Focussing now on $\Mscr_{D}\times \Cscr_{D}$ and letting $j_{\hbox{\bf A}}$
be one of the Lie algebra contractions (2.22)-(2.23), one sees that the effect
on the function $Y$ of such a contraction is equivalent to an infinite
translation of $\Mscr_D\sim diag(3,R)\sim R^3$.
The value of the function for the original point $C^a{}_{bc}$ therefore 
approaches its value for
the contracted points as one approaches infinity in $\Mscr_D \sim
diag(3,R)\sim R^3$ in the negative $\beta^0$ direction. 
$e^{-2\beta^0} V^{\ast}$ is such a scalar function and hence as one approaches 
infinity
in the $\beta^+\beta^-$ plane along the directions parametrized in Figure 3, 
the density $V^{\ast}$ approaches a rescaled version of the potential of the
corresponding contracted points of $\Cscr_D$.  Furthermore the difference 
between
$V^{\ast}$ and its contracted value at the same point of the $\beta^+\beta^-$ 
plane as one gets far from the origin becomes very small compared to the value
itself.
 
\begin{figure}[p!] 
\begin{center}
\includegraphics[width=.9\textwidth]{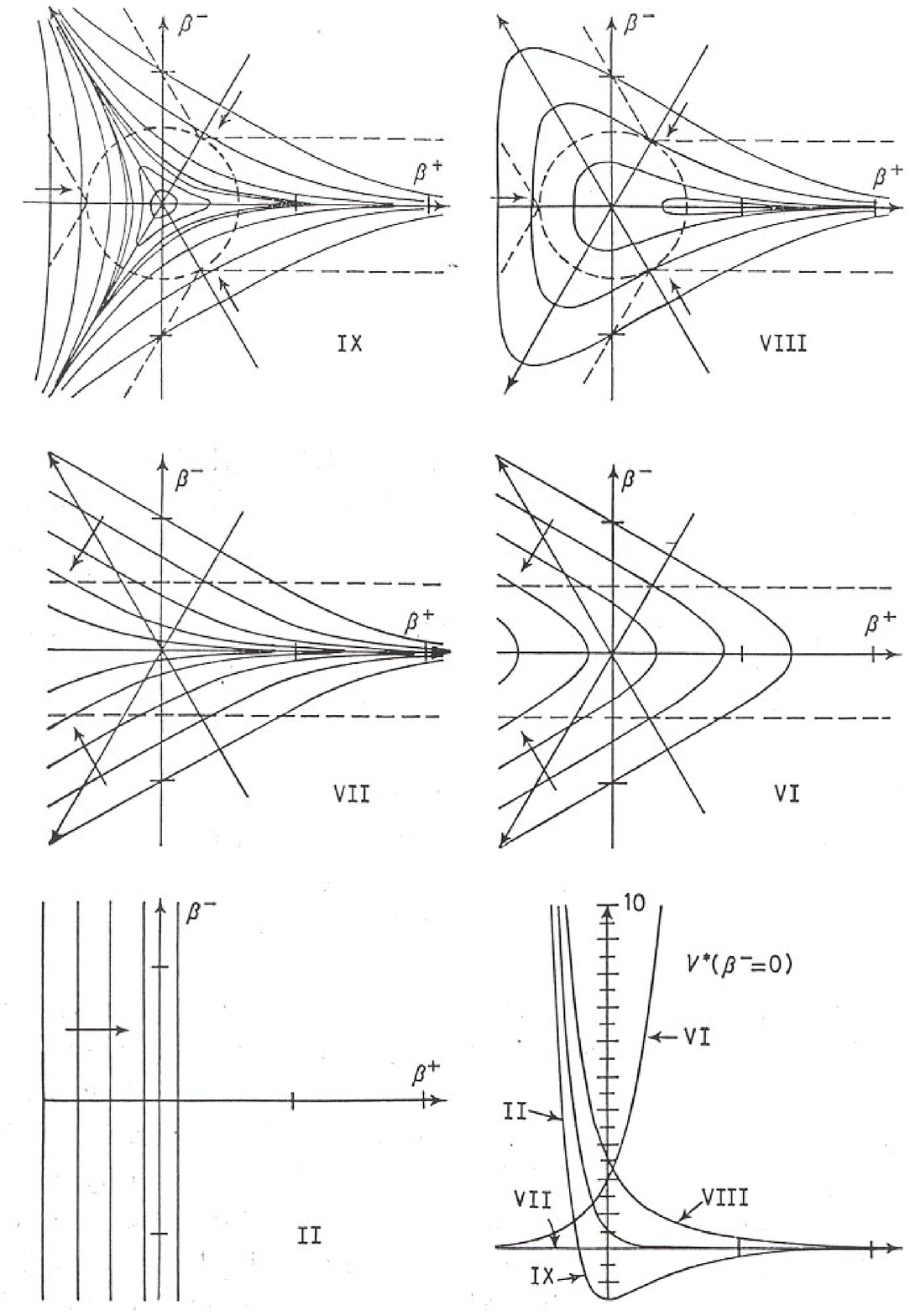}
\end{center}
\end{figure}
\begin{figure}[b]
\caption{
Suggestive contours of the 
potential function $V^{\ast}$ on the $\beta^+\beta^-$ plane are shown for the
canonical points of $\Cscr_D$ (not shown is the type IV case which is related
to type II by an active rotation of $4\pi/3$) and values of $V^{\ast}$ are 
plotted for
the Taub submanifold $\overline{\Mscr}_{T(3)}$, namely the $\beta^+$-axis.
Unit distances are marked on
the coordinate axes.  Arrows indicate the direction of the associated force 
field, pointing toward directions along which the potential decreases.  Note
that $V^{\ast}$ is nonnegative except in the type IX case where the contours
are closed for $V^{\ast}<0$ and open for $V^{\ast}\geq0$, the minimum value
-1.5 occurring at the origin or isotropic submanifold $\overline{\Mscr}_I$.  
}
\end{figure}

     This can be seen in the diagrams of Figure 4 which show suggestive
contours of the potential $V^{\ast}$ for canonical points of $\Cscr_D$. 
Contours of 
the same five function values are shown for each type, together with two 
additional closed contours for the type IX region where $V^{\ast}$ is negative.
As one proceeds
along any of the positive $\beta^{+}_{a}$ axes in the type IX case or the
$\beta^+$ axis in the type VIII case, the potential quickly approaches that
of a permutation of the canonical type VII potential, while along the
positive $\beta^+_1$ and $\beta^+_2$ axes the type VIII potential
approaches a permutation of the type VI potential.  For directions in
between these three positive axes the potential quickly approaches that
of a permutation of the type II potential.  Similarly the type VII
and VI potentials approach permutations of the type II potential for
directions between the positive $\beta^+$ and $\beta^{+}_{1}$ axes and the
positive $\beta^+$ and $\beta^{+}_{2}$ axes, but type I in the remaining
sector.  Finally the type II potential approaches type I along all
directions in the positive $\beta^+$ half plane.  Note further that for all
the nonsemisimple types the potential $V^{\ast}$ simply scales under 
translation
along $\beta^+$, so the contours are simply translates of each other, a 
consequence of the existence of the additional diagonal automorphism generated 
by the matrix $\hbox{\bf I}^{(3)}\equiv \hbox{diag}(1,1,0)$, except for
type II where the matrix is instead diag$(1,1,2)$.  The
reflection and permutation symmetries of all of the potentials reflect the
existence of discrete symmetries.  The reflection
symmetry about the $\beta^+$ axis for all types but IV is connected with a
discrete automorphism whose existence motivated the choice of canonical type 
II components. 

\begin{figure}[t!] 
\begin{center}
\includegraphics[width=.9\textwidth]{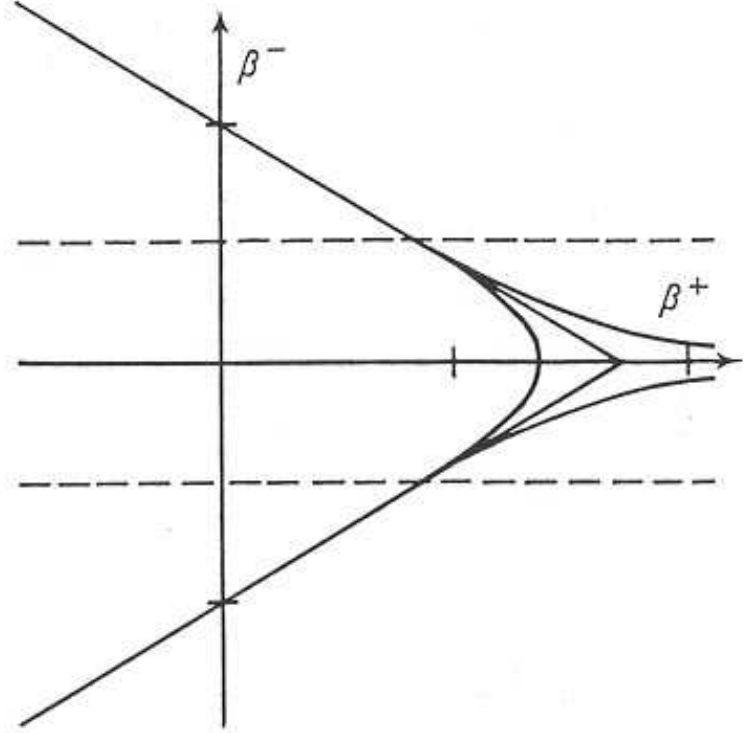}
\end{center}
\caption{
Open and closed channel 
contours and their asymptotes.}
\end{figure}

The dashed straight lines in Figure 4 indicate ``channels" of width 1 outside
of which the contours are essentially the same as the corresponding asymptotic 
Bianchi type II potential (the difference becoming exponentially small with
distance from the origin).  These channels themselves are either open or
closed; outside of the dashed circles of diameter 1 in  the type IX and type
VIII figures, the open and closed channels are essentially the same of the 
corresponding type VII and VI channels respectively, where these latter
channels are rotated by $\pm {2\pi/3}$ for comparison.  (Corresponding 
contours
are separated by a distance of less that .01 as one exits the circle and the
difference decreases exponentially with distance from the origin.) Figure 5 
shows the type VII and VI contours of the same function value together with
the asymptotic type II contours of the same function value, showing that the
deviation of the open and closed channel contours from the type II asymptotes   
only becomes important within the channel itself.  The only region of the
type IX and VIII potentials which is essentially different from those
of the remaining
Bianchi types (arising from them by contraction) is the interior of the 
dashed circle which occurs at the intersections of the three channels; 
similarly only the channels of the type VII and VI potentials 
are different from the potentials of the contracted types I and II.  

The potentials for the noncanonical points of $\Cscr_D$ are obtained merely
by translating the origin of coordinates in the $\beta^+\beta^-$ plane and
rescaling the potential.  Let $C^a{}_{bc}(can)\sim (\hbox{\bf n}(can),
a_a(can))$ be the canonical point of $\Cscr_D$ in the same orbit as
$C^a{}_{bc}\sim(\hbox{\bf n},a_a)\in\Cscr_D$.  One can then define the
nonsingular matrix $\bff{\gamma}=\gamma^A\hbox{\bf e}_A\in diag(3,R)$ 
by 
$$C^a{}_{bc} \equiv j^{-1}_{\hbox{$\bfs{\Gammait}$}}(C^a{}_{bc}(can))\ ,\quad 
\bff{\Gammait}\in Diag(3,R)\ ,\quad e^{\bfs{\gamma}}\equiv \hbox{diag}(|
\Gammait^1
{}_{1}|, |\Gammait^2{}_{2}|, |\Gammait^3{}_{3}|)\ .
\eqno(2.29)$$
Then one has the identity
$$ V^{\ast}(\hbox{\bf n}, \beta^{\pm}) = e^{2\gamma^0}V^{\ast}(\hbox{\bf n}
(can), \beta^{\pm}-\gamma^{\pm}) \ ,\eqno(2.30)$$
showing that $V^{\ast}$ is obtained from its canonical value by an active
translation of the $\beta^+\beta^-$ plane by $(\gamma^+, \gamma^-)$ and a
rescaling to its function values, leaving the shape of its contours unchanged.
A Lie algebra contraction then corresponds to an infinite translation.
All of the Lie algebra contractions parametrized by Figure 3 act on the 
potentials of Figure 4 to reduce the more complicated
ones to successively simpler ones. 	 

     Having exhausted the essential points regarding homogeneous 3-spaces,
the discussion may proceed to the spacetime level, introducing the 
$\Cscr_D$-parametrized spacetime $(M_{_{\Cscr_D}}, 
   ^4\hbox{\sl g}_{_{\Cscr_D}})\ .$  The spacetime manifold
is $M=R\times G_{_{\Cscr_D}}$, with the natural coordinate $t$ on the real 
line $R$
parametrizing the 1-parameter family of orbits of the natural left action
of $G_{\Cscr_D}$ on $M_{\Cscr_D}$, namely $t$-independent left translation
 of each copy
of $G_{\Cscr_D}$ in the product manifold.  The copies of $R$ in 
$M_{\Cscr_D}$ are the
$t$-lines which are interpreted as the normal geodesics to the family of 
orbits,
with $t$ coinciding with the proper time along these geodesics.  The spacetime
metric may therefore be written in the following form referred to as
synchronous gauge (zero shift and unit lapse)
$$ ^4\hbox{\sl g}_{_{\Cscr_D}} = -dt\otimes dt + g_{ab}(t)\,\omega^a
  _{_{\Cscr_D}}\otimes\omega^b_{_{\Cscr_D}}\ .  \eqno(2.31)$$
The vector field $e_{_{\perp}} = e_0 = \partial / \partial t$ is the unit
normal 
to the slicing of $M$ by spatially homogeneous hypersurfaces.  Proper time
derivatives will be denoted by a small circle $\ ^{\circ}\ $.  A 
reparametrization
of the time $t \rarrow \overline{t}(t)$ may be accomplished by introducing a 
nontrivial
spatially homogeneous lapse function $dt = N(\overline{t})d\overline{t};$ 
barred time derivatives
will be denoted by a dot $\ ^{\cdot}\ $, so one has the relation $\dl\alpha 
= N\odl\alpha$
for the time derivatives of a function $\alpha$ only of time.  Eq.(2.31)
with nontrivial lapse but zero shift will be referred to as almost synchronous
gauge.

     In synchronous gauge, or almost synchronous gauge as long as the lapse
function is an explicit function of $g_{ab}$ and $\dl g_{ab}$ or other known
quantities, the spacetime metric is completely determined by the parametrized
curve $\hbox{\bf g}(\overline{t})$ in $\Mscr$.  Acting on the curve by a 
parametrized curve  {\bf A}$(\overline{t})$ in the
matrix automorphism group
$$ \hbox{\bf g}(\overline{t})\rarrow \overline{\hbox{\bf g}}(\overline{t})
= f_{\hbox{\bf A}\hbox{$(\overline{t})$}}(\hbox{\bf g}(\overline{t}))
\eqno(2.32)$$
is equivalent to the introduction of a shift vector $\b N(\overline{t}) $
belonging to
$\xc(g_{_{\Cscr_D}})$ and satisfying the matrix equation
$$ 
\hbox{ad}_{\overline{e}}(\b N(\overline{t})) 
= (\overline{\omega}^a(
\sterling_{\b N(\overline{t})}\overline{e}_b)) 
= \dot{\hbox{\bf A}}(
\overline{t})\hbox{\bf A}^{-1}(\overline{t})\ ,
\eqno(2.33)
$$
and of a new spatial frame and off-hypersurface frame vector
$$ \overline{e}_a = A^{-1b}_{\hskip 13pt a}(\overline{t})e_b\ ,\quad
   \overline{e}_0 = N(\overline{t})e_{_{\perp}} + \b N(\overline{t})
    \eqno(2.34)$$
leading to the expression for the metric in a general symmetry compatible
gauge
$$ ^4\hbox{\sl g} = -N(\overline{t})^2d\overline{t}\otimes d\overline{t}
+ \overline{g}_{ab}(\overline{t})(\overline{\omega}^a + \overline{N}^a
d\overline{t})\otimes(\overline{\omega}^b + \overline{N}^bd\overline{t})
\ .\eqno(2.35)$$
Here $\overline{N}^a = \overline{\omega}^a (\b N)$ are the barred shift 
components, while eq.(2.33) follows from the comoving condition 
$[\overline{e}_0 , \overline{e}_a]=0$.  In this gauge
the spacetime metric is determined by the parametrized curve $\overline{
\hbox{\bf g}}(\overline{t})$ in $M$
together with the parametrized curve 
$\dot{\hbox{\bf A}}(\overline{t})
\hbox{\bf A}(\overline{t})^{-1}$ 
in the Lie algebra of
the matrix automorphism group, again provided the lapse is known.

     In synchronous or almost synchronous gauge, the extrinsic curvature
is simple and its matrix of mixed components
$$ \hbox{\bf K} = (K^a{}_{b}) = -\half\hbox{\bf g}^{-1}\odl{\hbox{\bf g}}
=-(2N)^{-1}\hbox{\bf g}^{-1}\dl{\hbox{\bf g}}\eqno(2.36)$$
represents a matrix-valued function on the gravitational velocity phase
space $T\Mscr$ , namely the tangent bundle of the gravitational configuration
space $\Mscr$.  Here $\{ g_{ab} ,\dl g_{ab}\}$ are the 
natural
``coordinates" on $\Mscr$ lifted from the ``coordinates" $\{ g_{ab}
\}$ on $\Mscr$.  The ADM gravitational Lagrangian density is a
Langrangian function $L_G$ on the velocity phase space
$$ \eqalign{L_G &= N(\Tscr - U_G)\cr 
N\Tscr &=N\Gscr^{abcd}K_{ab}K_{cd} = (4N)^{-1}\Gscr^{abcd}
\dl g_{ab}\dl g_{cd} = N\,g^{1\over2}\left< \hbox{\bf K,K} \right> _{_{DW}}\ .
\cr}   \eqno(2.37)$$ 
The definition of the momentum canonically conjugate to {\bf g} is simply the
associated Legendre transformation between the velocity and momentum phase
spaces
$$ \pi^{ab}= \partial L_G/\partial\dl g_{ab}\ ,\qquad \bfpi = (\pi^a{}_{b})
=-g^{{1\over2}}(\hbox{\bf K}- (\hbox{Tr {\bf K}})\hbox{\bf 1})\ .\eqno(2.38)$$
The gravitational phase space is just the cotangent bundle $T^{\ast}\Mscr$
on which $\{ g_{ab} , \pi^{ab}\}$ are the natural 
``coordinates"
lifted from the ``coordinates" $\{ g_{ab}\}$ on $\Mscr$.  Note
that different choices of lapse change the Legendre map.  The Hamiltonian
function on momentum phase space which is associated with $L_G$ is defined
in the usual way
$$ \eqalign{H_G &= \pi^{ab}\dl g_{ab} - L_G =N(\Tscr+U_G)= N \Hscr_G\cr
\Hscr_G &= \Gscr^{-1}_{abcd}\pi^{ab}\pi^{cd} + U_G = g^{{1\over2}}
(\hbox{Tr}\ \bfpi^2 -\half\hbox{Tr}^2\bfpi) + U_G =2g^{1\over2}\ ^4
G^{^{\perp}}_{\quad _{\perp}}\ .\cr}\eqno(2.39)$$
The scalar density $\Hscr_G$ is the gravitational super-Hamiltonian.  
Both $L_G$ and $H_G$ are
$\Cscr_D$-parametrized functions due to the potential $U_G$ .  The kinetic
energy has no parameter dependence and is just a rescaling of the square
of the DeWitt norm of the velocity vector of the system, where the DeWitt
metric $\Gscr$ on $\Mscr$ is given by [44]
$$ \eqalign{\Gscr &= \Gscr^{abcd}dg_{ab}\otimes dg_{cd}\cr
\Gscr^{abcd} &= g^{{1\over2}}(g^{a(c}g^{d)b}- g^{ab}g^{cd})\ ,\quad
\Gscr^{-1}_{abcd}= g^{1\over2}(g_{a(c}g_{d)b} - \half g_{ab}g_{cd})\ .
\cr}\eqno(2.40)$$

     The kinetic energy $\Tscr$
 generates the ``free dynamics" (the vacuum type I case:
$C^a{}_{bc}=0$) whose solutions are just the geodesics of the DeWitt metric
which are affinely parametrized by the time $t$ in synchronous gauge.  A
nontrivial lapse function $N$ which is an explicit function on $\Mscr$
corresponds to conformally rescaling the DeWitt metric $\Gscr \rarrow 
N^{-1} \Gscr$ so that $\overline{t}$ is an affine parameter with respect to 
the rescaled
metric. However, only the null geodesics are relevant to the free dynamics
due to the free super-Hamiltonian constraint $\Tscr = 0$ which requires
the tangent vector to the geodesic to be a null vector with respect to the
DeWitt metric.  A general lapse function leads to an arbitrary parametrization
of these null geodesics.

     The general linear group $GL(3,R)$ acting on $\Mscr$ through (2.16) is a
group of homothetic motions of $(\Mscr,\Gscr)$ and the special linear group
$SL(3,R)$ is the identity component of its isometry subgroup.  For each
$\hbox{\bf B}\in gl(3,R)$, the corresponding homothetic Killing vector 
field (simply
Killing vector field if $\hbox{\bf B}\in sl(3,R)$) is given by
$$ \xi(\hbox{\bf B}) = -g_{c(a}B^c{}_{b)}\partial/\partial g_{ab}\ ,
  \eqno(2.41)$$
where the vector fields $\partial / \partial g_{ab}$ on $\Mscr$ are defined by
$d g_{cd}(\partial / \partial g_{ab}) = \delta^a_{\,(c}\delta^b{}_{d)}\ .$  
The Lie
bracket of two such fields satisfies $[\xi (\hbox{\bf A}), \xi(\hbox{\bf B})]
 = -\xi ([\hbox{\bf A,B}])\ .$
The corresponding generator of the lifted canonical action on the momentum
phase space and the Poisson brackets of two such generators are given by
$$ P(\hbox{\bf B}) = -2\hbox{Tr}\ \hbox{\bf B}\bfpi\ , \quad \{
P(\hbox{\bf A}), P(\hbox{\bf B})\} = P([\hbox{\bf A,B}])\ ,
\eqno(2.42)$$
where the only nonvanishing Poisson brackets of the ``coordinates" $(g_{ab} ,
\pi^{ab})$ are defined by $\{ g_{ab} , \pi^{cd}\}
= \delta^c_{\, (a}\delta^d{}_{b)}\ .$  The corresponding function on the 
velocity phase space when $N = 1$ (and dot becomes circle)
$$ P(\hbox{\bf B}) = \Gscr^{-1abcd}\xi(\hbox{\bf B})_{ab}\odl g_{cd}
\eqno(2.43)$$
is just the inner product of the corresponding homothetic Killing vector
field with the velocity of the system.  However, only the natural action
of $SL(3,R)$ on the gravitational velocity and momentum phase spaces is
a symmetry of the kinetic energy for arbitrary lapse and so in general
only its canonical generators are conserved by the (unconstrained) free
dynamics.  Null geodesics on the other hand conserve $P(\hbox{\bf 1})$ as 
well, so the
free super-Hamiltonian constraint restores $GL(3,R)$ as a symmetry group of
the allowed solutions.

     In Misner's supertime gauge [15--18] 
$ N=12g^{1\over2}$, the
restriction of the metric $N^{-1}\Gscr$ to $\Mscr_D$ is just the flat Lorentz
metric induced on $\Mscr_D$ by its identification (2.17) with the inner
product space $(diag(3,R), \langle\langle , \rangle\rangle_{_{DW}})$, in 
terms of which $\{
\beta^A\}$ are inertial coordinates
$$ \eqalign{N^{-1}\Gscr\relv_{_{\Mscr_D}} &= \eta_{AB} d\beta^A\otimes d\beta^B
\cr N\Tscr &= \half\eta_{AB}\du \beta^A\du\beta^B = \half\eta
^{AB}p_Ap_B\ ,\qquad p_A = \eta_{AB}\du\beta^B \ .\cr} 
\eqno(2.44)$$
Here $\{\beta^A ,\du \beta^A\}$ and $\{\beta^A , p_A
\}$ are the natural lifted coordinates on $T\Mscr_D$ and $T^\ast
\Mscr_D$.  The null geodesics on $\Mscr_D$ are just null lines in $diag(3,R)$ 
$$ \eqalign{\beta^A(\overline{t}) 
&= \eta^{AB}p_B\overline{t} + \beta^A(0)
\ , \qquad \dl p_A = 0 = \eta^{AB}p_Ap_B \cr
\hbox{\bf g}^{\prime}(\overline{t}) 
&=e^{2\bfsbeta(\overline{t})} , \quad
\du{\bfbeta}(\overline{t}) 
= \eta^{AB}p_Ae_B = \hbox{\bf b} \ .\cr}\eqno(2.45)$$
In fact since $GL(3,R)$ maps null geodesics into null geodesics (and $\Mscr
_D$ is a totally geodesic submanifold of $\Mscr$), a general null geodesic is 
of the form
$$ \hbox{\bf g}(\overline{t}) = f^{-1}_{\hbox{\bf A}}(\hbox{\bf g}^{\prime}
(\overline{t}))\ , \qquad \hbox{\bf A}\in GL(3,R)\ .\eqno(2.46)$$

     The Kasner exponents $(p_1 , p_2 , p_3)$ are defined to be the
eigenvalues of the matrix $\hbox{\bf k} = (\hbox{Tr}\ \hbox{\bf K})^{-1} 
\hbox{\bf K}$ 
for a null geodesic $(\,\left<\hbox{\bf k,k}\right>_{_{DW}} =
\ketl\hbox{\bf k,k}\ketr-1=
0\,)$, therefore satisfying $p_1 + p_2 + p_3 = p_1^{\ 2} + p_2^{\ 2} + 
p_3^{\ 2} = 1$.
For the diagonal null geodesics (2.45) the extrinsic curvature matrix is
$\hbox{\bf K}(\overline{t})^{\prime} =-\fraction{1}{12}
 g^{-{1\over2}} (\overline{t})\du{\bfbeta} (\overline{t}
)$ so one has
$$ 
\eqalign{ \hbox{diag}(p_1,p_2,p_3)
&= (\hbox{Tr}\ 
\hbox{\bf b})^{-1}
\hbox{\bf b}\cr 
&\hskip-15pt
= (-3p_0)^{-1}\hbox{diag}(-p_0+p_++\sqrt3 p_-,-p_0+p_+-\sqrt3 
p_-, -p_0-2p_+)\cr  
&\hskip-15pt
= \fraction{1}{3}\hbox{\bf 1} - 
\fraction{1}{3} \hbox{sgn}\,p_0\ \hbox{diag}(\A p_++\sqrt3\,\A p_-,
\A p_+-\sqrt3\,\A p_-, -2\A p_+)\ ,\cr}
\eqno(2.47)$$
where the unit vector $(\A p_+ , \A p_- )
\equiv (p_+^2 + p_-^2 )^{-{1\over2}} (p_+ , p_-)$ coincides with\\ 
$|p_0|^{-1}
(p_+, p_-)$ for these null geodesics.
This gives the Kasner exponents as functions on the unit circle in the $p_+p_-
$ plane, representing the $S^1$-parametrized family
of null directions in the diagonal cotangent spaces.

\begin{figure}[t!] 
\begin{center}
\includegraphics[width=.9\textwidth]{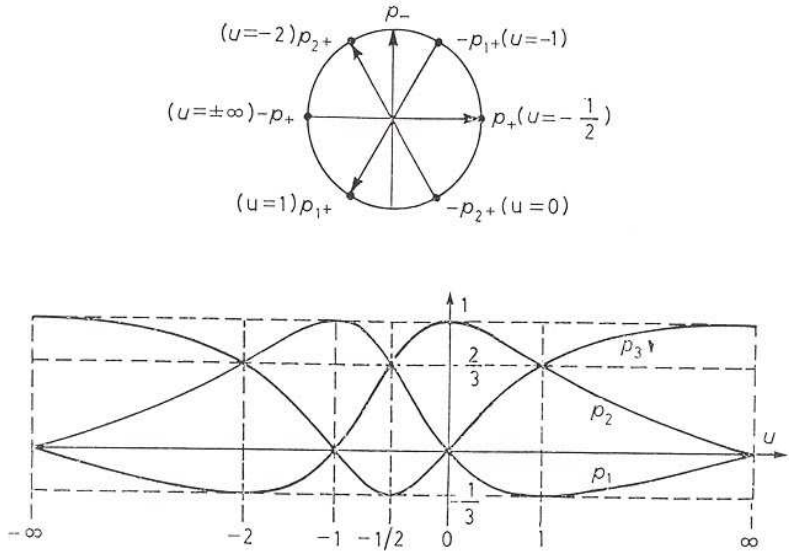}
\end{center}
\caption{
The Lifshitz-Khalatnikov parametrization of
the Kasner exponents and of the unit circle in the $p_+p_-$ plane.}
\end{figure}

     A very useful parametrization of this circle was given by Lifshitz
and Khalatnikov [54] 
$$\eqalign{
(p_1(u), p_2(u), p_3(u)) &= (u^2 + u + 1)^{-1} (-u, 1 + u, u(1 + u))\cr
\, (\A p_+, \A p_- ) &= -(u^2 + u +1)^{-1} (u^2 + u - \half, 
                        \sqrt3\, [u + \half])\ . \cr}\eqno(2.48)$$
The Kasner exponents as functions of the variable $u$ are shown in Figure 6,
together with the correspondence with the unit circle in the $p_+p_-$ plane
established by the second equality, which may be inverted to give $u_{\pm} =
 -{1\over2}(1\mp[3(1-\A p_+)/(1+\A p_+)]^{1\over2}$, where the upper (lower)
sign applies in the upper (lower) half circle.  Each of the six sectors into
which this circle is divided by the $p_{1+}$, $p_{2+}$, and $p_{3+}$
axes (where the coordinates $\{ p_{1+}, p_{1-}\}$ and
$\{ p_{2+}, p_{2-}\}$ are obtained from $\{ p_+, p_-
\}$ by rotations by $2\pi/3$ and $-2\pi/3$ respectively and 
$\{ p_{3+}, p_{3-}\}\equiv\{ p_+, p_-\}$)
represents a different ordering of the same interval 
of eigenvalues.  The intersections of these axes with the circle represent the
three permutations of the two inequivalent Taublike case null geodesics
whose associated type I spacetime metrics (2.31) are locally rotationally
symmetric.  If $(a,b,c)$ is a cyclic permutation of (1,2,3) then an 
interchange
of the basis vectors $e_b$ and $e_c$ leads to reflection across the $\beta^+
_a$ axis in $\Mscr_D$ and hence reflection  about the $p_{a+}$ axis in
each cotangent space, under which the unit circle is invariant.  In terms
of the variable $u$ parametrizing this circle, the reflections across the
$p_{1+}$, $p_{2+}$ and $p_{3+}$ axes respectively correspond to the discrete
transformations $P_{23} (u) = u^{-1}$, $ P_{31} (u)=-u(u+1)^{-1}$ and $P_{12} 
(u) = -(u+1)$ , representing transpositions of the basis vectors $\{
e_a\}
.^{(15)}$  The transformations corresponding to cyclic permutations of these
basis vectors leading to rotations of the $\beta^{\pm}$ and $p_{\pm}$ planes
by $\pm 2\pi/3$ may be obtained by combining two transpositions.  The index
permutation $(1,2,3)\rarrow(2,3,1)$ is represented by $P_{231}(u)=P_{23}\circ 
P_{31}(u)= -u^{-1}(1+u)$, corresponding to a positive rotation by $2\pi/3$
while $(1,2,3)\rarrow(3,1,2)$ is represented by $P_{312}(u)= -(1+u)^{-1}$
, corresponding to a negative rotation by $2\pi/3$ [15].

     The diagonal geodesics are characterized by the vanishing of the
moment function (2.42) for any offdiagonal matrix {\bf B} since both {\bf K} 
and $\bfpi$ are
diagonal for these geodesics.  Such values for the moment function will be
referred to as angular momentum.  The constant transformation $f_{\hbox{\bf A
}}$ applied
to the geodesic (2.46) ``transforms away" its constant nonzero angular
momentum.  For the spacetime metric (2.31) corresponding to (2.46), the new
spatial frame $e^{\prime}$ obtained from $e$ by this transformation
$$ e_a^{\ \prime} = A^{-1b}_{\hskip 13pt a}e_b\eqno(2.49)$$
is an orthogonal frame of eigenvectors of the mixed extrinsic curvature tensor
(both {\bf g}$^{\prime}$ and {\bf K}$^{\prime}$ are diagonal) .  Such a 
spatial frame is called a Kasner
frame and its elements are called Kasner axes [55].  
Note, however, that unless $\hbox{\bf A}\in Aut_e(g)$, the new
structure constant tensor components will differ from the old ones and will
in general not belong to $\Cscr_D$.  In the analogy with the central force 
problem, changing the spatial frame by a constant linear transformation
corresponds to changing the origin with respect to which the orbital angular
momentum is defined.   

     In the supertime gauge $GL(3,R)$ is the isometry group of the rescaled
DeWitt metric so all of its generators are conserved by the (unconstrained)
free dynamics.  However, invariance of the rescaled curvature potential $NU
_G$ under the full automorphism group $Aut_e(g)$ requires an incompatible
time gauge $N\propto g^{-{1\over2}}$, so $SAut_e(g)$ is the largest 
simultaneous symmetry group of both the kinetic and potential energies.

\noindent \hskip 18.05pt
     For a nonvacuum spatially homogeneous spacetime with spatially homogeneous
energy-momentum components $T^{\alpha}{}_{\beta}$ in the synchronous gauge, 
matter variables may usually be chosen so that the matter super-Hamiltonian
$\Hscr_M=-2k g^{1\over2} T^{^{\perp}}_{\quad_{\perp}}$ acts as a potential 
function for
the matter driving force that appears in the evolution equations.  If
$d_{\Mscr}$ is the exterior derivative on $\Mscr$, then this potential must
satisfy
$$ T^{\ast} = k g^{1\over2}T^{ab}dg_{ab} = -d_{\Mscr}\Hscr_M\ .
\eqno(2.50)$$
If this is not possible, one can simply introduce a matter component $Q_M$
of the nonpotential force [42].  
The matter supermomentum components
are $\Hscr^M_a= -2k g^{1\over2} T^{^{\perp}}_{\hskip 8pt a}$ .

     The components of the gravitational supermomentum in almost synchronous
gauge are defined by [2,42,43] 
$$ \eqalign{\Hscr^G_a &= 2g^{1\over2}\ ^4G^{^{\perp}}_{\quad a}
= P(\bfdelta_a)\ ,
  \qquad\bfdelta_a = \hbox{\bf k}_a -2a_b\bfdelta^c{}_{a}
  \hbox{\bf e}^b{}_{c}
\cr \hbox{Tr}\ \bfdelta_a &= 0\ ,\qquad [\bfdelta_a, \bfdelta_b] = 
  (\epsilon_{abd}n^{cd} + 3a_f\delta^{fc}_{ab})\bfdelta_c\ ,\cr}\eqno(2.51)$$
where the adjoint matrices $\hbox{\bf k}_a=C^b{}_{ac}\hbox{\bf e}^c{}_{b}$
are introduced in appendix A and given explicitly by (A.3).
The matrices {$\bfdelta_a$} generate a subgroup of $SL(3,R)$ which coincides
with the linear adjoint group in the class A case and a projective
automorphism subgroup in the class B case [2].  
These matrices are 
linearly
dependent for Bianchi types I $(\bfdelta_a=0)$, type II ($\bfdelta_1=0$ when
$n^{(1)}\neq 0$, etc.) and type VI$_{-1/9}$ $(|n^{(1)}|^{1\over2}
\bfdelta_1+
|n^{(2)}|^{1\over2}\bfdelta_2=0)$ where the supermomentum constraints are
degenerate, imposing additional constraints on the matter supermomentum.

     The Einstein evolution equations in almost synchronous gauge are then
the equations of motion of the total Lagrangian/Hamiltonian system with
the nonpotential force Q and total Lagrangian and Hamiltonian
$$ L= L_G-N\Hscr_M\ ,\qquad H= H_G+N\Hscr_M =N\Hscr\ ,\eqno(2.52)$$
namely
$$\eqalign{-\delta L/\delta g_{ab}&\equiv (\partial L/\partial\dl g_{ab})
\dl{\ } -\partial L/\partial g_{ab}= NQ^{ab}\cr
\dl g_{ab} &= \{ g_{ab}, H\}\ ,\qquad \du{\pi}^{ab} = 
\{\pi^{ab}, H\} +NQ^{ab}\ .
\cr}\eqno(2.53)$$
These are subject to the constraint equations
$$ \Hscr = \Hscr_G+\Hscr_M =0, \qquad \Hscr^G_a+\Hscr^M_a =0\ .\eqno(2.54)$$
For the present paper the source of the gravitational field will be assumed
to be a perfect fluid.  The appropriate choice of variables and expressions
for the fluid super-Hamiltonian, supermomentum and equations of motion are
discussed in appendix C.

\section{Diagonal Gauge as an Almost Synchronous Gauge Change of
 Variables}

     The Einstein equations for a spatially homogeneous spacetime in almost
synchronous gauge have been put in the form of a $\Cscr_D$-parametrized 
constrained
classical mechanical system driven by the matter variables which themselves
satisfy certain equations of motion.  The combined equations of motion are
invariant under the action of constant elements of the automorphism matrix
group $Aut_e(g)$ representing the freedom remaining in the choice of the
spatially homogeneous spatial frame in almost synchronous gauge for each
fixed point $C^a{}_{bc}\in \Cscr_D$.  However, only the special automorphism
matrix group $SAut_e(g)$ acts naturally on the Lagrangian/Hamiltonian system
as a symmetry group, corresponding to the additional restriction that the
3-form $\omega^1\wedget\,\omega^2\wedget\,\omega^3$ remain invariant under 
change of
frame, so in this context it is $SAut_e(g)$ rather than $Aut_e(g)$ which
plays an important role [3].
The supermomentum constraint functions are
associated with this symmetry, although the direct connection is not so
obvious when the nonpotential force is nonzero.

     Whenever a dynamical system has a symmetry group, one may simplify the
system by choosing new variables adapted to the action of the symmetry group
on this system, a very instructive example being the central force problem.
Exactly how to adapt the variables depends on the particular way in which
the symmetry group acts on the system.  For all points of $\Cscr_D$ except 
the set
of measure zero occupied by the type I, II and V orbits, the group $\A 
G\equiv SAut_e(g)$ is 3-dimensional, closely connected to the three linearly
independent supermomentum constraint functions, and has an offdiagonal matrix
Lie algebra $\A g$ which is such that any point $\hbox{\bf g}\in\Mscr$ may be
diagonalized by one or more elements of this group acting as in (2.16). 
For the remaining three Bianchi types, $SAut_e(g)$ has a larger dimension and
for types I and II there are fewer linearly independent supermomentum
constraint functions, but there do exist families of 3-dimensional subgroups
$\A G$ with offdiagonal Lie algebras which may be used to diagonalize an 
arbitrary
point of $\Mscr$.  Recalling the advantages of diagonal metric component
matrices suggests that $\Mscr_D$ should be used to parametrize the orbits of
the action of these 3-dimensional groups $\A G$ on $\Mscr$, leading to the 
following class of parametrizations $(\beta, \hbox{\bf S})\in diag(3,R)
\times \A G\rarrow\Mscr$  
$$ \hbox{\bf g} = f^{-1}_{\hbox{\bf S}}(e^{2\bfsbeta}) = f^{-1}_{\hbox{\bf S}}
(\hbox{\bf g}^{\prime}) =\hbox{\bf S}^T e^{2\bfsbeta}\hbox{\bf S}\ ,
\eqno(3.1)$$
which decompose the metric matrix variables into ``diagonal" and 
``offdiagonal"
variables as described in the introduction.  The space $diag(3,R)\sim 
Diag(3,R)^+$ $\sim\Mscr_D$ representing the diagonal variables has already
been parametrized in various ways, leaving to be discussed the parametrization
of $\A G$ as well as its choice for the type I, II and V cases.  For these 
latter
types and the remaining nonsemisimple types, some of the diagonal variables 
are associated with additional diagonal automorphisms.

     The property that the orbit of $\Mscr_D$ under the action of $\A G$ be 
$\Mscr$
(in order that any point $\hbox{\bf g}\in\Mscr$ may be represented in the 
form (3.1))
requires that the offdiagonal matrix Lie algebra $\A g$ have an ordered basis
$\{\bfkappa_a\}$ with the property that for each cyclic 
permutation $(a,b,c)$ of
$(1,2,3)$, then $\bfkappa_a\in \hbox{span}\{ \hbox{\bf e}^b{}_{c},
\hbox{\bf e}^c{}_{b}\}$.  Consider the following
Lie algebra basis valued function $\{\bfkappa_a\}$ on $\Cscr_D$,
defined by
$$ \eqalign{\bfkappa_a &= e^{-\alpha^a}\hbox{\bf k}^0_a\ ,\qquad
\hbox{\bf k}^0_a = -n^{(b)}\hbox{\bf e}^c{}_{b} +n^{(c)}\hbox{\bf e}
^b{}_{c}\ ,\cr
e^{\alpha^a} &= 2^{-{1\over2}}\ketl\hbox{\bf k}^0_a, \hbox{\bf k}^{0\ T}_a
\ketr^{1\over2} = 2^{-{1\over2}}[(n^{(b)})^2 + (n^{(c)})^2]^{1\over2}\cr}
\eqno(3.2)$$
and satisfying
$$ \eqalign{[\bfkappa_a, \bfkappa_b] &= {\A C}\null^a{}_{bc}\bfkappa_c\ ,
\qquad {\A C}\null^a{}_{bc} = \epsilon_{bcd}\A n^{ad}\cr
\A {\hbox{\bf n}} &= \hbox{diag}(\A n^{(1)}, \A n^{(2)}, \A n^{(3)})\ ,\qquad
\A n^{(a)} = n^{(a)}e^{\alpha^a-\alpha^b-\alpha^c}\ ,\cr}\eqno(3.3)$$
where it is clear from the context when $(a,b,c)$ is to be interpreted as a
cyclic permutation of (1,2,3).  This is well defined everywhere on $\Cscr_D$
except for the type I, II, IV and V orbits (precisely those points of $\Cscr_D$
where $\hbox{rank}\,\hbox{\bf n}<2$ and the scale matrix $e^{\bfsalpha} =
\hbox{diag}(e^{\alpha^1}, e^{\alpha^2}, e^{\alpha^3})$ is singular) 
where it has direction dependent limits. Consider those points of $\Cscr_D$
for which $n^{(3)}=0$, for example.  One then has
$$
\eqalign{ 
e^{\bfsalpha}&
=2^{-{1\over2}} \hbox{diag}(|n^{(2)}|, |n^{(1)}|, |(n^{(1)})^2 +(n^{(2)})^2|
^{1\over2})\cr 
\{\bfkappa_a\}&= \{ -\sqrt2\,\hbox{sgn}\,
n^{(2)}\ \hbox{\bf e}^3{}_{2}, \sqrt2\,\hbox{sgn}\,n^{(1)}\ 
\hbox{\bf e}^3{}_{1}, \sqrt2(-\cos\phi\,\hbox{\bf e}^2{}_{1} +
\sin\phi\,\hbox{\bf e}^1{}_{2})\}\cr
\A{\hbox{\bf n}}& 
= \sqrt2\,\hbox{sgn}\,n^{(1)} n^{(2)}
\hbox{diag}(\sin\phi, \cos\phi\, 0)
\cr 
&\qquad(\cos\phi, \sin\phi)\equiv 
|(n^{(1)})^2 +(n^{(2)})^2|^{-{1\over2}}(n^{(1)}, n^{(2)})\ .\cr}
\eqno(3.4)$$   
For type IV only the sign of either $\bfkappa_1$ or $\bfkappa_2$ does not
have a well defined limit for a given orbit, so $\{\bfkappa_a\}$
has two values (but a unique $\A G$) for each orbit.  In addition to this
sign indeterminacy (which does not make $\A G$ multivalued),  an 
$S^1$-parametrized family of limits exists for each type II and V orbit 
component
(associated with the indeterminancy of the matrix $\bfkappa_3$ for the
type V orbit and of the matrix $\bfkappa_a$ for the type II orbit component
on which $n^{(a)}$ is the only nonvanishing structure constant tensor 
component),
while an $S^2$-parametrized family of limits exists for the single type I
point containing all well-defined values of this function.  Thus $\{
\bfkappa_a\}$ is a multivalued Lie algebra basis valued function on 
$\Cscr_D$ whose values at
each point determine the offdiagonal diagonalizing automorphism matrix 
subgroup Lie algebras $\A g$.  $\A G$ is then a multivalued matrix Lie group 
valued
function on $\Cscr_D$.  (The space of values of this function is diffeomorphic 
to $S^2$ modulo reflection about the origin, namely $P^2$.)  
For the types I, II, IV 
and V where the multivaluedness occurs, one may pick any value to describe the
dynamics.

At the canonical Bianchi type IX point of $\Cscr_D$, $\A G$ has the value
$SO(3,R)$ and the parametrization (3.1) reduces to the one introduced by Ryan
for all Bianchi types [5].
The latter parametrization arose from
considerations completely unrelated to symmetries but by a fortunate
coincidence agrees with the correct choice for the most interesting case:
Bianchi type IX with canonical structure constant components.  Given any
frame $e$ with metric component matrix {\bf g}, there is a natural
orthonormal frame related to $e$ by the symmetric square root of {\bf g}
(unique up to ordering when the eigenvalues of {\bf g} are nondegenerate)
$$ \hbox{\bf g} = \hbox{\bf B}^2\ , \qquad B^a{}_{b}= B^b{}_{a}\ ,
\qquad e_a^{\prime\prime\prime}=B^{-1b}_{\hskip 13pt a}e_a\ .\eqno(3.5)$$
Since any symmetric matrix can be diagonalized by an orthogonal transformation,
($\hbox{\bf B}= \hbox{\bf O}^T\hbox{\bf B}_D \hbox{\bf O}\ , \hbox{\bf O}
\in SO(3,R)\ , \hbox{\bf B}_D\in Diag(3,R)^+$), using the property $
\hbox{\bf O}^T=\hbox{\bf O}^{-1}$ one obtains the result $\hbox{\bf g}=
\hbox{\bf O}^T\hbox{\bf B}_D^{\ 2}\hbox{\bf O}$.  By setting $\hbox{\bf B}_D
=e^{\bfsbeta}$, one arrives at the Ryan parametrization and its associated
orthonormal frame
$$\hbox{\bf g}=\hbox{\bf O}^T e^{2\bfsbeta}\hbox{\bf O}\ ,\qquad
e_a^{\prime\prime\prime}=(\hbox{\bf O}^T e^{\bfsbeta}\hbox{\bf O})^{-1b}_{
\hskip 13 pt a} e_b = O^{-1c}_{\hskip 13pt a}(e^{\bfsbeta}\hbox{\bf O})
^{-1b}_{\hskip 13pt c}e_b\ .\eqno(3.6)$$
The incompatibility of this parametrization with the symmetry at all points
of $\Cscr$ except for the type I point (where it is unnecessary for perfect
fluid spacetimes) and the type IX points where {\bf n} is proportional to 
its canonical value made its application to other points of $\Cscr$ 
ineffective.  (Compare the expressions of the first paper of ref.(23) for the
nonpotential force with (3.29).) 

     If needed one may use the following parametrization of $\A G$ 
$$ \hbox{\bf S}= e^{\theta^1\bfskappa_{\sst 1}} e^{\theta^2\bfskappa_{\sst 2}} 
e^{\theta^3\bfskappa_{\sst 3}}\ .\eqno(3.7)$$
Suitably restricting the domain of the parameters $\{\theta^a\}$
when $\A G$ has
compact directions leads to local canonical coordinates of the second kind
on $\A G$ (which are always valid in an open neigborhood of the identity if
not globally)
 and hence through (3.1) to local coordinates on each 3-dimensional 
orbit of $\A G$ on $\Mscr$.  These local coordinates on $\Mscr$ become 
singular on $\A G$-orbits of dimension less than three, which represent fixed 
points of the
action of the compact subgroups of $\A G$.  The basis $\{\bfkappa_a
\}$ has the
property that when the single parameter $\theta^a$ for some fixed value
of $a$ is nonzero in (3.7), then (3.1) parametrizes the symmetric case
submanifold $\Mscr_{S(a)}=\{\hbox{\bf g}\in\Mscr \relv g_{ab}=g_{ac}=0
\}$, where as usual
$(a,b,c)$ is a cyclic permutation of (1,2,3).  These three submanifolds are
associated with discrete spacetime symmetries [43].

     Through the parametrization (3.1), the action of $\A G$ on $\Mscr$ 
$$ \hbox{\bf g} = f^{-1}_{\hbox{\bf S}}(\hbox{\bf g}^{\prime})\rarrow
f^{-1}_{\hbox{\bf A}}(\hbox{\bf g}) = f^{-1}_{\hbox{\bf SA}^{-1}}(
\hbox{\bf g}^{\prime})\ ,\qquad\hbox{\bf A}\in\A G\ ,\eqno(3.8)$$
becomes inverse right translation on $\A G$ itself.  Since the DeWitt metric
is $SL(3,R)$ invariant and $\A G\subset SL(3,R)$ by virtue of having a 
tracefree
Lie algebra, if a right invariant frame on $\A G$ is employed, the components 
of the restriction of the DeWitt metric to the orbit (identified with $\A G$) 
will have components which can at most depend on the diagonal variables.

The relations 
$$\eqalign{ \hbox{\bf S}^{-1}d\hbox{\bf S} &=\bfkappa_aW^a\ ,\qquad
W^a(E_b)=\delta^a{}_{b}\cr
d\hbox{\bf S S}^{-1}&= \bfkappa_a {\s W}\null^a\ ,\qquad {\s W}\null
^a({\s E}_b) = \delta
^a{}_{b}\cr}\eqno(3.9)$$
define the left and right invariant frames $\{E_a\}$ and 
$\{ {\s E}_a\}$ on $\A G$ with respective
dual frames $\{W^a\}$ and $\{{\s W}\null^a\}$ and 
structure functions ${\A C}\null^a{}_{bc} = W^a([E_a, E_b])
=-{\s W}\null^a([{\s E}_b, {\s E}_c])$ which are naturally associated with the
basis $\{\bfkappa_a\}$ of its matrix Lie algebra.  They satisfy
$[E_a, {\s E}_b]=0$.  If $\hbox
{\bf S}(\overline{t})$ is a parametrized curve in $\A G$, then 
$$ \hbox{\bf S}(\overline{t})^{-1}\du {\hbox{\bf S}}(\overline{t})= \bfkappa_a
\du W \null^a(\overline{t})\quad,\quad 
\du {\hbox{\bf S}}(\overline{t})\,\hbox{\bf S}(\overline{t})^{-1} =
\bfkappa_a {\dot{W}{}}\null^a(\overline{t})\eqno(3.10)$$ 
defines the component functions $\du W\null^a(\overline{t})$ and $\dot{W}{}\null^a
(\overline{t})$ of the curve's tangent vector with respect to these frames.
Let the corresponding functions on the cotangent bundle $T^{\ast}{\A G}$
be denoted by $P_a$ and $\s P_a$ respectively.  If $\{\theta^a\}$
are local
coordinates on $\A G$, then one has the following coordinate expressions for
the right invariant frame quantities
$$\eqalign{{\s E}_a &= {\s E}\null^b{}_{a}(\theta)\partial/\partial\theta^b\ ,
\qquad {\s W}\null^a = {\s W}\null^a{}_{b}(\theta)d\theta^b\ ,\qquad 
({\s W}\null^a{}_{b})= ({\s E}\null^a{}_{b})^{-1}\ ,\cr
{\dot{W}{}}\null^a &= {\s W}\null^a{}_{b}\du {\theta}^b\ ,\qquad \s P_a = {\s E}
\null^b{}_{a}p_b\ ,\cr}\eqno(3.11)$$
where $\{\theta^a, {\du {\theta}}\null^a\}$ and $\{
\theta, p_a
\}$ are the natural lifted coordinates on the tangent and cotangent
bundles.  Dropping the tildes leads to the corresponding left invariant
frame quantities which are related by the linear adjoint transformation on
$\A G$  
$$\eqalign{ \hbox{\bf S}\bfkappa_a\hbox{\bf S}^{-1} &= \bfkappa_b {\A R}\null
^b{}_{a}
\ ,\qquad \A {\hbox{\bf R}} =  e^{\theta^1 \A {\hbox{\bf k}}_1}
e^{\theta^2 \A {\hbox{\bf k}}_2}
e^{\theta^3 \A {\hbox{\bf k}}_3}\ ,\qquad \A {\hbox{\bf k}} 
_a = {\A C}\null^b{}_{ac}\hbox{\bf e}^c{}_{b}\cr
{\dot{W}{}}\null^a &= {\A R}\null^a{}_{b}{\du W}\null^b\ ,\qquad {\s P}_a = P_b 
{\A R}\null^{-1b}_{\hskip 13pt a}\ ,\qquad etc.\ .\cr}\eqno(3.12)$$
One may also use the local canonical coordinates $(\theta^a, p_a)$ to 
evaluate the following Poisson brackets 
$$\eqalign{ &\{{\s P}_a, {\s P}_b\} = {\A C}\null^c{}_{ab} 
{\s P}_c\qquad
\{\hbox{\bf S}, {\s P}_a\} = \bfkappa_a\hbox{\bf S}\cr
&\{P_a, P_b\}= -\A C\null^c{}_{ab}P_c \qquad
\{\hbox{\bf S}, P_a\}= \hbox{\bf S}\bfkappa_a
\qquad \{P_a, {\s P}_b\}=0 \ .\cr}\eqno(3.13)$$
Since the basis $\{\bfkappa_a\}$ is closely related
to the standard basis of one of the standard diagonal form adjoint
matrix Lie algebras with rank {\bf n} $>1$, explicit expressions  
may easily be obtained from the formulas of appendix A for 
{\bf S}, $\A {\hbox{\bf R}}$ and the component matrices of the invariant
fields in terms of the parametrization (3.7). The case $n^{(3)}=0$
is given as an example in that appendix. 

The parametrization (3.1) has the following meaning. Given the spacetime
metric (2.24) is some almost synchronous gauge, i.e. given the parametrized 
curve $\hbox{\bf g}(\overline{t})$ in $\Mscr$ and the lapse function $N(
\overline{t})$, then $\hbox{\bf A}(\overline{t}) = \hbox{\bf S}(\overline{t})
$ is the matrix in (2.27) which affects the change to diagonal gauge, where
the new metric matrix $\overline{\hbox{\bf g}}(\overline{t}) = \hbox{\bf g}
^{\prime}(\overline{t})$ is diagonal and the new spatial frame
$$ e_a^{\ \prime} = S^{-1b}_{\hskip 13 pt a}(\overline{t})e_b\eqno(3.14)$$
is orthogonal, with the associated shift vector field satisfying
$$ \hbox{ad}_{e^{\prime}}(\b N(\overline{t})) = \bfkappa_a {\dot{W}{}}\null^a\ ,
\eqno(3.15)$$
according to (2.33). The matrix $e^{\bfsbeta}\in Diag(3,R)^+$ then normalizes
this orthogonal spatial frame, leading to the natural symmetry adapted 
orthonormal spatial frame
$$e_a^{\prime\prime}= (e^{\bfsbeta})^{-1b}_{\hskip 13pt a} e_b^{\prime}=
(e^{\bfsbeta}\hbox{\bf S})^{-1b}_{\hskip 13pt a}e_b \eqno(3.16)$$
which may be used to introduce spinor fields on the spacetime in ``time
gauge" [42,56,57]
or to introduce a natural Newman-Penrose null tetrad
or to put the Einstein equations in a simple form without a 
Lagrangian/Hamiltonian formulation [58].  
The offdiagonal velocities
${\dot{W}{}}\null^a$ are linearly related to the angular velocity of this natural
spatial triad relative to one which is parallelly propagated along the
normal congruence [42].
Note that in the canonical type IX case, this
orthonormal spatial frame differs from the frame (3.6) which is associated
with the Ryan parametrization by an additional rotation, and like that
frame, is unique modulo ordering of the frame vectors and barring degeneracies
for each value of $\A G$.

Let a prime indicate components with respect to the diagonal gauge spatial
frame (3.14).  As noted in the previous section, spatial curvatures have
simpler expressions in diagonal gauge; in particular the scalar curvature
potential function $U_G$ is independent of the offdiagonal variables and is
simply given by the formula (2.27).  The extrinsic curvature (2.36) and the
kinetic energy have the following expressions when evaluated with respect to
the primed frame
$$\eqalign{\hbox{\bf K}^{\prime} &= \hbox{\bf SKS}^{-1} = -N^{-1}(\du 
{\bfbeta}+
\bfkappa_a^{\#\ \prime} {\dot{W}{}}\null^a)\ , \qquad \bfkappa_a^{\#\ \prime} = 
\half(\bfkappa_a + e^{-2\bfsbeta}\bfkappa_a^{\ \ T}e^{2\bfsbeta})\cr
N\Tscr &= e^{3\beta^0}\left<\hbox{\bf K}^{\prime}, \hbox{\bf K}^{\prime}
\right>_{_{DW}} = N^{-1}e^{3\beta^0}(6\eta_{AB}\du {\beta}^A\du {\beta}^B + 
\overline{\Gscr}_{ab}\dot{W}{}\null ^a \dot{W}{}\null ^b)\cr
\overline{\Gscr}_{ab} &=\ketl\bfkappa_a^{\#\ \prime}, \bfkappa_b^{\#\ \prime}
\ketr_{_{DW}} = \ketl\bfkappa_a^{\#\ \prime}, \bfkappa_b^{\#\ \prime}\ketr\ ,
\cr}
\eqno(3.17)$$
indicating that the DeWitt metric itself is
$$ 
\fraction{1}{4}\Gscr = e^{3\beta^0}(6\eta_{AB}d\beta^A\otimes d\beta^B + 
\overline{\Gscr}_{ab}{\s W}\null^a\otimes{\s W}\null^b)\ .
\eqno(3.18)$$
The components of the rescaled DeWitt metric $\overline{\Gscr}\equiv
{1\over4} g^{-{1\over2}}\Gscr$ along the
orbit directions are functions on the $\beta^+\beta^-$ plane and are diagonal
due to the choice of basis (3.2) of $\A g$ 
$$ \overline{\Gscr}_{aa} = \half e^{-2\alpha^a}(n^{(b)}e^{\beta^{bc}} - 
n^{(c)}e^{-\beta^{bc}})^2 \equiv (\overline{\Gscr}\null^{-1\ aa})^{-1}\ .
\eqno(3.19)$$
When $(e^{-\alpha^a}n^{(b)}, e^{-\alpha^a}n^{(c)})$ equals respectively
$(\sqrt2,0)$, $(1,-1)$ and $(1,1)$, as occurs at canonical points of $\Cscr_D$,
this expression has the values 
$e^{2\beta^{bc}}$, $2\hbox{cosh}^2\beta^{bc}$ and $2\hbox{sinh}^2\beta^{bc}$.
When $\bfkappa_a$ is a compact generator, which means that $e^{-\alpha^a}n^{(b)}
$ and $e^{-\alpha^a}n^{(c)}$ are nonzero and of the same sign, then 
$\overline{\Gscr}_{aa}$ vanishes for 
$$\beta^-_a =\beta^-_{a0}\equiv  -(4\sqrt3)^{-1}\ln |e^{-
\alpha^a}n^{(b)}/e^{-\alpha^a}n^{(c)} | \eqno(3.20)$$
which for canonical points of $\Cscr_D$ represents the 2-dimensional orbit
of the Taub submanifold $\Mscr_{T(a)}$. Such points of $\Mscr$ represent
singularities of the parametrization (3.1). This same condition on $\beta^-_a$
picks out the submanifold of $\Mscr_D$ for which $d\overline{\Gscr}_{aa}= 0= 
d\overline{\Gscr}\null^{-1aa}$ when $e^{-\alpha^a}n^{(b)}$ and 
$e^{-\alpha^a}n^{(c)}$ are both nonzero.

The velocity-momentum relations following from the above kinetic energy are
$$\eqalign{p_A &= \partial(N\Tscr)/\partial\du {\beta}^A = 
12e^{3\beta^0}N^{-1}
\eta_{AB}\du {\beta}^B\ , \qquad \du {\beta}^A = (12e^{3\beta^0})^{-1}N\eta
^{AB}p_B\cr
\s P_a &= \partial(N\Tscr)/\partial\dot{W}{} ^a = 2e^{3\beta^0}N^{-1}\overline{\Gscr}
_{ab} {\dot{W}{}}\null^b\ , \qquad \dot{W}{}\null^a = \half e^{-3\beta^0}N\overline{\Gscr}
\null^{-1ab}
\s P_b\ .\cr}\eqno(3.21)$$
Re-expressing the gravitational momentum (2.38) in terms of these momenta using
(3.15) leads to
$$ \bfpi^{\prime} = \hbox{\bf S}\bfpi\hbox{\bf S}^{-1} = \fraction{1}{12}
(\eta^{AB}p_A\hbox{\bf e}_B + 3p_0\hbox{\bf e}_0) + \half\overline{\Gscr}\null
^{-1ab}{\s P}_a\bfkappa_b^{\#\ \prime}\ .\eqno(3.22)$$
Defining
$$ P^{\prime}(\hbox{\bf A}) = -2\hbox{Tr}\,\bfpi^{\prime}\hbox{\bf A}\ ,
\qquad \hbox{\bf A}\in gl(3,R)\ ,\eqno(3.23)$$
one finds
$$p_A = -P^{\prime}(\hbox{\bf e}_A)\ ,
\qquad{\s P}_a = -P^{\prime}(\bfkappa_a)\ ,\eqno(3.24)$$
the minus sign arising from the inverse action used in the parametrization
(3.1) to conform with other conventions.  Note that the canonical generators
of the action (3.8) of $\A G$ on $\Mscr$ are instead (using (2.43), (3.12),
(3.22) and (3.23))
$$ P(\bfkappa_a) = P^{\prime}(\hbox{\bf S}\bfkappa_a\hbox{\bf S}^{-1}) =
P^{\prime}(\bfkappa_b){\A R}\null^b{}_{a} =- P_a\ ,\eqno(3.25)$$
namely the left invariant frame momenta.    Finally, re-expressing the
kinetic energy as a function on momentum phase space leads to
$$ N\Tscr = \half N(12e^{3\beta^0})^{-1}(\eta^{AB}p_Ap_B + 6 \overline{
\Gscr}\null^{-1ab}{\s P}_a{\s P}_b)\eqno(3.26)$$
The action (3.8) of $\A G$ is a symmetry of the free Hamiltonian dynamics 
generated by the kinetic energy function alone and hence the canonical
generators $P_a$ are conserved; this is clear from (3.26) since $P_a$
commute with $N\Tscr$ provided that $N$ is $\A G$-invariant.  The momenta
$\s P_a$ are related to the conserved momenta by the time dependent adjoint
transformation.

To summarize, $\{\partial/\partial\beta^A, {\s E}_a\}$ represents
through the parametrization (3.1) an orthogonal frame on $\Mscr$ adapted to
the three-plus-three decomposition of $\Mscr$ into orbits of the action of
$\A G$ (offdiagonal or ``angular" variables) and their orthogonal submanifolds
(diagonal or ``radial" variables).  In Misner's supertime time gauge $N=12
e^{3\beta^0}$, the diagonal part of the kinetic energy is simply the standard
kinetic energy of the flat Lorentz geometry of the metric space $(diag(3,R), 
\ketl\ketl,\ketr\ketr_{_{DW}})$ 
expressed in inertial
coordinates $\{\beta^A\}$, while the offdiagonal part of the
kinetic energy is just the one associated with the right invariant metric
$I = {1\over6}\overline{\Gscr}_{ab}{\s W}\null^a\otimes{\s W}\null^b$ on 
$\A G$, the two
pieces coupled together by the $\beta^{\pm}$ dependence of the components
$I_{ab} = {1\over6} \overline{\Gscr}_{ab}$.

The offdiagonal dynamics is thus governed by the natural generalization  of 
the rigid body dynamics at the canonical type IX point of $\Cscr_D$ where
$\A G=SO(3,R)$ to the
(multivalued) matrix group $\A G$ at each point of $\Cscr_D$.  This is 
discussed at length for a general Lie group by Abraham
and Marsden in a more abstract notation [46].
The spatial frames $e$ and 
$e^{\prime}$
are respectively identified with the space-fixed and body-fixed axes, related
by the passive transformation {\bf S} which corresponds to Goldstein's matrix
{\bf A} of his equation (4.46) [45],
the value assumed by {\bf S} for the 
canonical
type IX point of $\Cscr_D$ when expressed in his Euler angle parametrization 
of $SO(3,R)$.  Here $(-\du W\null^a,{-\dot{W}{}}\null^a)$ and $(-P_a,-{\s P}_a)$ 
play the roles of the space-fixed and body-fixed components of the angular 
velocity and spin angular
momentum of the rigid body, and in a general time gauge, $I_{(a)}= 2 N^{-1}
e^{3\beta^0}\overline{\Gscr}_{aa}$ play the role of the principal moments of 
inertia,
namely the  eigenvalues of the moment of inertia tensor $I$.  Notice that 
it is the space-fixed components of the angular momentum which are conserved 
by
the free dynamics, exactly as in the force free rigid body case; however, it
is the principal (body-fixed) axes of the moment of inertia tensor will allow
one to solve the equations of motion for that case.  It is this fact which
motivates diagonal gauge in the present problem. 

The identification of {\bf S}
with the passive coordinate transformation from space-fixed to body-fixed
coordinates leads to the minus sign which arises here and in (3.24).  The
parametrization (3.1) with $f$ rather than $f^{-1}$ would lead to the
identification of {\bf S} with the active transformation from the space-fixed 
basis
vectors to the body-fixed basis vectors (Goldstein's matrix {\bf A}$^{-1}$),
eliminating the minus sign and interchanging left and right in the above
discussion.  For the type IX case, this rigid body analogy is simply the
restriction to symmetry compatible diffeomorphisms and subsequent translation
into frame language of the Fischer-Marsden discussion for general spatially
compact spacetimes.  The finite dimensional situation here allows the analogy
to be carried much further.

     On the other hand, the left invariant frame momenta $P_a$ generate the
canonical action of $\A G$ on the momentum phase space and thus act like 
orbital
angular momenta in the central force problem, leading to the analogy in which
the radial and angular variables correspond to the diagonal and offdiagonal
variables of the present system.  The fact that this action of $\A G$ arises 
from
the action of spatial diffeomorphisms on the spacetime metric (2.31) makes
the decomposition into diagonal and offdiagonal variables relevant to the
Einstein equations.

A key feature of this decomposition is the resulting trivialization of the
supermomentum constraints (which are in general intimately connected with
the spatial diffeomorphism group) 
and of the nonpotential force.  These involve the
matrices $\{\bfdelta_a\}$ introduced in (2.51) and given 
explicitly by
$$ \vcenter{\halign{$\lft{#}$&$\lft{#}$\cr
\bfdelta_1 = -n^{(2)}\hbox{\bf e}^3{}_{2} +n^{(3)}\hbox{\bf e}^2{}_{3}
 -3a\hbox{\bf e}^3{}_{1} &=e^{\alpha^1}\bfkappa_1 -3a 2^{-{1\over2}}
\,\hbox{sgn}(n^{(1)})
\bfkappa_2 \cr 
\bfdelta_2 = -n^{(3)}\hbox{\bf e}^1{}_{3} +n^{(1)}\hbox{\bf e}^3{}_{1}
 -3a\hbox{\bf e}^3{}_{2} &=e^{\alpha^2}\bfkappa_2 +3a2^{-{1\over2}}
 \,\hbox{sgn}(n^{(2)})
 \bfkappa_1\cr
\bfdelta_3 = -n^{(1)}\hbox{\bf e}^2{}_{1} +n^{(2)}\hbox{\bf e}^1{}_{2}
 +a\hbox{\bf e}_+ &=e^{\alpha^3}\bfkappa_3 +a\hbox{\bf e}_+\ ,\cr}}
\eqno(3.27)
$$
which leads to the definition of a matrix $\bfrho$ by 
$$ \bfdelta_a =\bfkappa_b \rho^b{}_{a} + a_a\hbox{\bf e}_+\ .
\eqno(3.28)$$
The nonpotential force is then evaluated as follows
$$ \eqalign{Q&= 2e^{3\beta^0}a^c\ketl\bfdelta_c,\hbox{\bf g}^{-1}
 d\hbox{\bf g}\ketr 
= 2ae^{\beta^0+4\beta^+}\ketl\bfdelta_3,\hbox{\bf Sg}
^{-1}d\hbox{\bf gS}^{-1}\ketr\cr
&=4ae^{\beta^0+4\beta^+}(6ad\beta^+ +e^{\alpha^3}\overline{\Gscr}_{33}{\s W}
\null^3) \equiv Q_+d\beta^+ + Q_3\s W\null^3\ .\cr}
\eqno(3.29)$$
Note that the $\beta^+$ component of the nonpotential force exactly cancels
the $\beta^+$ component of the Einstein force arising from the exterior
derivative of the $a^2$ term in the expression (2.27) for $U_G$ 
$$ Q_+ - \partial/\partial\beta^+(6a^2e^{\beta^0+4\beta^+}) = 0\ ,
\eqno(3.30)$$
i.e., only the first term $e^{\beta^0}V^{\ast}$ is relevant to the equations 
of motion for the $\beta^{\pm}$ variables.  (Subtraction of this second term 
from $U_G$ in the Lagrangian or Hamiltonian would require the introduction
of a $\beta^0$ component of $Q$; if $\beta^0$ is determined by integrating
the super-Hamiltonian constraint rather than by equations of motion,
this is irrelevant, leaving only the component $Q_3$ as relevant to the 
remaining equations of motion and which vanishes for Bianchi type V.)

The primed gravitational supermomentum components are given by
$$ \Hscr_a^{G\ \prime} = P^{\prime}(\bfdelta_a) = -{\s P}_b\rho^b{}_{a}
-a_ap_+\ .\eqno(3.31)$$  
As already noted, these are linearly dependent for Bianchi types I, II and
VI$_{-1/9}$.  This is reflected by the vanishing of $\det\bfrho =
[e^{\alpha^1}e^{\alpha^2} +9a^2 \hbox{sgn}(n^{(1)}n^{(2)})]e^{\alpha^3}$ which
has the class A value $e^{\alpha^1+\alpha^2+\alpha^3}$ and the class B value
$|n^{(1)}n^{(2)}|(1+9h)e^{\alpha^3}$.  This also vanishes for Bianchi type V
where $e^{\alpha^3}=0$.

For the remaining types one may solve the supermomentum constraints for the
offdiagonal momenta or velocities
$$\eqalign{\s P_a &= -\rho^{-1b}_{\hskip 13pt a}(\Hscr_b^{G\ \prime} +a_bp_+)
=\rho^{-1b}_{\hskip 13pt a}(\Hscr_b^{M\ \prime}-a_bp_+)\cr
{\dot{W}{}}\null^a &= \half e^{-3\beta^0}N\overline{\Gscr}\null^{-1ab}\rho^{-1c}
_{\hskip 13pt b}(\Hscr_c^{M\ \prime} -a_cp_+)\ ,\cr}\eqno(3.32)$$
and thus determine the diagonal gauge shift vector field (3.15) modulo
time dependent elements of $\s g$ (spatial Killing vector fields).  This is
essentially just an application of the thin sandwhich formulation of the 
initial value problem [59,60];
initial values of $(\hbox{\bf g}^{\prime}, 
\dl {\hbox{\bf g}}^{\prime}, \Hscr_a^{M\ \prime})$ are specified arbitrarily 
and the supermomentum constraints are used to determine the diagonal gauge
shift variables.  For Bianchi type V the diagonal/offdiagonal decomposition
is not compatible with the thin sandwhich interpretation since the diagonal
momentum $p_+$ rather than the offdiagonal momentum $\s P_3$ is constrained,
while for types I, II and VI$_{-1/9}$ certain shift components (i.e.
offdiagonal momenta) remain freely specifiable initial data due to the
degeneracy of the constraints.  The important distinction between the 
present case and the thin sandwhich interpretation is that here the shift
variables have been made velocities associated with dynamical variables and
the supermomentum constraints allow one to determine the shift variables
directly at each moment of time rather than by integrating their (first
order) equations of motion.

     The idea of incorporating the shift variables into the dynamics as the
velocities associated with gauge variables was developed by Fischer and
Marsden [46].
Their construction when adapted to the spatially
homogeneous case extends the configuration space $\Mscr$ to $\Mscr\times 
R\times Aut_e(g)$ with natural variables $(\hbox{\bf g},t,\hbox{\bf A})$ and 
velocities $(\dl {\hbox{\bf g}}, \du t =N,
\du {\hbox{\bf A}}\hbox{\bf A}^{-1}= \hbox{ad}_{\overline{e}}(\b N))$ whose 
interpretation is the same as in (2.31)-(2.35).
Imposition of the $\A G$-diagonalization gauge condition $(\hbox{\bf g},
\hbox{ad}_{\overline{e}}(\b N))\in
\Mscr_D\times \A g$ on this extended system then leads to the results of the
present approach.

     Thus the diagonal/offdiagonal decomposition of the almost
synchronous gauge gravitational variables leads automatically to the diagonal
gauge equations of motion, while the diagonal gauge supermomentum constraints
then determine the diagonal gauge shift variables which appear in these
equations of motion, except in the degenerate cases (Bianchi types I, II, V
and VI$_{-1/9}$) where first order equations of motion for the
undetermined shift variables must be used to evolve them.  These first order
equations may be obtained directly in Hamiltonian form by using the
Poisson bracket relations (3.13).  (For the vacuum dynamics these are just
generalized Euler equations with time dependent principal moments of inertia,
driven by the additional nonpotential force component $Q_3$ in the class B
case.)  To complete this scheme one needs to 
introduce the diagonal gauge matter variables by a time dependent change of
variables involving the transformation matrix {\bf S}, leading to the
appearance of shift terms in the new matter equations of motion through 
(3.15).  These terms are analogous to the centrifugal force which appears
in the rotating body-fixed frame of the rigid body problem. 
 
The combined Einstein and matter equations then involve only diagonal gauge
variables and the diagonal gauge shift variables which are in general
determined by the supermomentum constraints (or by first order equations
of motion if not).  The spacetime metric may then be represented directly
in diagonal gauge form, without integrating the equations which determine
{\bf S}, the transformation back to almost synchronous gauge.  The 
offdiagonal part of the kinetic energy then acts like a time dependent
effective potential for the diagonal gauge gravitational variables, which
from the central force analogy has been called the centrifugal potential
by Ryan [20,21].  
The matter super-Hamiltonian acts as another 
time dependent potential for these variables.
  
For a perfect fluid source as discussed in appendix C, the appropriate 
variables are $(n,l,v_a)$; the diagonal gauge variables are defined by (C.3)
with $\hbox{\bf A}\,=\,\hbox{\bf S}$, namely
$$ (n^{\prime},l^{\prime},v_a^{\prime}) = (n,l,v_bS^{-1b}_{\hskip 13pt a})\ .
\eqno(3.33)$$
The primed components of the matter supermomentum and the matter 
super-Hamiltonian then have the expressions
$$\eqalign{\Hscr^{M\ \prime}_a &= -2klv_a^{\prime}\cr
\Hscr_M &= 2kl(\mu^2 + g^{\prime ab}v_a^{\prime}v_b^{\prime})^{1\over2}
 -2kpe^{3\beta^0}\cr
g^{\prime ab}v_a^{\prime}v_b^{\prime} &= e^{-2\beta^0}(e^{4\beta^+_1}
 v_1^{\prime\ 2} + e^{4\beta^+_2}v_2^{\prime\ 2} +e^{4\beta^+_3}v_3^
 {\prime\ 2})\ .\cr}\eqno(3.34)$$
In the Bianchi type I dust (zero pressure) case, Ryan has called the first
term in the matter super-Hamiltonian the rotation potential since it directly
influences the gravitational variables only when the fluid is tilted 
($v_a\neq 0$) [61]
and tilt is equivalent to rotation of the fluid in this
case.  However, for the remaining types tilt and nonzero rotation ($C^{ab}
v_b\neq 0$) are no longer synonymous [43], 
so tilt potential is a 
better name.  Let $\Hscr^{tilt}_M$ designate this potential.  The second
term or pressure potential only affects the equations of motion for $\beta^0$.

To evaluate the equations of motion for the offdiagonal momenta, one needs
the following Poisson brackets which are a consequence of (3.13) and (3.33)
$$ \{ {\s P}_a, v^{\prime}_b\} =v_c^{\prime}\kappa_{a\hskip 4pt 
b}^{\hskip 4pt c} \qquad \{ P_a, v^{\prime}_b\} =
v^{\prime}_c\kappa^{\hskip 4pt c}_{d\hskip 4pt b}{\A R}\null^d{}_{a}\ .
\eqno(3.35)$$
One then finds the result
$$ \eqalign{
({\s P}_a) {\dl{\ }} 
&= \half Ne^{-3\beta^0} \s P_c 
{\hat C}\null^c{}_{ba} \overline{\Gscr}\null^{-1bd}{\s P}_d  +N(Q_a +F_a)\cr
&={\s P}_c{\hat C}\null^c{}_{ba}\dot{W}{}\null^b +N(Q_a +F_a)\cr
(P_a){\dl{\ }} &=({\s P}_b\A R\null^b{}_{a}){\dl{\ }} 
=N(Q_a+F_b{\A R}\null
^b{}_{a})\cr
\qquad&
F_a \equiv \{{\s P}_a, \Hscr^{tilt}_M\} 
=2kl(v^{\sst \bot})^{-1} v^{\prime}_c\kappa^{\hskip 4pt c}_{a\hskip 4pt b}
v^{\prime b} \cr}
\eqno(3.36)$$
For the free dynamics and class A vacuum dynamics ($Q_a=0=F_a$), these 
equations generalize the Euler equations for the body-fixed components of
the angular momentum of a rigid body, and like those equations conserve the 
space-fixed components
of the angular momentum. For the class B vacuum dynamics the component $Q_3$
appears as a driving term and the conserved quantities are instead the 
unprimed supermomentum components
$$\eqalign{ \Hscr^G_a &= P(\bfdelta_a)=P^{\prime}(\bfdelta_b)S^b{}_{a}
=P^{\prime}(\hbox{\bf S}\bfdelta_a\hbox{\bf S}^{-1})\cr
&=P^{\prime}(\hbox{\bf S}\bfkappa_b\hbox{\bf S}^{-1})\rho^b{}_{a}+ a_a 
P^{\prime}(\hbox{\bf Se}_+\hbox{\bf S}^{-1})\cr
&=P^{\prime}(\bfkappa_c)R^{-1c}_{\hskip 13pt b} \rho^b{}_{a}
+a_aP^{\prime}(\hbox{\bf e}
_+-3(\theta^1\bfkappa_1
+\theta^2\bfkappa_2))\cr
&=-P_b\rho^b{}_{a} -a_a(p_+-3(\theta^1{\s P}_1+\theta^2{\s P}_2))\ .\cr}
\eqno(3.37)$$  
The momenta $P_1$ and $P_2$ are conserved by the vacuum dynamics, being 
constant linear 
combinations of $\Hscr^G_1$ and $\Hscr^G_2$, but $P_3$ is not unless $Q_3=0$.
Because of the supermomentum constraints (3.32), the equations of motion for 
the offdiagonal momenta $\s P_a$ are very closely related to the primed
fluid equations of motion which may be obtained by transforming (C.5)
$$ \du l = Nl(v^{\sst \bot})^{-1}\,2a^{\prime c}v^{\prime}_c\qquad
(v^{\prime}_a)\dl{\ }\ =Nv^{\prime}_b[(v^{\sst \bot})^{-1}C^b{}_{ca}v^{\prime c}
-\kappa^{\hskip 4pt b}_{c\hskip 4pt a}\dot{W}{}\null^c]\eqno(3.38)$$

The lapse $N$ is still arbitrary and the super-Hamiltonian constraint remains 
to be dealt with.  Several options are available.  Introduce the function $I
_{\s h}$ depending on an arbitrary real parameter $\s h$ by
$$ I_{\s h} = -(\hbox{sgn}\,p_0)[24e^{3\beta^0}(\Hscr + \s h)+p_0^{\ 2}]
^{1\over2}\ ,\eqno(3.39)$$ 
enabling the super-Hamiltonian constraint to be expressed in the compact form
$$\Hscr = \fraction{1}{24}e^{-3\beta^0}
(-p^{\ 2}_0 + I_0^{\ 2})=0\ ,\eqno(3.40)$$  
which suggests that it be used to solve for $p_0$, the result being $p_0=
-I_0$.  ($p_0$ is negative for expansion or increasing $\beta^0$ with 
increasing proper time and is positive for contraction.)  

Suppose one initially assumes unit lapse function, so that the time of the
classical mechanical system coincides with the proper time $t$ on the 
spacetime.  The super-Hamiltonian constraint may then be used to eliminate 
another degree of freedom from the system by a standard technique of 
classical mechanics [1]
which replaces the time $t$ as the integration
variable by some function $\overline{t}$ of $\bfbeta$ (accomplished by 
changing the independent variable in the action).  This is called an 
``intrinsic time reduction" of the system [62].
The usual and obvious
choice is $\overline{t}=\beta^0=-\Omega$ or negative $\Omega$-time.  
($\Omega$-time is useful for approaching the initial singularity.)  This
is clearly equivalent to specifying a new lapse function $N$.  The reduction
technique leads to the reduced Hamiltonian $I_0$ and equations of motion
$$ {\du {\beta}}\null^{\pm} = \{ \beta^{\pm}, I_0\}\ ,\qquad
{\dl p}_{\pm} = \{ p_{\pm}, I_0\} + \delta^+{}_{\pm}NQ_+\ ,
\eqno(3.41)$$
where the new lapse function $N$ is determined by the first order equation for
the original time $t$ 
$$ N= \du t = \partial I_{\s h} /\partial {\s h}\relv_{\s h =0} = 12 e^{3
\beta^0}I_0^{\ -1}\eqno(3.42)$$
and the dot refers to the new time derivative.

This same result can also be obtained by simply setting $N=-12e^{3\beta^0}
p_0^{\ -1} $ in the original Hamiltonian $H = N\Hscr$ 
$$\eqalign{H &= \half(p_0 - p_0^{\ -1}I_0^{\ 2})\cr
  {\du {\beta}}\null^0 &= \{\beta^0,H\} = \half(1+p_0^{\ -2}I_0
^{\ 2})\ ,\cr}\eqno(3.43)$$
the result ${\du {\beta}}\null^0 =1$ following from the imposition of the 
constraint
$p_0=-I_0$.  For any variable $y$ other than $p_0$, one has $\partial H/
\partial y = (-p_0^{\ -1}I_0)\partial I_0/\partial y$, showing that $I_0$ acts
as the Hamiltonian for all variables but $\beta^0$ when the constraint $p_0=
-I_0$ is imposed.  Misner's supertime time gauge  $N=12e^{3\beta^0}$ differs
differs only by eliminating the factor $-p_0$ in the $\beta^0$-time lapse
function, leading to a Hamiltonian of familiar form rather than a square root
Hamiltonian as above
$$ H= -\half p_0^{\ 2}+\half I_0^{\ 2}\ .\eqno(3.44)$$
In this time gauge, ${1\over2}I_0^{\ 2}$ acts as the Hamiltonian for the
variables other than $(\beta^0,p_0)$ and the super-Hamiltonian constraint
${\du {\beta}}\null^0=-p_0=I_0$ may be used to evolve $\beta^0$.  Note that
$$\eqalign{\half I_0^{\ 2} &= H_{eff} + 72a^2e^{4(\beta^0+\beta^+)} -
24e^{6\beta^0}kp\cr
H_{eff} &= \half (p_+^{\ 2} + p_-^{\ 2}) + 3\overline{\Gscr}\null^{-1ab}
{\s P}_a{\s P}_b + 12 e^{4\beta^0}V^{\ast} +12 e^{3\beta^0}\Hscr^{tilt}_M\ 
.\cr}\eqno(3.45)$$
The term $H_{eff}$ acts as an effective Hamiltonian for the variables other 
than
$(\beta^0,p_0)$; the corresponding Hamiltonian equations of motion for the
$\beta^{\pm}$ degrees of freedom contain no nonpotential force.

Whatever the time gauge, one may introduce an effective Hamiltonian for the
$\beta^{\pm}$ degrees of freedom by subtracting out the $p_0^{\ 2}$ term,
the pressure potential and the $a^2$ term in the scalar curvature potential
$$\eqalign{H_{eff} &= H-N[-\fraction{1}{24}e^{-3\beta^0}p_0^{\ 2} +6a^2e^
{\beta^0+4\beta^+} -2kpe^{3\beta^0}]\cr
 &= N(12e^{3\beta^0})^{-1}[\half(p_+^{\ 2} +p_-^{\ 2}) 
+3\overline{\Gscr}
\null^{-1ab} {\s P}_a{\s P}_b +12e^{4\beta^0}V^{\ast} + 
24e^{3\beta^0}\Hscr_M^{tilt}]\ .\cr}\eqno(3.46)$$
As discussed above, the use of $H_{eff}$ rather than $H$ eliminates the need
to consider the nonpotential force component $Q_+$, so $H_{eff}$ alone
generates the correct equations of motion for the $\beta^{\pm}$ degrees of
freedom.  The component $Q_3$ is still relevant to the equations of motion 
for the offdiagonal variables, when needed, for which either $H$ or $H_{eff}$
may be used.  A nonvanishing value of $Q_3$ is connected with the single
nonvanishing offdiagonal component of the spatial Ricci tensor in diagonal
gauge (see (2.25)). 

One advantage of the supertime time gauge is that the kinetic energy for the
diagonal variables corresponds to a relativistic particle in the flat 
3-dimensional spacetime ($diag(3,R)$, $\eta_{AB}d\beta^A\otimes d\beta^B$).
Neglecting the remaining terms in the Hamiltonian (the free system with
zero orbital angular momentum), one obtains affinely parametrized geodesics 
for the system trajectories and the Hamiltonian constraint just requires
that the particle velocity be null.  Except for small anisotropy in the
type IX case, the remaining terms are all positive (although the pressure
potential is negative, the total fluid Hamiltonian is positive) and their
effect is therefore to make the particle velocity timelike.  It is future
pointing in the case of expansion ($\beta^0$ increasing) and past pointing
in the case of contraction ($\beta^0$ decreasing), reversing direction
only in the type IX case where recollapse occurs.

The $\beta^0$-time gauge, on the other hand, uses the natural time variable
$\beta^0$ of the 3-dimensional flat spacetime to parametrize the system 
trajectories; for the geodesics it is also an affine parameter.  This has the
advantage of making the reduced Hamiltonian for the $\beta^{\pm}$ degrees of
freedom explicitly rather than implicitly time dependent. 

The curvature potential $V^{\ast}$ has been described and diagrammed in the 
previous section and enters the Hamiltonian in the supertime time gauge as the 
time dependent potential  $U_g=12e^{4\beta^0}V^{\ast}$, which apart from 
constants has been called the gravitational potential by Ryan [20,21].
The centrifugal potential 
$$U_c=3\overline{\Gscr}\null^{-1ab}{\s P}_a{\s P}_b =
 3\sum_{n=1}^3\overline{\Gscr}\null^{-1ab}{\s P}^{\ 2}_a \equiv \sum_{n=1}
^3U_c^{(a)}\eqno(3.47)$$
is the sum of three terms whose $\beta^{\pm}$ dependence is given by (3.19) 
and is repeated in a more suggestive form here.

$$\vcenter{\halign{$\rt{#}$&\quad$\lft{#}$\cr
(i)& e^{-2\alpha^a}n^{(b)}n^{(c)}>0:\hskip 15pt
\overline{\Gscr}\null^{-1aa} = \half|e^{-2\alpha^a}n^{(b)}n^{(c)}|^{-1}
\hbox{sinh}^{-2}(2\sqrt3(\beta^-_a -\beta^-_{a0}))\cr
(ii)& e^{-2\alpha^a}n^{(b)}n^{(c)}<0:\hskip 15pt
\overline{\Gscr}\null^{-1aa} = \half|e^{-2\alpha^a}n^{(b)}n^{(c)}|^{-1} 
\hbox{cosh}^{-2}(2\sqrt3(\beta^-_a-\beta^-_{a0}))\cr
(iii)& \left\{\vcenter{\vbox{\hbox{$
e^{-\alpha^a}n^{(b)}\neq 0, e^{-\alpha^a}n^{(c)}=0:\hskip 15pt
 \overline{\Gscr}\null^{-1aa} = 2|e^{-2\alpha^a}n^{(b)\,2}|^{-1}e^{-4\sqrt3
 \beta^-_a}$}\hbox{$
e^{-\alpha^a}n^{(b)}=0, e^{-\alpha^a}n^{(c)}\neq 0:\hskip 15pt
 \overline{\Gscr}\null^{-1aa} = 2|e^{-2\alpha^a}n^{(c)\,2}|^{-1}e^{4\sqrt3
 \beta^-_a}$}}}\right.\cr
&\hskip 40pt 2\sqrt3\beta^-_{a0} \equiv -\half
\ln|e^{-\alpha^a}n^{(b)}/e^{-\alpha^a}
n^{(c)}|\ .\cr}}\eqno(3.48)$$

\begin{figure}[t!] 
\begin{center}
\includegraphics[width=.9\textwidth]{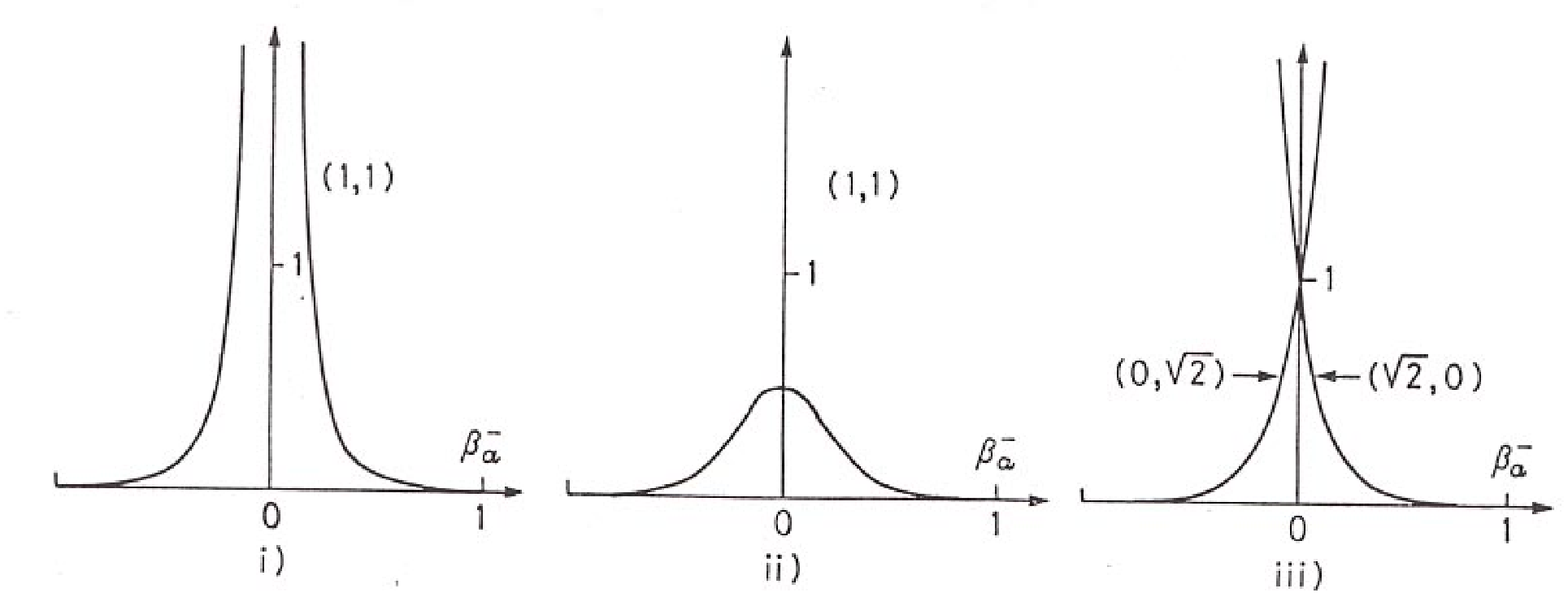}
\end{center}
\caption{Plots of 
$\overline{\Gscr}\null^{-1aa}$
for canonical points of $\Cscr_D$, at which the pair $e^{-\alpha^a}
(n^{(b)},n^{(c)})$ assumes the values shown.}
\end{figure}

These are illustrated for the canonical points of $\Cscr_D$ in Figure 7.
The potentials for noncanonical points arise from these by a combined 
rescaling and translation, infinite translations corresponding to Lie algebra
contractions. All of these potentials are exponentially cut off in one or
both directions.  Case $(i)$ corresponds to a compact generator $\bfkappa_a$
in which the potential behaves like $(\beta^-_a)^{-2}$ at the origin, exactly
as in the central force problem where the centrifugal potential has the 
dependence $r^{-2}$ on the radial coordinate $r$.  Thus the singularities of
the parametrization (3.1) are shielded by angular momentum barriers exactly 
like the spherical coordinate singularity at the origin in the central force 
problem.  Case $(ii)$ also acts like an angular momentum barrier but can be
overcome if the energy of the system is sufficiently large.  In case $(iii)$
the barriers have been translated out to infinity.  In the supertime time 
gauge, all of the time dependence of the centrifugal potentials arises from
the angular momentum factors.  Increasing ${\s P}\null_a^{\ 2}$ increases 
$U_c^{(a)}$ and causes a given value of the potential to move in the direction
of decreasing values.  The barrier $(i)$ may be crossed provided that the
relevant angular momentum component vanishes at the moment of crossing.

\begin{figure}[t!] 
\begin{center}
\includegraphics[width=.9\textwidth]{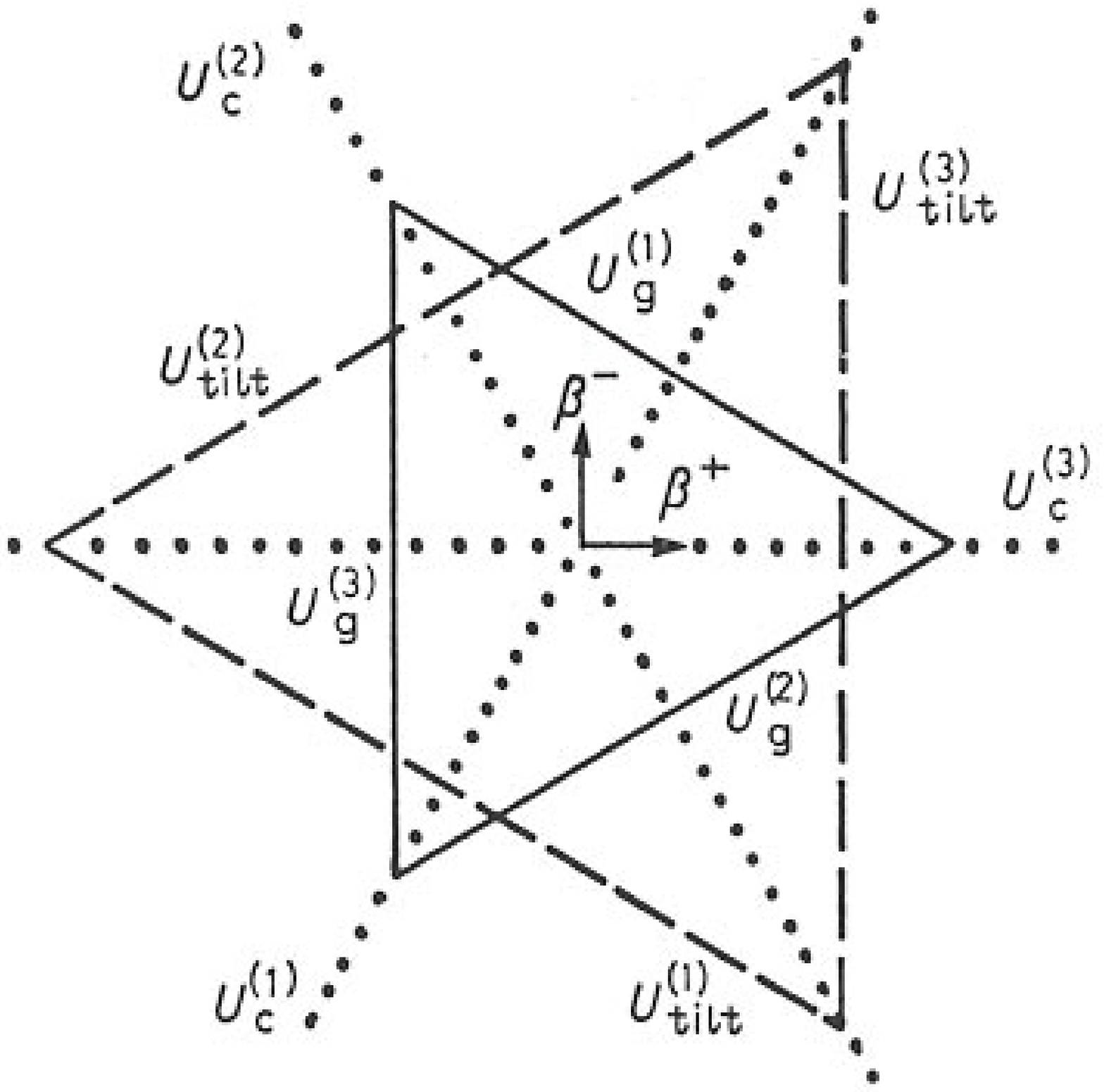}
\end{center} 
\caption{
The key to Figure 9.  
Together
with Figure 2, this shows the $\beta^{\pm}_a$ coordinate associated with each
potential represented in Figure 9 by wall contours.  In general the wall
contours associated with tilt and type $(iii)$ centrifugal potentials will be
translated from their symmetrical positions.  The dotted lines for the type 
$(i)$ and $(ii)$ centrifugal potential contours in Figure 9 have double arrows
since they actually represent a pair of contours symmetrically located with
respect to the point at which the potential becomes infinite or assumes its
maximum value. }
\end{figure}

\begin{figure}[p!] 
\begin{center}
\includegraphics[width=.9\textwidth]{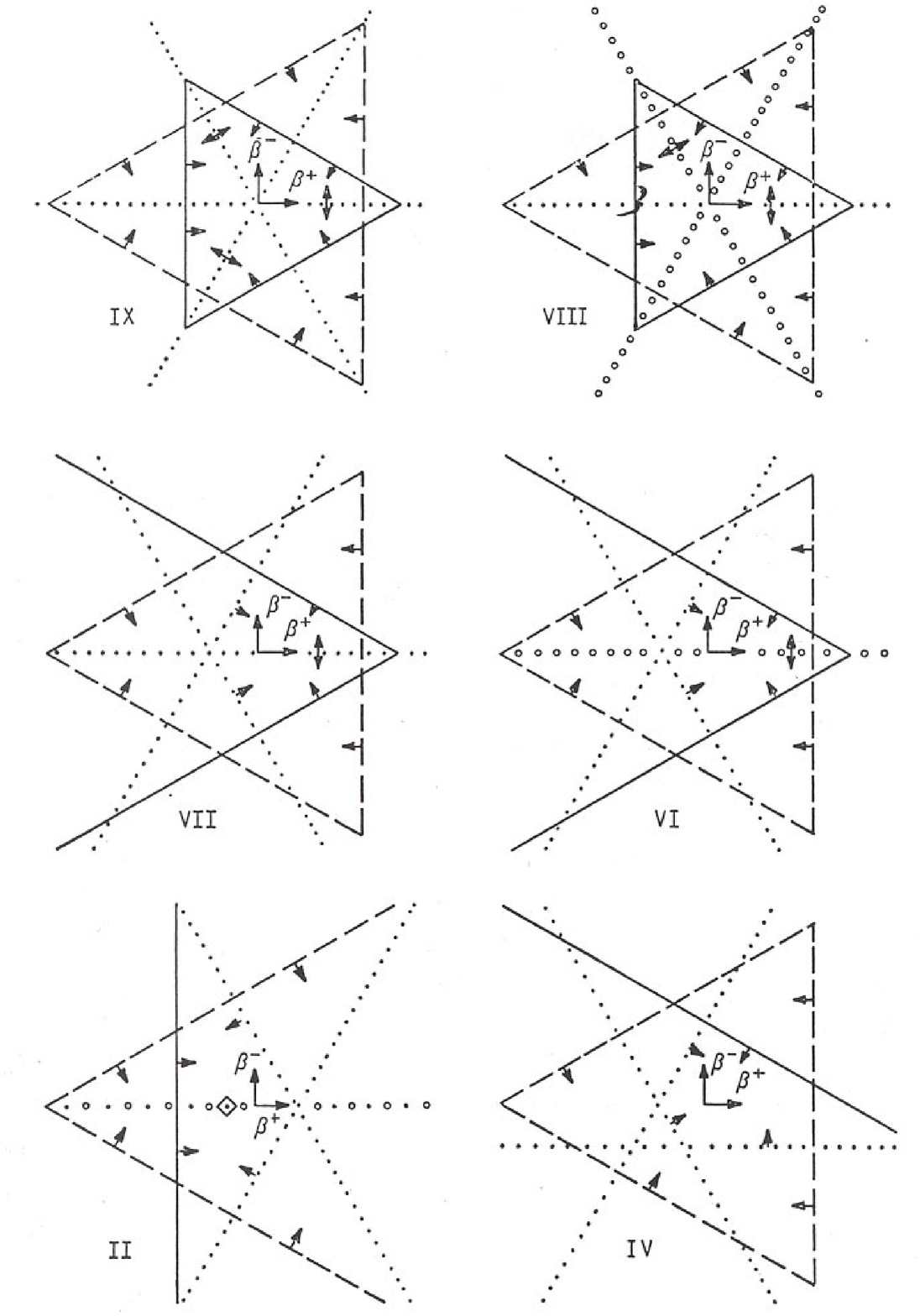}
\end{center}
\caption{
The walls associated with the 
gravitational, tilt and centrifugal potentials for the canonical points of 
$\Cscr_D$.  Arrows indicate the direction of decreasing values of each
potential.  Although shown as symmetrical about the origin for simplicity, 
the tilt and type $(iii)$ centrifugal walls will in general assume
unsymmetrical positions. Figure 8 explains the conventions for this figure.}
\end{figure}

For the canonical points of $\Cscr_D$, the gravitational potential (in 
supertime time gauge) is closely approximated by the simple time dependent
Bianchi type II exponential potential $U_g^{(a)}
 = 12n^{(a)\,2}e^{4\beta^0-8\beta^+_a}$ between the positive $\beta^+_b$- and
$\beta^+_c$-axes and outside the channels. 
Because of the simple exponential dependence of this potential,
a given value of this potential has a constant value of $\beta^0-2\beta_a^+$,
i.e. its contours move with velocity $d\beta^+_a/d\beta^0={1\over2}$ in the
flat 3-dimensional spacetime $(diag(3,R),\eta_{AB}d\beta^A\otimes 
d\beta^B)$.  Similarly the tilt potential (in supertime time gauge)
in the limit of large anisotropy
or extreme tilting along the diagonal axis $e_a^{\ \prime}$ is closely 
approximated by the exponential potential $U^{(a)}_{tilt} = 24kle^{2(\beta^0 +
\beta^+_a)} |v_a^{\ \prime}|$ between the negative $\beta^+_b$- and 
$\beta^+_c$-axes, and hence
its contours move with velocity $d\beta^+_a/d\beta^0=-1$ plus whatever
additional velocity is due to the time variation of $v^{\prime}_a$ 
(and $l$ in the class B case).
 These
exponential potentials are indicated by single equipotential lines in Figures
8 and 9, the direction of decreasing values (the direction of the associated 
force) indicated by arrowheads on these lines. Figure 8 provides a key which
indicates the labeling of the various potentials of Figure 9. 
Although the tilt equipotential 
locations are shown as symmetrical, decreasing $v_a^{\ \prime}$ translates 
$U^{(a)}_{tilt}$ in the positive $\beta^+_a$ direction, so that when an 
individual (primed) component of the fluid current vanishes, the corresponding
tilt potential moves out to infinity.  A similar statement holds for the 
exponential gravitational potentials and their dependence on $|n^{(a)}|$.
One must actively translate these canonical potentials by  the translation
(and rescaling) (2.30) to obtain the noncanonical potentials,
with Lie algebra contraction causing the potentials to move out to infinity.
Equipotential lines of these potentials move with $\beta^0$-velocity ${1\over
2}$ in the direction of decreasing values, while the tilt potentials in the
exponential regime move with 
$\beta^0$-velocity 1 in the direction of decreasing values
neglecting the additional time dependence due to the fluid variables.
The centrifugal potentials in supertime time gauge have no explicit
dependence on $\beta^0$ and hence their motion is entirely due to the 
$\A G$-angular momentum
factor.  When this latter factor is constant as occurs in
the class A symmetric cases to be described shortly, these potentials are 
stationary.
The type VI and VII potentials in each sector between the negative 
$\beta^+_b$- and $\beta^+_c$-axes, namely $U_g(\hbox{VII}, n^{(a)}=0) =
2|n^{(b)}n^{(c)}| e^{4(\beta^0+\beta^+_a)}\hbox{sinh}^2 2\sqrt3(\beta^-_a
-\beta^-_{a0})$ and
$U_g(\hbox{VI}, n^{(a)}=0) = 2|n^{(b)}n^{(c)}| e^{4(\beta^0+\beta^+_a)}
\hbox{cosh}^22\sqrt3(\beta^-_a-\beta^-_{a0})
$ simply translate along the $\beta^+_a$ 
axis with velocity $d\beta^+_a/d\beta^0=-1$ in supertime time gauge, so
the channels move with unit $\beta^0$-velocity in this time gauge
(outside their common intersection at semisimple points of $\Cscr_D$).
 
In the $\beta^0$-time gauge, all of the potentials become multiplied by an
additional factor of $(-p_0)^{-1}$, leading to additional implicit 
time dependence and an additional contribution to the velocities of their
contours.

Before discussing the qualititive effect of the various potentials on the
dynamics, it is important to have in mind a classification of the 
specializations which can occur.  Because of the existence of various 
additional continuous  (local rotational symmetry and isotropy) and discrete
spacetime symmetries, the dynamics admits various special cases where one may 
consider the phase subspace associated with a submanifold of the configuration
space such that initial data in this phase subspace remains in it when evolved
by the equations of motion.  These special cases are important since they are
governed by much simpler systems of differential equations which often admit
exact solutions.  These exact solutions are the basis of analytic 
approximation schemes or ``diagrammatic solutions".
 
Since $Aut_e(g)$ is a symmetry group of the equations of motion, any such
phase subspaces which are related by the action of an automorphism are 
equivalent and it suffices to consider only one representative from each
equivalence class.  These have been described elsewhere for Dirac 
spinor [42] 
and perfect fluid [43] 
sources for the canonical points of $\Cscr_D$
and are summarized for the perfect fluid case in Table 3.   

\begin{table} 
\begin{center}
\vrule width0pt\boxit{\hbox{\qquad General case \qquad}}
\\
\vrule width0pt
\hbox{
\vrule width0pt height 12pt
\hskip 1cm
\vrule width1pt height 12pt
\hskip 1cm
\vrule width0pt height 12pt
\hskip 1cm
\vrule width1pt height 12pt
}
\\
\vrule width0pt\hbox{\boxit{\vbox{
\halign{\lft{#}\cr
Symmetric cases:  {\bf g} $\in\Mscr_{S(a)}$ , $v_b=v_c=0$\cr
\quad class A:  all types $(a=1,2,3)$\cr
\quad class B:  all types $(a=3)$\cr}}}
\hfill
\boxit{\vbox{
\halign{\lft{#}\cr
Generalized Taub-like cases:  {\bf g} $\in\Mscr_{GT(a)}$ , $v_b=v_c$\cr
\quad class A:  I $(a=1,2,3)$, V $(a=3)$ \cr
\quad class B:  V,\  VI$_h$ $(a=3)$\cr}}} }
\\
\vrule width0pt\hbox{\vrule width1pt height 12pt
\hskip 3cm
\vrule width1pt height 12pt
\hskip 3cm
\vrule width1pt height 12pt}
\\ 
\vrule width0pt\vbox{\hbox{\boxit{\halign{\lft{#}\cr
Diagonal case:  {\bf g} $\in\Mscr_D$\cr
\quad class A:  all types; $v_a=0$\cr
\quad class B:  type V; $v_1=v_2=0$\cr}} 
\hfill
\boxit{\halign{\lft{#}\cr
Taublike symmetric cases: {\bf g} $\in\Mscr_{TS(a)}\ ,\ v_b=v_c=0$\cr
\quad class A:  type VI$_0\ (a=3)$ , type I $(a=1,2,3)$\cr
\quad class B:  types VI$_h$,V $(a=3)$\cr}} }}
\\
\vrule width0pt\hbox{\vrule width1pt height 12pt
\hskip 3cm
\vrule width0pt height 12pt
\hskip 3cm
\vrule width1pt height 12pt}
\\
\vrule width0pt\boxit{\vbox{
\halign{\lft{#}\cr
Taublike cases:  {\bf g} $\in\Mscr_{T(a)}$\cr
\quad class A:  $\{\vcenter{\vbox{\hbox{types IX, I $(a=1,2,3)$}
\hbox{others $(a=3)$}}}\}$; $v_1=v_2=v_3=0$\cr
\quad class B: $\{\vcenter{\vbox{\hbox{types VII$_h$, V $(a=3)$; $
v_1=v_2=0$}\hbox{type III$\equiv$VI$_{-1}\ (a=2)$; $v_1=v_2=v_3=0$}}}
\}$\cr}} }
\\
\vrule width1pt height 12pt
\\ 
\vrule width0pt\boxit{\vbox{
\halign{\lft{#}\cr
Isotropic case:  {\bf g} $\in\Mscr_I$, $v_a=0$\cr
\quad class A:  types IX, VII$_0$, I\cr
\quad class B:  types VII$_h$, V\cr}}} 
\end{center}

\caption{
Specialization diagram for the
perfect fluid dynamics at the canonical points of $\Cscr_D$ , letting
$(a,b,c)$ be a cyclic permutation of $(1,2,3)$ when appropriate.  For
noncanonical points of $\Cscr_D$ , the submanifolds analogous to $\Mscr
_{T(a)}$, $\Mscr_{TS(a)}$, and $\Mscr_{GT(a)}$ 
are characterized by the condition (3.20) 
rather than $\beta^-_a=0$ .  For noncanonical class A points of $\Cscr_D$ ,
the type VI$_0$ Taublike symmetric cases occur for the index value coinciding
with the index associated with the vanishing diagonal component of {\bf n}
and the Taublike cases occur for the index $a$ such that $n^{(b)}$ and
$n^{(c)}$ are of the same sign or both vanish.  The manifold analogous to
the isotropic submanifold is obtained by applying to $\Mscr_I$ the diagonal
transformation which transforms the canonical point into the given 
noncanonical point of $\Cscr_D$.}
\end{table}
For types I and IX all symmetric and Taublike cases (and for type I all 
Taublike symmetric cases) are equivalent, while only two inequivalent symmetric
cases exist for the remaining class A types.  The existence of nontrivial
continuous isotropy subgroups of the action of $Aut_e(g)$ on $\Mscr$ for
types I, II and V leads to the following additional equivalences for perfect 
fluid sources.  For type I the general case is equivalent to the diagonal case.
For types II and V the symmetric case $\Mscr_{S(3)}$ is equivalent to the
diagonal case, while the Taublike type V case is equivalent to the isotropic
case when $v_3=0$. All Taublike cases except for Bianchi types VI$_0$ and 
VI$_h$ are associated with local rotational symmetry and all isotropic cases
with spatial isotropy.   There
also exist equivalences between different Bianchi types [42].
The type
VII$_0$ and type I Taublike cases are equivalent, as are the type VII$_h$ and
type V Taublike cases.
The symmetric cases correspond to the situation
in which the $\A G$-angular momentum is aligned with one of the body-fixed 
axes, 
i.e. ${\s P}_a$ has only one nonvanishing component (a class A constant of the
motion); this preferred body-fixed axis coincides with one of the space-fixed
axes and is a Kasner axis.  The remaining two Kasner axes move in the plane
orthogonal to this axis.  
The diagonal case corresponds to vanishing angular momentum, in which case the
frame $e=e^{\prime}$ is a Kasner frame. The general case is characterized by 
the fact that the $\A G$-angular momentum is not aligned with any of the 
body-fixed axes so that the Kasner axes are distinct from the body-fixed 
axes.

For the class B case, certain cases are characterized by the vanishing of the
nonpotential force, so that the system becomes an ordinary Hamiltonian system.
The symmetric case submanifold $\Mscr_{S(3)}$ is described by the conditions
$\theta^1=\theta^2=0$ in terms of the combined parametrization (3.1) and (3.7).
The restriction of ${\s W}\null^a$ to this submanifold is simply 
${\s W}\null^a\relv_{S(3)}
= \delta^a{}_{3}d\theta^3$, using an obvious abbreviated notation for the
restriction, so the restriction of the nonpotential force is
$$\eqalign{
Q\relv_{S(3)} 
&= 4a^2 e^{\beta^0+4\beta^+}( 6d\beta^+ 
+ a^{-1}e^{\alpha^3}\overline{\Gscr}_{33}d\theta^3)\cr
\overline{\Gscr}_{33} &= \half e^{-2\alpha^3}(n^{(1)} e^{2\sqrt3 \beta^-}
- n^{(2)}e^{-2\sqrt3\beta^-})^2\ .\cr}\eqno(3.49)$$
For this symmetric case the fluid must satisfy $v_1=v_2=0$ or equivalently
$v_1^{\ \prime}=v_2^{\ \prime}=0$.  For type V, one has $e^{\alpha^3}=0$
and $Q\relv_{S(3)}$ is proportional to $d\beta^+$; when $v_3=0$, the third
supermomentum constraint becomes ${\du \beta}\null^+=0$ and the constant 
value of $\beta^+$ may be transformed to  zero by the action of an 
automorphism, leaving
an ordinary Hamiltonian system in the remaining degrees of freedom.  (This
case is equivalent to the diagonal case.)  The Taublike type VII$_h$ case,
being equivalent to the Taublike type V case, is also Hamiltonian. 
$(\overline{\Gscr}_{33}=0.)$  For type
VI$_h$ (or type V) the Taublike symmetric case is described by the 
additional condition $d\overline{\Gscr}_{33} =0$ (namely (3.18) which is 
equivalent to $n^a{}_{a}=0\ ^{(14)}$)
 or $e^{-\alpha^3}n^{(1)} e^{2\sqrt3\beta^-} 
=-e^{-\alpha^3}n^{(2)}e^{-2\sqrt3\beta^-} = e^{-\alpha^3}q$,
where $q$ is defined by the two conditions $q^2=-n^{(1)}n^{(2)}$ and $\hbox{sgn}
\,q = \hbox{sgn}\,n^{(1)}$, so that 
$\overline{\Gscr}_{33} = e^{-2\alpha^3}q^2$ 
and the restriction of the nonpotential force to the Taublike symmetric case
configuration space is 
$$
Q\relv_{TS(3)} 
= -4a^2e^{\beta^0+4\beta^+}d \ln u\ ,\qquad
\ln u \equiv -6\beta^+ -2\lambda qe^{-\alpha^3}\theta^3\ , 
\eqno(3.50)
$$
while the third supermomentum constraint becomes
$$
\Hscr^G_3\relv_{TS(3)} = 2ae^{2(\beta^0+\beta^+)}(\ln u)\dl{\ } = 2klv_3\ .
\eqno(3.51)$$
When $v_3=0$, then this constraint requires one to restrict the Taublike 
symmetric case configuration space by the condition $du=0$, leading to the
vanishing of the nonpotential force and an ordinary Hamiltonian system.  
Setting $q=0$ gives the type V limit which is again equivalent to the 
diagonal case.

The Taublike symmetric case $\Mscr_{ST(3)}$ is diagonalized by the following
linear transformation {\bf AB}, which transforms the structure constant
tensor out of the space $\Cscr_D$ except for type V $(q=0)$ 
$$
\eqalign{
\bfkappa_3  
&= e^{-\alpha^3}(-n^{(1)}\hbox{\bf e}^2{}_{1} 
+ n^{(2)}\hbox{\bf e}^1{}_{2})\ ,\qquad \hbox{\bf n} 
= \hbox{diag}(n^{(1)}, n^{(2)}, 0)\cr
\overline{e}_a 
&= (\hbox{\bf AB})^{-1b}_{\hskip 13pt a}e_b\ , \qquad 
\hbox{\bf B} = e^{\beta^-\hbox{\bf e}_-}\ , \qquad 
\hbox{\bf A} 
= e^{{1\over4}\pi(\hbox{\bf e}^1{}_{2} - \hbox{\bf e}^2{}_{1})}\cr
\overline{\bfkappa}_3 
&= \hbox{\bf AB}\bfkappa_3(\hbox{\bf AB})^{-1} 
=
qe^{-\alpha^3}(\hbox{\bf e}_-/\sqrt3)\ ,\qquad \overline{\hbox{\bf n}} 
= 
\hbox{\bf ABn(AB)}^T = q(\hbox{\bf e}^1{}_{2} 
+ \hbox{\bf e}^2{}_{1})\cr
\overline{\hbox{\bf g}} 
&= (\hbox{\bf AB})^{-1\,T}\hbox{\bf g(AB)}^{-1} 
= 
e^{2(\beta^0\hbox{\bf e}_0 + \beta^+\hbox{\bf e}_+ 
+\overline{\beta}\null^-
\hbox{\bf e}_-)}\ , \qquad 
\overline{\beta}\null^- \equiv 
qe^{-\alpha^3}\theta^3/\sqrt3\ .\cr}
\eqno(3.52)$$
The generalized Taublike case $\Mscr_{GT(3)}$ is mapped onto the symmetric case $\Mscr_{S(1)}$ by this linear transformation.
The spatial frame $\overline{e}$ is a Kasner frame consisting of eigenvectors
of the spatial Ricci curvature.

When $v_3=0$, then $u=-3(\beta^+ +\lambda\sqrt3\overline{\beta}\null^-)$ has a
constant value which may be transformed to zero by a translation in $\theta^3$.
For Bianchi type III $=$ VI$_{-1}$, $\lambda = \hbox{sgn}\,n^{(1)} = \pm 1$
holds and $u=5\overline{\beta}\null^-_2 \quad (\lambda=1)$ or $u=-6\overline{
\beta}\null^-_2\quad(\lambda=-1)$ and one has a Taublike case 
$\Mscr_{T(2)}$ (i.e.
$\overline{\beta}\null^-_2=0$) or $\Mscr_{T(1)}$ (i.e. $\overline{\beta}\null
^-_1=0$)
respectively when referred to the barred frame.  This Taublike case for 
Bianchi type III is exactly the case where the fourth linearly independent
Killing vector field which exists for all spatial metrics of this 
type [10]
becomes a spacetime Killing vector field, leading to local
rotational symmetry.  Since the (positive curvature) Kantowski-Sachs metrics
are related to these by a simple analytic continuation in a particular
coordinate system, as discussed in appendix D, the Kantowski-Sachs case is 
also Hamiltonian.   

\section{Qualitative Considerations}

With the machinery that has been developed in the preceding sections, one can
qualitatively or numerically study the time evolution of the general spatially
homogeneous perfect fluid spacetime. Three regimes in this evolution are of 
particular interest, namely the limit $\beta^0\rarrow-\infty$ towards the
initial singularity in the class of initially expanding models ($p_0<0$ as $
\beta^0\rarrow-\infty$), the limit $\beta^0\rarrow\infty$ away from this 
singularity in the same class of models [38]
(omitting the Bianchi type IX 
subclass, which reach a maximum of $\beta^0$ and recollapse) and
the small anisotropy limit in the subclass of models which are compatible with
spatial isotropy [63]
The first regime is especially simple to treat 
qualitatively, the key ideas for which will now be sketched.

Lifshitz and Khalatnikov, later joined by Belinsky, were led to the qualitative
study of spatially homogeneous models by their investigation of the nature of a
generic initial cosmological singularity.  It seems that the approach to the 
singularity in adapted comoving coordinates induces a sort of contraction of 
the Einstein equations which essentially decouples the evolution of the 
three-plus-one variables at different spatial points while mimicking the 
behavior of a particular spatially homogeneous model at any given point.  This
work is concisely summarized and updated in a recent article by these 
authors [64].
At about the same time as the main thrust of the BLK work
began, Misner approached the problem of the dynamics of spatially homogeneous
models from an entirely different point of view, paving the way for Ryan to
study the qualitative behavior of the approach to the initial singularity in
general Bianchi type IX models.  This is the same class of models to which
BLK confined most of their attention, the relation between the two approaches
being described in ref.(24).  The formalism of the present article allows one
to extend this work to the general spatially homogeneous model.
 
In studying the regime $\beta^0\rarrow-\infty$, it is convenient to use the 
variable $\Omega\equiv-\beta^0$ which increases towards the singularity.  As
one approaches the singularity, the ``anisotropy" increases and the scale over
which the motion of the $\beta^{\pm}$ variables occurs becomes much larger 
than the one over which the curvature of the valleys and corners associated
with the channels of Figure 4 is apparent.  The various potentials reduce
to time dependent exponential or exponentially cutoff potentials in the 
positive or negative $\beta^+_a$-sectors (defined with respect to a shifted
origin) which are relevant to each potential (as indicated in Figures 8 and 
9).  These potentials move with either constant or time varying velocity on
the $\beta^+\beta^-$ plane in supertime time gauge.  Because of the sharp 
rise of an exponential, a given potential has little effect on the motion
of the universe point until a ``collision" or ``bounce" occurs, during which
the universe point is significantly affected by the potential and then returns
to a state in 
which that potential has little effect on the motion.  Again due to the 
exponential nature of the potentials, essentially only one potential 
usually affects
the universe point at any given time.  When no potential exerts a noticable
effect on the universe point, the evolution reduces to the free dynamics whose
exact solutions are the Kasner solutions, characterized by straight line 
motion in the $\beta^+\beta^-$ plane with unit $\beta^0$-velocity (a null 
curve in the flat 3-dimensional spacetime $diag(3,R)$ ).

Exact solutions also exist for each of the cases in which only one potential
is present.  These may be viewed as scattering problems and used to relate
the asymptotic Kasner solutions before and after the collision with that
potential.  When the velocity of a given potential is constant and timelike
($\beta^0$-velocity less than one), one can always transform to its rest
frame where the problem reduces to an ordinary 1-dimensional scattering
problem in a fixed potential at constant energy [17].
If the velocity
is constant but null, one can transform to null coordinates and solve the
problem.  Because of the super-Hamiltonian constraint, the system trajectory
is always a timelike or null curve in $diag(3,R)$ except near the point of
maximum expansion in Bianchi type IX models where the spatial curvature is
positive and the trajectory becomes spacelike as it makes the transition 
from expansion to contraction (see Figure 3 of ref.(17)).   

One must also consider the possibility that the system trajectory is affected
by one of the channels, although this becomes increasingly less likely as the
scale of the $\beta^{\pm}$ motion increases (simply because the width of the 
channel is fixed).  In this case, referred to as the case of small oscillations
by BLK and as a mixing bounce by Ryan and others, only a qualitative solution
exists for the exit of the universe point from the channel [15,29],
apart from a numerical calculation of a particular open channel mixing 
bounce [5,22].  
This case can be complicated by the presence of a
centrifugal
potential in the channel, as discussed by Chitre and Matzner [65] and
others [40].

Misner introduced the extremely useful idea of associating a moving ``wall"
with each potential by selecting a particular contour line which marks the 
point at which the potential has a large enough value to significantly affect 
the motion of the universe point.  For a given potential and a given moment of
time in supertime time gauge, one can consider the simplified super-Hamiltonian
constraint in which only the expansion energy and that potential appear and 
solve it for $\beta^{\sst wall}$, the value of the particular $\beta^+_a$ or
$\beta^-_a$ coordinate on which the potential depends at which the constraint
is satisfied.   This locates the ``wall contour" or simply ``wall" associated
with this potential
$$  0=-\half p_0^{\ 2} + U(\beta^{\sst wall}) \ . \eqno(4.1)$$
This contour is the one at which a turning point of the motion would occur
if $U$ were time independent, $p_0$ were constant, all other potentials were
zero and the motion were orthogonal to the contour.  In fact since the terms
omitted in the super-Hamiltonian constraint are positive (away from the point
of maximum expansion in the type IX case), the ``turning point" of the 
orthogonal component of the motion must occur at a smaller value of the 
potential, namely before the system point can make contact with the wall
contour.  (Because of the motion of the walls, the ``turning point" may only
be an actual turning point in the rest frame of the potential.)  However, this
gap between the wall and ``turning point" is on a scale which is usually
much smaller than the scale over which the $\beta^{\pm}$ motion is occurring
and so provides a useful marker of where a collision occurs between the
potential and the universe point.  

Since $p_0$ is a constant far from the wall contour during a Kasner phase, the
value of the potential at the wall contour remains fixed and the wall contour
coincides with a given contour of the potential, thus moving with the same
velocity as the potential.  However, $p_0$ may change during the course of a
collision, a decrease in $|p_0|$ requiring a decrease in the potential value
of the wall contour and causing an additional inward motion of the wall
contour towards the universe point.  In supertime time gauge, the equation of
motion for $|p_0|=-p_0$ in the expanding phase is given by
$$\eqalign{
|p_0|\dl{\ } &=-\{p_0, H\} 
=\partial H/\partial\beta^0 
\cr
&
=
12[4e^{4\beta^0}(V^{\ast} +6a^2e^{4\beta^+}) + e^{3\beta^0}\Hscr_{\sst tilt}
(3-(v^{\sst \bot})^{-2}g^{\prime ab}v^{\prime}_av^{\prime}_b) 
-12kpe^{6\beta^0}
]\ .\cr}
\eqno(4.2)$$
Using the definitions of Appendix C, the fluid term inside the brackets of
eq.(4.2) may be expressed in the obviously nonnegative form
$$ 6\gamma(l/u^{\sst \bot})^{\gamma} e^{3\beta^0(2-\gamma)}[2-\gamma+
\fraction{1}{3}\gamma\mu^{-2}v^av_a] \ .\eqno(4.3)$$
Thus all of the terms inside the brackets of eq.(4.2) are nonnegative 
 (apart from the positive curvature regime at Bianchi
type IX points),  indicating that $|p_0|$ increases
with supertime and therefore decreases as $\Omega\rarrow\infty$,
except in the vacuum type I case where $|p_0|$ is a constant.
Note that the equation of motion for $\Omega$ in this time gauge, namely 
$\dott{\Omega} = d\Omega/d\tb =p_0$, determines $\Omega$-time as a function
of supertime; these are affinely related only when $p_0$ is a constant.

The walls associated with the various gravitational, centrifugal and tilt
potentials and their $\Omega$-velocities are as follows [20,21] 
$$
\eqalign{
&U^{(a)}_g= \half (n^{(a)})^2e^{-8(\beta^+_a-{1\over2}\beta^0)}\cr
&\qquad (\beta^+_a)^{\sst wall}_g 
= -\half\Omega -\fraction{1}{4} \ln |p_0/n^{(a)}|\ , \cr
&\qquad d(\beta^+_a)^{\sst wall}_g /d\Omega
=-\half- \fraction{1}{4}d\,\ln|p_0|/d\Omega
\phantom{AAAAAAAAAAAAAAAAA}\cr
}
$$
$$
\eqalign{
&U^{(a)}_c(i/ii)
= \fraction{3}{2}{\s P}\null^{\ 2}_a |e^{-2\alpha^a}n^{(b)}
n^{(c)} |^{-1} 
[\left(\vcenter{\vbox{\hbox{sinh}\hbox{cosh}}}\right)\,
2\sqrt3(\beta^-_a-\beta^-_{a0})]^{-2}\cr
&\qquad(\beta^-_a)^{\sst wall}_c 
= \beta^-_{a0} \pm(2\sqrt3)^{-1}
\left(\vcenter{\vbox{\hbox{sinh$^{-1}$}\hbox{cosh$^{-1}$}}}\right)
\zeta^{-1}\ ,\cr
&\qquad\zeta\equiv\null |p_0/{\s P}\null_a|\null 
|e^{-2\alpha^a}n^{(b)}n^{(c)}/3|^{1\over2}\cr
&\qquad (d\beta^-_a)^{\sst wall}_c/d\Omega 
=\pm(2\sqrt3)^{-1} 
\left(\vcenter{\vbox{ \hbox{$(\zeta^{-2}+ 1)^{-{1\over2}}$}
\hbox{$(\zeta^{-2}-1)^{-{1\over2}}$} }}\right)
d\,\ln|\s P_a\null/p_0|/d\Omega\ ,\cr
&\cr
&U^{(a)}_c(iii) 
= 6\s P_a\null^2|e^{-2\alpha^a}\left(\vcenter{\vbox{\hbox{$
(n^{(b)})^2$}\hbox{$(n^{(c)})^2$}}}\right)|^{-1}e^{\mp 4\sqrt3\beta^-_a}
\ , \cr
&\qquad 
\iota\equiv |p_0/{\s P}_a|\null
|(2\sqrt3)^{-1} e^{-\alpha^a}\left(\vcenter{\vbox{\hbox{$n^{(b)}$}
\hbox{$n^{(c)}$}}}\right)|\cr
&\qquad(\beta^-_a)^{\sst wall}_c 
= \pm(2\sqrt3)^{-1}\ln\iota^{-1}\ , \cr
&\qquad 
d(\beta^-_a)^{\sst wall}_c/d\Omega 
= \pm(2\sqrt3)^{-1}d\,\ln|{\s P}_a/p_0|
/d\Omega\cr
}
\eqno(4.4)
$$
$$
\eqalign{
&U^{(a)}_{\sst tilt} 
= 24kle^{2(\beta^0+\beta^+_a)}|v^{\prime}_a|\cr
&\qquad(\beta^+_a)^{\sst wall}_{\sst tilt}
=\Omega +\half\ln[p_0\null^2 
(48kl|v^{\prime}_a|)^{-1}]\ ,\cr
&\qquad d(\beta^+_a)^{\sst wall}_{\sst tilt}
/d\Omega =1+ \half d\,\ln[p_0\null^2/|v^{\prime}_a|]/d\Omega
\phantom{AAAAAAAAAAAAAAA}\cr
}$$  
Note that there are two walls symmetrically located about $\beta^-_a  =
\beta^-_{a0}$ for the centrifugal potentials of type $(i)$ and $(ii)$.
(For the latter case these walls coincide when $\zeta=1$ and then disappear
for $\zeta>1$.)
These pairs are represented in Figures 8 and 9 by single dotted lines at 
$\beta^-_a =0$ ($\beta^-_{a0}=0$ at the canonical points when well defined).
More detailed figures indicating these pairs would generalize Figure 1 of
ref.(24).  Although these
walls move away from each other as $|p_0|$ decreases for fixed $\s P_a$,
their separation decreases on a scale which expands with $\Omega$ to keep 
up with the other walls.  All of the centrifugal walls move in the
direction of decreasing values of the associated potential as $|p_0|$
decreases. 

Between collisions the universe point has unit $\Omega$-velocity  while the 
gravitational walls have $\Omega$-velocity ${1\over2}$ and the component of the
$\Omega$-velocity of the centrifugal walls associated with the explicit
dependence on $\beta^0$ is 0, so the universe point will overtake and collide
with these walls, the length of Kasner segments of the $\beta^+\beta^-$ motion
increasing with $\Omega$ as $\Omega\rarrow\infty$.  The tilt wall on the other
hand also moves with unit $\Omega$-velocity  between collisions neglecting 
the additional
component of the velocity due to the time dependence of $|v^{\prime}_a|$
(and $l$ in the class B case), so
the universe point can never catch this wall unless significant changes in its
motion occur due to the time dependence of $|p_0|$, $|v^{\prime}_a|$ and $l$. 
 This
means that the tilt potential becomes less and less important as $\Omega\rarrow
\infty$, as discussed in detail by Ryan [20,21].  
All of the walls for 
each canonical point of $\Cscr_D$ are shown in Figures 8 and 9.

A bounce from a single curvature potential $U^{(a)}_g$ is described by the
exact type II diagonal vacuum solution found by Taub [11];
the supertime
$\overline{t}$ is affinely related to Taub's time function (by the factor of
12).  Consider the case $a=1$ for comparison with the Lifshitz-Khalatnikov
notation; the remaining cases follow by cyclic permutation.  The Hamiltonian
in supertime time gauge is expressed in terms of the coordinates $\{
\beta^{\pm}_1, p_{1\pm}\}$ by
$$ H=\half (-p_0\null^2+p_{1+}\null^2+p_{1-}\null^2) +6(n^{(1)})^2e^{4(\beta
^0-2\beta^+_1)}\ ,
\eqno(4.5)$$
which is the  Hamiltonian of a relativistic particle (with timelike momentum
due to the super-Hamiltonian constraint) in an exponential potential which
is moving uniformly in the positive $\beta^+_1$-direction with constant
$\beta^0$-velocity of magnitude $v={1\over2}$.  A boost in this direction with
$\gamma$-factor $(1-v^2)^{-{1\over2}}=2/\sqrt3$ transforms the inertial
coordinates to the rest frame of the potential [16]
$$\eqalign{&(\overline{\beta}\null^0,\overline{\beta}\null ^+
,\overline{p}_0,\overline
{p}_{1+})= 3^{-{1\over2}}(2\beta^0-\beta^+_1, 2\beta^+_1-\beta^0,2p_0+p_+,
2p_++p_0)\cr &\hbox{\bf g}\null^{\prime} =e^{2\bfsbeta}\ , \quad \bfbeta
=2\sqrt3 \hbox{diag}(-\overline{\beta}\null^+_1, \overline{\beta}\null^0+
\overline{\beta}\null^+_1+ \beta^-_1, \overline{\beta}\null
^0 +\overline{\beta}\null^+_1 -\beta^-_1)\cr
& H=\half(-\overline{p}_0\null^2 +\overline{p}_{1+}\null^2 +p_{1-}
\null^2) +6(n^{(1)})^2e^{-4\sqrt3\, \overline{\beta}\null^+_1} \ .
\cr}
\eqno(4.6)$$
The problem therefore reduces to free motion in the $\overline{\beta}\null^0
\beta^-_1$ plane and an ordinary 1-dimensional scattering 
problem
in the $\overline{\beta}\null^+_1$-direction with positive energy $\Escr_0=
{1\over2}(\overline{p}_0\null^2 -p_{1-}\null^2)\equiv {1\over2}
(\overline{p}^{\infty}_{1+})^2$ where $\overline{p}^{\infty}_{1+}>0$ is a 
constant.  The super-Hamiltonian constraint
$$ (\dott{\overline{\beta}}\null^+_1) \null^2 + 12(n^{(1)})^2 e^{-4\sqrt3\,
\overline{\beta}^+_1}=(\overline{p}^{\infty}_{1+})^2
\eqno(4.7)$$
may be integrated to obtain $\overline{t}$ as a function of $\overline{\beta}
\null^+_1$; inverting the result yields the solution
$$ 
\eqalign{&(\overline{\beta} \null^0(\tb),\beta^-_1(\tb))=
(-\overline{p}_0\tb+\overline{\beta}\null^0_0,p_{1-}\tb+(\beta_1^-)_0)\cr&
e^{2\sqrt3\,\overline{\beta}\null^+_1(\tb)} =4\sqrt3|n^{(1)}|(\overline{p}\null
^{\infty}_{1+})^{-1} \hbox{cosh}\,4\sqrt3\overline{p}\null^{\infty}_{1+}
(\tb-\tb_0)\cr 
&p_0(\tb)=3^{-{1\over2}}(2\overline{p}_0 -\overline{p}\null^{\infty}_{1+}
\hbox{tanh}\, 4\sqrt3\, \overline{p}\null^{\infty}_{1+}(\tb-\tb_0))\cr
}\eqno(4.8)$$
where $\overline{t}_0$ is the turnaround time.  This solution interpolates
between two asymptotic Kasner solutions characterized by the null momenta
$$ (p_0(\pm\infty), p_{1+}(\pm\infty), p_{1-})=3^{-{1\over2}}(2\overline{p}
_0\mp\overline{p}\null^{\infty}_{1+}, \pm2\overline{p}\null^{\infty}_{1+}-
\overline{p}_0, \sqrt3p_{1-}) \eqno(4.9)$$
The scattering problem is therefore most naturally interpreted as a map of the
unit circle in the $p_+p_-$ plane into itself, which is the result of the 
Lorentz transformation of ordinary reflection from the static potential in its
rest frame.

\begin{figure}[t!] 
\begin{center}
\includegraphics[width=.9\textwidth]{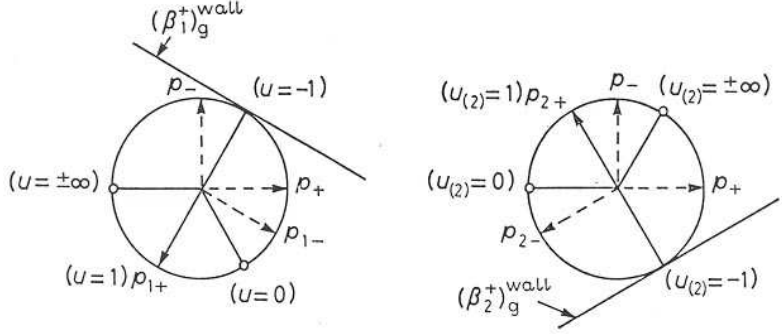}
\end{center}
\caption{
The Lorentz transforms of
simple reflection by the gravitational potentials $U^{(1)}$ and $U^{(2)}$
in their rest frames.  The asymptotic Kasner states are related by a change
of sign of the parameter $u$ in the first case and of $u_{(2)}$ in the 
second case. 
} 
\end{figure}

The Lifshitz-Khalatnikov parametrization of the unit circle in the $p_+p_-$
plane leads to the following representation of the unit momentum coordinates
$$ ({\A p}_{1+},{\A p}_{1-}) = \half(1+u+u^2)^{-1}((1+u)^2,\sqrt3(u^2-1))\ ,
\eqno(4.10)$$
which enables one to define $u_{\pm\infty}$ for the asymptotic values (4.11)
of this unit vector at $\overline{t}=\pm\infty$.  The final states with
$u_{\infty}\in[1,\infty)$ correspond to initial states with $u\in (-\infty,
-1]$; similarly $u_{\infty}\in(0,1]$ corresponds to $u_{-\infty}\in[-1, 0)$.  
No asymptotic states exist with $u=0$ or $u=\pm\infty$ which represent the
Lorentz transforms of directions parallel to the equipotential lines of the
static potential.  Figure 10 illustrates this situation.
 
Since 
$$ p_{1-}/|\overline{p}_0|= (p_{1-}/|p_0|)[2-p_{1+}/|p_0|]^{-1}\eqno(4.11)$$
is a constant of the motion, its asymptotic values 
$$\A{\overline{p}}_{1-}(\pm\infty) ={\A p}_{1-}(\pm\infty)[2-{\A p}_{1+}
(\pm\infty)]^{-1}=3^{-{1\over2}}(1-u_{\pm\infty}^{\ 2})\eqno(4.12)$$
must coincide, which can only be true if $u_{-\infty}=-u_{\infty}$.  
($u_{\infty}=u_{-\infty}$ is ruled out by the case $p_{1-}=0$ of normal
incidence.)  Thus scattering from the curvature wall $U^{(1)}_g$ simply leads
to a change of sign of the Lifshitz-Khalatnikov parameter $u_{(1)}=u$.
Similarly scattering from the rotated potentials $U^{(2)}_g$ and $U^{(3)}_g$
leads to a change of sign of the rotated Lifshitz-Khalatnikov parameters
$u_{(2)}=P_{231}(u)$ and $u_{(3)}= P_{312}(u)$.

Suppose $u_{\infty}\in[2,\infty)$ and hence $u_{-\infty}\in(-\infty, -2]$, so
that the Kasner exponents at $\overline{t}=\infty$ are ordered: $p_1(u_{\infty}
)\leq p_2(u_{\infty})\leq p_3(u_{\infty})$.  The transposition $P_{12}$ then
orders the Kasner exponents at $\tb =-\infty$ by reflecting the sector
$u\in(-\infty,-2]$ across the $p_+$-axis into the ordered sector
$$\eqalign{&P_{12}(u_{-\infty})=-(1+u_{-\infty})=u_{\infty}-1\cr &
(p_1(u_{\infty}-1), p_2(u_2(u_{\infty}-1), p_3(u_{\infty}-1)= (p_2(u_{-\infty}), 
p_1(u_{-\infty}), p_3(u_{-\infty})\cr &p_1(u_{\infty}-1)\leq p_2(u_{\infty}-1)
\leq p_3(u_{\infty}-1)\ .\cr}\eqno(4.13)$$
Thus the ``ordered" Lifshitz-Khalatnikov parameter decreases by 1 in the
negative time direction, i.e. as $\Omega$ increases.  On the other hand if
$u_{\infty}\in[1,2)$, so that $u_{-\infty}\in(-2,-1]$ and the Kasner
parameters at $\tb=\infty$ are ordered, then the cyclic permutation $P_{312}$ 
orders the
Kasner exponents at $\tb=\infty$ by rotating the sector $u\in[-2,-1]$ into
the ordered sector
$$ \eqalign{&P_{312}(u_{-\infty}) =-(1+u_{-\infty})^{-1} = (u_{\infty}-1)^{-1}
\cr&
p_1((u_{\infty}-1)^{-1})\leq p_2((u_{\infty}-1)^{-1}) \leq p_3((u_{\infty}-1)
^{-1})\ .\cr}\eqno(4.14)$$
The ``ordered" Lifshitz-Khalatnikov parameter thus decreases by 1, entering 
the interval [0,1) and then inverts.

The connection between a Lie algebra contraction and an anisotropic singularity
may be illustrated by introducing the two special turnaround times
defined by
$\tb\null^{\pm}_0=\pm(4\sqrt3)^{-1}$ $\ln (\sqrt3|n^{(1)}|/\overline{p}\null
^{\infty}_{1+})$.  For $\tb_0=\tb\null^+_0$, the future Kasner limit 
$\overline{\beta}\null^+_1(\tb)\rarrow-\overline{p}\null^{\infty}_{1+}\tb$ is
obtained as $\tb\rarrow\infty$; the same asymptotic behavior arises from the
contraction $n^{(1)}\rarrow0$ which sends the turnaround time to $-\infty$.
In other words long after the collision the solution approaches the solution
for the contracted Bianchi type corresponding to the anisotropic singular
scaling $\bfbeta(\tb)\rarrow\bfbeta(\infty)$.  Similarly for $\tb_0=\tb\null
^-$, the past Kasner limit $\overline{\beta}\null^+_1(\tb)\rarrow\overline{p}
\null^{\infty}_{1+}\tb$ is obtained either as $\tb\rarrow-\infty$ or $n^{(1)}
\rarrow 0$.

A collision with a single centrifugal potential $U^{(a)}_c$ is described by 
an exact Bianchi type I vacuum solution with nonzero angular momentum, namely
the symmetric case $\Mscr_{S(a)}$.  The solution may be found either by 
transforming the Kasner solutions or by solving the equations of motion. 
For definiteness consider the case $a=3$.   The Hamiltonian in supertime time
gauge is
$$ H=\half(-p_0\null^2+p_-\null^2+p_-\null^2) +3\overline{\Gscr}\null^{-1\,33}
{\s P}_3\null^2\eqno(4.15)$$
An $S^1$-parametrized family of potentials are possible, corresponding to the
possible values of $\bfkappa_3$ given by (3.4) with $\phi\in[0,2\pi)$.  
$\s P_3$ is a constant of the motion and since $H$ only depends on $\beta^-$,
both $\beta^0$ and $\beta^+$ obey the free dynamics with $p_0$ and $p_+$
constants of the motion, reducing the problem to a 1-dimensional scattering
problem for $\beta^-$ with constant energy $\Escr_0={1\over2}
(p_0\null^2-p_+\null^2) \equiv (p^{\infty}_-)^2$ and static potential $U^{(3)}
_c$.  The solution may be obtained by integrating the super-Hamiltonian 
constraint for the supertime as a function of $\beta^-$ and then inverting
the result (note that supertime and $\Omega$-time are affinely related here).

The following variables make this calculation simple
$$ 
\eqalign{
&v=2\sqrt3\beta^-\,, \quad u^{\pm}=2^{-{1\over2}}(\cos\phi \,e^v\pm
\sin\phi\,e^{-v})\,, \cr
&\qquad (u^+)^2-(u^-)^2 = \sin2\phi\,, \quad 
du^+=u^-dv\,,\cr 
&\qquad\half \overline{\Gscr}_{33} = (u^-)^2 =(u^+)^2-\sin2\phi \,,
\qquad\quad \Escr_0=\half(\du\beta\null^-)^2 + \fraction{3}{2}\s P_3\null^2
(u^-)^{-2} \,,\cr 
&\qquad (24\Escr_0)^{1\over2}(\tb-\tb_0)= \int du^+[(u^+)^2
-\epsilon b^2]^{-{1\over2}} =\left\{\vcenter{ \halign{$\lft{#}$
&\qquad$\lft{#}$
\cr\hbox{cosh}^{-1}|u^+/b| &\epsilon=1\cr \hbox{sinh}^{-1}(u^+/b)
&\epsilon=-1\cr
\ln|u^+| &\epsilon=0\cr}}\right. \cr
&\qquad \epsilon b^2\equiv \sin2\phi + \fraction{3}{2}\s P_3\null^2/\Escr_0 \,,
\qquad \qquad 2\sqrt3\beta^-_0 =\half\ln|\tan\phi| \,.\cr}
\eqno(4.16)$$
The solutions $\epsilon=0$ and $\epsilon=-1$ occur only for the type $(ii)$
potential when the energy equals or is greater than, respectively, the maximum
value of the potential $U_{\sst max}\equiv{3\over2}\s P_3\null^2
|\sin 2\phi|^{-1}$ for this case;
the latter solutions pass over the potential barrier while the former ones
approach or leave the point of unstable equilibrium $\beta^-=\beta^-_0$. 
The  $\epsilon =1$ bounce solutions are then
$$\eqalign{& |u^+|=b\,\hbox{cosh}(24\Escr_0)^{1\over2}(\tb-\tb_0)\qquad \quad
p_- = \pm|2\Escr_0|^{1\over2}[1-\fraction{3}{2}\s P_3/\Escr_0(u^-)^{-2}]\cr
&\beta^-=\left\{\vcenter{\halign{$\lft{#}$&\qquad
$\lft{#}$\cr \beta^-_0+ (2\sqrt3)^{-1}\hbox{cosh}^{-1}|u^+ |\sin2\phi|
 ^{-{1\over2}}| &\sin2\phi >0\cr
\beta^-_0 +(2\sqrt3)^{-1}\hbox{sinh}^{-1}(u^+|\sin2\phi|^{-{1\over2}}) 
&\sin2\phi<0\cr
(2\sqrt3)^{-1}(\cos2\phi)\ln|\sqrt2\,u^+| &\sin2\phi =0\cr}}\right.\cr
}\eqno(4.17)$$
Note that for a fixed value of $\s P_3$, once $\Escr_0$ decreases below 
$U_{\sst max}$, only the bounce solutions are relevant, i.e. for small enough
$\Escr_0$, all three types of potential reverse the orthogonal component of
the motion.  Similarly as $\Escr_0$ decreases, the bounce occurs far enough
from $\beta^-_0$ that the potentials $(i)$ and $(ii)$ essentially reduce to 
pure exponentials.  

These solutions have Kasner asymptotes as $\tb\rarrow\pm\infty$ which are
characterized by the momenta ($p_0(\pm\infty)$, $p_+(\pm\infty)$, $p_-(\pm\infty)$)
 = ($p_0$, $p_+$, $\pm p_-^{\infty}$) for a bounce solution and 
($p_0$, $p_+$, $p_-^{\infty}$) 
for an $\epsilon=-1$ solution, but $(p_-(\infty), p_-(-\infty))$ =
$(0, p_-^{\infty})$ or $(p_-^{\infty},0)$ for an $\epsilon=0$ solution.
The change in asymptotic momenta associated with the bounce solutions 
corresponds to reflection across the $p_-$-axis.

For this symmetric case one has $(\theta^1,\theta^2,\theta^3)=(0,0,\theta)$
and $ \dot{W}{}\null^a=\delta^a{}_{3}\du\theta$ so from (3.21) one has
$$ 
\eqalign{&\du\theta=6\s P_3\int d\tb \, \overline{\Gscr}\null^{-1\,33}\,,
\quad  
z\equiv\zeta\,\hbox{tanh}\, (24\Escr)^{1\over2}
(\tb-\tb_0) \,, \quad 
\zeta \equiv (\fraction{2}{3}\Escr_0|\sin2\phi|
(\s P_3)^{-2})^{1\over2}\,,\cr
& \theta-\theta_0= 3\s P_3 \int d\tb[(u^+)^2-\sin2\phi]^{-1}
= \half|\sin2\phi|^{-{1\over2}} \left\{\vcenter{\halign{\quad$\lft{#}$&
\qquad$\lft{#}$\cr  \tan ^{-1}z &\sin2\phi>0\cr 
\hbox{tanh}^{-1}z&\sin2\phi<0\cr
z&\sin2\phi=0\cr} }\right.\cr }
\eqno(4.18)$$
In the compact case $\sin2\phi>0$, let $\s\theta\equiv|\sin2\phi|^{1\over2}
\theta$; since $e^{\theta\bfskappa_{\sst 3}}=-\hbox{\bf 1}$ for $\s \theta
=\pi)$, ${\s\theta} \in [0,\pi)$ represents a closed path in $\Mscr_{S(3)}$,
although twice this interval represents a closed path in $\A G$. 
Note that for $\zeta\gg1$, $\tan2(\s\theta-\s\theta_0)$ begins at
a very large negative value at $\tb-\tb_0\rarrow-\infty$ and approaches the
sign reversed value at $\tb-\tb_0\rarrow\infty$, so that the total
change in $2\s\theta$ during one bounce is just slightly less that $\pi$.
As $\zeta$ decreases the total change in $\s\theta$ decreases from $\pi/2$
and approaches 0 as $\zeta\rarrow0$.  This is clearly evident in the 
numerical calculation for the canonical type IX symmetric case $\Mscr_{S(3)}$ [22], 
where $\s P_3$ is constant but $|p_0|$ and therefore $\Escr_0$
decreases as $\Omega\rarrow\infty$.  In all cases $(\Delta\theta)_{\sst bounce}
\rarrow0$ as $\zeta\rarrow0$.

\def\barg{v-v_0}
\def\bargp{(v-v_0)}

The rescaled DeWitt metric on $\Mscr_{S(3)}$ (set $(\theta^1,\theta^2,
\theta^3)=(0,0,\theta)$) is
$$ -d\beta^0\otimes d\beta^0 +d\beta^+\otimes d\beta^+ + {}^2h\ ,\qquad 
{}^2h=
d\beta^-\otimes d\beta^- +\fraction{1}{6}\overline{\Gscr}_{33}d\theta\otimes
d\theta\ .
\eqno(4.19)$$
The affinely parametrized null geodesics of this metric describe the solutions
of the collision with the single potential $U^{(3)}_c$ in supertime time gauge.
The $\beta^-$ scattering problem is then equivalent to finding the geodesics of 
$^2h$.  This is just the induced metric on the future hyperboloid
$$ H_{(2\sqrt3)^{-1}} = \{ (y^0,y^1,y^2)\in R^3 \relv (y^0)^2-(y^1)^2
-(y^2)^2 = (2\sqrt3)^{-2}\ , y^0>0\} \eqno(4.20)$$
in 3-dimensional Minskowski spacetime with metric $-dy^0\otimes dy^0 
+dy^1\otimes dy^1 +dy^2\otimes dy^2$.  Defining
$$ (e^{\theta\bfkappa_{\sst3}})^Te^{2\beta^-\hbox{\bf e}_-} e^{\theta\bfkappa
_{\sst3} } = (y^0+y^1)\bfe^1{}_1 +(y^0-y^1)\bfe^2{}_2 -y^2(\bfe^1{}_2 +\bfe^2{}_1) +\bfe^3{}_3
\eqno(4.21)$$
maps the coordinate surfaces of constant $(\beta^0,\beta^+)$ onto 
$H_{(2\sqrt3)^{-1}}$.  Letting $x\rarrow\theta$ in $(A.4)$ and letting 
$(c_3(2),s_3(2))$ be obtained from $(c_3,s_3)$ by doubling the argument, and
using the double angle identities
$$ c_3\null^2 = \half (c_3(2) +1) \qquad s_3\null^2 =-\half (\s m\null^{(3)})
^{-2} (c_3(2)-1) \ , 
\eqno(4.22)$$
one obtains explicitly for the case $\s m\null^{(3)}= 
(n^{(1)}n^{(2)}))^{1\over2} \neq 0$ (set $\epsilon = \hbox{sgn} (n^{(1)}
n^{(2)})$)
$$ \eqalign{&y^0+y^1 = e^{2\sqrt3\beta^-_0}(\half[e^{\barg}-\epsilon
e^{-\bargp}] c_3(2) +\half[e^{\barg}+\epsilon e^{-\bargp}])\ ,\cr
&y^0-y^1 
= \epsilon e^{2\sqrt3\beta^-_0}(\half[e^{\barg}-\epsilon e^{-\bargp}]
c_3(2) +\half e^{\barg}+ \epsilon e^{-\bargp}])\ ,\cr
&y^2= \hbox{sgn}\, n^{(1)} (\half[e^{\barg}-\epsilon e^{_\bargp}])s_3(2) 
\hskip 30pt v-v_0 =2\sqrt3(\beta^--\beta^-_0)\ .
\cr}\eqno(4.23)$$
The special linear subgroup $SL(2)_3$ acting on $\Mscr_{S(3)}$ maps onto the
action of the 3-dimensional Lorentz group on $H_{(2\sqrt3)^{-1}}$; the
coordinates $\{ \beta^-, \theta\}$ are comoving with respect to
the subgroup generated by $\bfkappa_3$.  This subgroup corresponds to null
rotations for $\phi=0,\pi$, rotations in the $y^1-y^2$ plane for $\phi=
\pi/4,5\pi/4$, boosts along $y^2$ for $\phi=3\pi/4,7\pi/4$ and combinations
of these for other values ($\hbox{\bf e}_-$ generates boosts along $y^1$).  

Geodesics with $\theta=\theta_0$ have zero angular momentum with respect to
this subgroup, i.e. $\s P_3=0$, and are diagonalized by the constant active
transformation $e^{\theta_0\bfskappa_{\sst3}}$.  However, these geodesics have
nonzero angular momentum with respect to the subgroups generated by other
values of $\bfkappa_3$ (modulo sign).  In the canonical type IX or VIII cases,
$\theta\rarrow\theta+\theta_0$ corresponds to a rotation or boost by the
angle or natural boost parameter $2\theta_0$ because of the double angle
functions.  For small angular momentum in the canonical type IX case, a 
geodesic has $\Delta(2\theta)\sim\pi$ as if the projection to the
$y^1-y^2$ plane were flat, but as the angular momentum increases, the 
curvature of the hyperboloid decreases this angle.  

The classic discussion of Lifshitz and Khalatnikov of an ``era" of successive
alternating Kasner regimes describes a sequence of bounces between the
gravitational wall $U^{(1)}$ in the upper half plane and $U^{(2)}$ in the
lower half plane as the universe moves out of the corner where these
walls intersect with increasing  supertime.  Reversing the direction of time
in order to approach the initial singularity, as in $\Omega$-time, one must
reverse the sign of the momenta; ``initial" and ``final" will be correlated
with this reversed direction of time in the following discussion.  An initial
Kasner phase with $u^{\sst ordered}=u\in [1,\infty)$ is headed for the wall
$\beta^+_1=(\beta^+_1)^{\sst wall}_g$, and then bounces from this wall into
the second Kasner phase with $u^{\sst ordered}=u-1$ so that it is headed for
the wall $\beta^+_2=(\beta^+_2)^{\sst wall}_g$.  It then bounces from this
wall into the Kasner phase with $u^{\sst ordered}=u-2$, etc.  When 
$u^{\sst ordered}$ enters the interval $[0,1]$, the final Kasner phase (or 
``epoch") heads away from both walls.  Figure 10 indicates the scattering
problem for each bounce.

This approximation is valid as long as the universe point does not penetrate
deep enough into the vertex or corner between the two asymptotic potentials to
feel the effects of the channel where the straight line contour approximation
of Figure 5 breaks down.  When the effect of the channel is important, the
regime is called a ``long era" or ``mixing bounce" and the type VII or VI
gravitational potential for an open or closed channel respectively may be used
as an approximation for the channels of types VIII and IX outside their 
common intersection.  A qualitative treatment of the exact type VII$_0$ and
VI$_0$ diagonal vacuum case in this context was first given by Lifshitz and
Khalatnikov [26] 
and then in a better approximation by Khalatnikov and
Pokrovsky [29]. 

The Hamiltonian for this case in supertime time gauge at the canonical points
of $\Cscr_D$ is
$$ \left(\vcenter{\vbox{\hbox{VII$_0$}\hbox{VI$_0$}}}\right): \hskip 30pt
H =\half(-p_0\null^2 +p_+\null^2 +p_-\null^2) 
  +24e^{4(\beta^0+\beta^+)}
\left(\vcenter{
\vbox{\hbox{$\sinh^2 2\sqrt3\beta^-$}\hbox{$\cosh^2 2\sqrt3\beta^-$}}}
\right)
\ .
\eqno(4.24)$$
The potential moves with constant $\Omega$-velocity $d\beta^+/d\Omega=-1$
(a null velocity in $diag(3,R)$) so a rest frame does not exist and one must
instead use null coordinates 
$$ 
\eqalign{ 
&(w,v,p_w,p_v) = (\beta^0+\beta^+, \beta^0-\beta^+, \half(p_0+p_+),
  \half(p_0-p_+)) \ ,\cr
&(\beta^0,\beta^+,p_0,p_+) = (\half(w+v),\half(w-v),p_w+p_v, p_w-p_v)\ ,\cr
&\hbox{\bf g} =\hbox{diag}(e^{2w}, e^{2w},e^{-w+3v}) \ ,\qquad
\hbox{\bf g} \rarrow f^{-1}_{\hbox{$e^{\alpha\hbox{\bf I}^{(3)}}$}} (\hbox{\bf 
g}) 
\hskip 2pt : \hskip 2pt (w,v) \rarrow (w+\alpha,v+\fraction{1}{3}\alpha)
\ ,\cr
& H=-2p_wp_v +24e^{4w}\left(\vcenter{\vbox{\hbox{sinh$^22\sqrt3\beta^-$}
\hbox{cosh$^22\sqrt3\beta^-$}}}\right) +\half p_-\null^2 \ ,\cr
&w-w_0=-2p_v\tb\ ,\quad {\dl p}_v=0\ , \quad p_v<0\quad (\hbox{expansion}) \ ,\cr
&\ddot{v} = -2{\dl p}_w = 8\cdot 24e^{4w}
\left(\vcenter{\vbox{\hbox{sinh$^22\sqrt3\beta^-$}
\hbox{cosh$^22\sqrt3\beta^-$}}}\right)\ . \cr}
\eqno(4.25)$$
The coordinate $w$ comoves with the potential function, while $v$ describes
the motion of the system relative to the potential.
 
The Taublike type VI$_0$ case with $\beta^-=0=p_-$ is an exact solution which
is useful to consider in detail and represents the path at the center of the
closed channel  
$$\eqalign{
& v-v_0-V\tb= 3p_v\null^{-2}e^{4w}=\fraction{1}{12}\xi^2 \ , \qquad
0=H=-2p_vV\rarrow V=0 \ ,\cr
&(\beta^0,\beta^+)  = (\half w+\fraction{3}{2}p_v\null^{-2}e^w \ , 
\half w-\fraction{3}{2}p_v\null^{-2}e^w) \ ,
 \qquad d\beta^+/d\Omega(\pm\infty)=\mp1 \ ,  
\cr
&(p_0,p_+)=
( -|p_v|(1+12p_v\null^{-2}e^{4w}),
-|p_v|(1-12p_v\null^{-2}e^{4w}))
\ .\cr}
\eqno(4.26)$$
The turnaround time $\tb_0$ is defined by $p_0(\tb_0)=0$ or $12p_v\null^{-2}
e^{4w(\tb_0)} =1$, while setting ${1\over2}p_0\null^2 = 
24e^{4(\beta^0 +\beta^+_{\sst wall})} = 24
e^{4w_{wall}} $ defines the location of the wall for this $\beta^-=0$  
case
$$\eqalign{&\beta^+_{\sst wall} =\Omega =\half \ln[ |p_0|(4\sqrt3)^{-1}] \,,
\qquad 
d\beta^+_{\sst wall}d\Omega =1-4(1+(12)^{-1}p_v\null^2e^{4w})^{-2} \,,\cr
&d\beta^+_{\sst wall}/d\Omega(\pm\infty)=1\,, \qquad d\beta^+_{\sst wall}
/d\Omega(\tb_0)=0\ .\cr}
\eqno(4.27)$$ 
The system has the past Kasner limit at $\tb=-\infty=w$ characterized by the
Lifshitz-Khalatnikov parameter $u={1\over2}$ (free motion in the positive 
$\beta^+$-direction towards the wall), and the Kasner limit at $\tb=\infty=w$
characterized by the L-K parameter $u=\infty$ (free motion away from the
wall) 
long after the collision when it has reversed direction and is again moving
along a null curve in $diag(3,R)$.  The future Kasner limit does not look like
free motion in the supertime time gauge since the supertime remains sensitive
to the small curvature in this limit and does not reach a stage where it is 
affinely related to $\Omega$-time and instead diverges from $\Omega$-time.

From the point of view of $\Omega$-time,
both the universe point and the wall initially have unit $\Omega$-velocity
and hence the universe point cannot overtake the wall.  (This is the origin
of the term ``long era", since this behavior will persist for a long period 
of $\Omega$-time.)  However, since the universe point and potential have no 
relative motion, the very small value of the potential at the location of the
universe point has a continued effect and over a long time begins to reverse
the motion and decrease the expansion energy which in turn slows down the
wall enough so that an actual collision occurs at the turnaround point when
both the wall and universe point have zero $\Omega$-velocity and the same
position, after which the wall resumes its unit $\Omega$-velocity in the
same direction and the universe point returns to its unit $\Omega$-velocity
but in the opposite direction.  Modulo constants this same solution describes
a head on collision with the approximate tilt potential $U^{(3)}_{\sst tilt}$
for constant $|v^{\prime}_3|$.

By introducing the new time variable $\xi=|6/p_v|e^{2w}=|6/p_v|e^{2w_0 +4|p_v|
\tb}$ and the abbreviation $q=4\sqrt3\beta^-$, the equation of motion for
$\beta^-$ is given by [29] 
$$ d^2q/d\xi^2 +\xi^{-1}dq/d\xi + \hbox{sinh}\,q =0 \ .\eqno(4.28)$$
For $q\ll 1$ where $\hbox{sinh}\,q \approx q$, this is just Bessel's equation
of order zero [16,26], while for $\xi\gg 1$ (at very late supertime), this
has an exact solution in terms of Jacobi elliptic functions [29]. 
Khalatikov and Pokrovsky use a WKB-like approximation involving this 
approximate solution to obtain a qualitative solution valid for all times
and whose limit at $\tb\rarrow-\infty$ specifies the asymptotic Kasner phase
which precedes the channel corner run in supertime (but follows it in 
$\Omega$-time).  

Consider the case $\beta^-\ll1$ where the approximation $\hbox{sinh}\,4
\sqrt3\beta^-\approx 4\sqrt3\beta^-$ is valid and the part of the Hamiltonian
which governs the $\beta^-$-motion is just that of a harmonic oscillator with
the supertime frequency $\omega\equiv24 e^{2w}= 4|p_v|\xi$ 
$$ H=- p_wp_v + 24e^{4w}\delta^Z_{VIII} + H_- \qquad H_- =\half p_-\null^2
+\half\omega^2(\beta^-)^2 \ .\eqno(4.29)$$ 
In the adiabatic limit which occurs at late supertime ($\xi\gg1$) when the
change in frequency per period of ocsillation goes to zero, one has the
approximate solution
$$ 
\eqalign{
\beta^- &= A\cos(\int \omega d\tb ) =A\cos(\xi-\xi_0)\,,\cr 
A&=(be^{-2w}/12)^{1\over2} =|2p_v\xi/b|^{-{1\over2}} 
= [3\pi\xi/C]^{-{1\over2}}
\,,\cr}
\eqno(4.30)$$
where the expression for the amplitude follows from the constancy of the 
adiabatic invariant $H_-/\omega={1\over2}A^2\omega = b$, as discussed by
Misner [15].  
The constant $C$ is the constant introduced by Khalatnikov
and Pokrovsky [29].

The equation of motion for $v$ may be averaged, using $\ketl(\beta^-)^2\ketr
={1\over2}A^2$ 
$$ 
\eqalign{&\ddot{v}=192e^{4w}\delta^Z_{VIII} +(48)^2e^{4w}(\beta^-)^2\ ,\qquad
\ketl\ddot{v}\ketr= 192e^{4w}\delta^Z_{VIII} +96be^{2w} \ ,\cr
&\ketl\dott{v}\ketr = 24|p_v|^{-1}e^{4w}\delta^Z_{VIII} +12be^{2w} +V\ ,\cr
&\ketl v-v_0\ketr+V\tb = 3|p_v|^{-2}e^{4w}\delta^Z_{VIII} +3be^{2w}  
=\half b|p_v|\xi +\fraction{1}{12}\xi^2\delta^Z_{VIII} \ .\cr}
\eqno(4.31)$$
The averaged super-Hamiltonian constraint requires $V=0$.  Note that the
leading approximation to the $(w,v)$ motion in the closed channel case is
just the Taublike type VI$_0$ vacuum solution.

At early $\Omega$-time, the universe point in its initial asymptotic Kasner 
phase and the potential both have unit 
$\Omega$-velocity and keep pace with each other, so $w=\beta^+-\Omega$
changes only slowly in $\Omega$-time.  (Supertime and $\Omega$-time diverge
at $\tb\rarrow\infty$.)  As the frequency decreases with increasing 
$\Omega$-time, the amplitude increases and the $\beta^+$ motion slows down
relative to the potential until the small $\beta$ approximation breaks down,
eventually reversing its direction, and after a final bounce goes into
an asymptotic Kasner phase which leaves the channel potential behind.

One can also consider a mixing bounce with a centrifugal wall present in the
channel.  This corresponds to the Bianchi type VII$_0$ or VI$_0$ 
symmetric case $\Mscr_{S(3)}$.  Adding the term $3{\s P}_3\null^2\overline{
\Gscr}\null^{-1\,33}$ to (4.24) only effects the $\beta^-$ motion directly,
while the constant $\s P_3$ enables one to find the variable $\theta$ defined
as in (4.18).  Matzner and Chitre [65] 
discuss this case in the adiabatic limit for $\beta^-\ll1$.
The mixing bounce is confined to the region between the channel gravitational
wall and one centrifugal wall in the open channel case where the $\beta^-$
motion corresponds to a 2-dimensional oscillator with nonzero angular
momentum.  In the closed channel case the situation is much different.

When the curvature of the channel is not important, one simply has a sequence 
of bounces between one gravitational wall and a centrifugal wall.  During the
bounces from the latter wall the ordered Lifshitz-Khalatnikov parameter 
$u^{\sst ordered}$
does not change.  Such a sequence of bounces with an exponential centrifugal
wall is described in terms of transformations of the first and second type 
in ref.(24).

The importance of the $a^2$ scalar curvature term on the dynamics also can be
described by a wall
$$ 
\eqalign{&U^{\sst class B}_g =72a^2e^{4(\beta^0+\beta^+)}\ ,\cr
&\beta^+_{\sst wall} = \Omega +\half\ln|p_0/12a|\qquad d\beta^+_{\sst wall}
/d\Omega = 1 +\half d\ln|p_0|/d\Omega\ .\cr}
\eqno(4.32)$$
This potential becomes important only during collisions with its associated
wall, but since this wall has unit $\Omega$-velocity away from the location
of the universe point when it is in free motion, it becomes less and less 
important as $\Omega\rarrow\infty$.   The exact diagonal Bianchi type V
solution describes a collision with this wall
$$ H=\half (-p_0\null^2+p_+\null^2+p_-\null^2) +\half(12a)^2e^{4(\beta^0+
\beta^+)}\ .\eqno(4.33)$$
$\beta^{\pm}$ obey the free equations of motion (since $V^{\ast}=0$) 
with constant $p_{\pm}$, leaving
the super-Hamiltonian constraint to determine $\beta^0$. This is easily
integrated if one imposes the supermomentum constraint $p_+=0$, thus setting   
$\beta^+=\beta^+_0$.  (The equations of motion for $p_+\neq 0$ are
integrable in null coordinates, but the super-Hamiltonian constraint then
forces $p_+=0$.)  
$$
\eqalign{
&\beta^-=p_-(\tb-\tb_0)+\beta^-_0 \ ,\qquad \dott{\beta}\null^{0\,2}-
(12a)^2e^{4(\beta^0+\beta^+_0)} = p_-\null^2 \ ,\cr
&e^{2(\beta^0+\beta^+_0)}
= -|p_-/12a|\hbox{csch}\,|p_-|(\tb-\tb_0) \ ,\qquad -p_0= -|p_-|
\hbox{coth}\,2|p_-|(\tb-\tb_0)\ ,\cr
&\beta^+_{\sst wall}=\beta^+_0 + \half\ln\hbox{cosh}\,2|p_-|(\tb-\tb_0)\ ,\qquad
d\beta^+_{\sst wall}/d\Omega =\hbox{tanh}^22|p_-|(\tb-\tb_0)\ .\cr}
\eqno(4.34)$$   
The evolution of $\beta^0$ corresponds to motion in the 1-dimensional potential\hfill\newline
$-{1\over2}(12a)^2e^{4(\beta^0+\beta^+)}$ with constant energy $\Escr_0={1\over
2}p_-\null^2$.
The expanding solution has $\tb-\tb_0\in (-\infty,0)$; at $\tb\rarrow-\infty$
the solution has a Kasner limit but it runs out of supertime at $\tb=\tb_0$
when a collision takes place.  The wall moves in from $\beta^+=\infty$ at $\tb=
-\infty$ when it has unit $\Omega$-velocity to  $\beta^+=\beta^+_0$ where its
velocity vanishes and the wall makes contact with the universe point at a 
finite value of $\beta^-$.  However, the proper time goes to $\infty$ at
$\tb=\tb_0$ since near $\tb\approx\tb_0$ 
$$ 
t-t_0 =\int  e^{3\beta^0}d\tb \sim\int  |\tb-\tb_0|
^{-{3\over2}}d\tb
\sim |\tb-\tb_0|^{-{1\over2}}\ .\eqno(4.35)$$
Thus even though $\beta^-$ obeys the free dynamics in supertime time gauge,
it runs out of supertime at a finite value $\beta^-_0$, i.e. its value
becomes frozen in proper time or $\Omega$-time. 

It is basically an effect of this kind which seems to result in isotropization 
of type VII$_h$ models with ``frozen in" values of $\beta^-$ as 
$\beta^0\rarrow\infty$ [40].  
A similar effect occurs for all Bianchi types
when the matter contributions to the Hamiltonian dominate the dynamics as
$\beta^0\rarrow\infty$, as noted by Doroshkevich, Lukash and Novikov [40].
For example, consider the exact diagonal type I case where $v_a=0=P_a$ and the
Hamiltonian is
$$ H = \half (-p_0\null^2 +p_+\null^2 +p_-\null^2) + 12e^{3\beta^0}\Hscr_M\,,
\qquad 12e^{3\beta^0}\Hscr_M = 24kl^{\gamma}e^{3\beta^0(2-\gamma)}\ .
\eqno(4.36)$$
Here $p_{\pm}$ are constants and $\beta^{\pm}$ undergo free motion, reducing
problem of the evolution of $\beta^0$ to motion in the potential $-24kl^{\gamma}
e^{3\beta^0(2-\gamma)}$ with constant energy $\Escr_0={1\over2}(p_+\null^2
+p_-\null^2)$.  One may directly integrate the super-Hamiltonian constraint
$$
\eqalign{ &\half(\dott{\beta}\null^0)^2 -24kl^{\gamma} e^{3\beta^0(2-\gamma)} 
 =\Escr_0 \ ,\cr
&\qquad e^{\beta^0}=\left\{\vcenter{
 \halign{$\lft{#}$&\qquad $\lft{\gamma\null#}$\cr
|(24kl^{\gamma}/\Escr_0)^{1\over2}\hbox{sinh}\,(2\Escr_0)^{1\over2}
(\tb_0-\tb)|^{-2(2-\gamma)
^{-1}/3}& \neq 2\cr  e^{[2(\Escr_0 + 24kl^2)]^{1\over2}(\tb-\tb_0)}
& =2 \cr}}\right.\ .\cr}
\eqno(4.37)$$
The growth of the matter Hamiltonian as $\beta^0\rarrow\infty$ again leads to
a singularity in supertime and hence a ``freezing in" of the values of $\beta
^{\pm}$ occurs in proper time or $\Omega$-time when $\gamma\neq 2$.  However,
these effects are important for the direction $\beta^0\rarrow\infty$ but 
diminish in importance as $\Omega\rarrow\infty$, since the solution quickly
approaches its Kasner asymptote.
 
On the other hand when $\gamma=2$, the system again experiences free motion
but along a timelike direction in $diag(3,R)$
$$ \eta^{AB}p_Ap_B =   -48kl^2\,,
\quad d\beta/d\Omega= [(d\beta^+/d\Omega)^2 +(d\beta^-/d\Omega)^2]^{1\over2}
 =[1-48kl^2p_0\null^{-2}]^{1\over2}\ .
\eqno(4.38)$$  
A stiff perfect fluid with $v_a=0$ only changes the supertime time gauge
Hamiltonian by the addition of a positive constant; the only change one must
make in the description of a collision with a gravitational or centrifugal wall
is $\Escr_0\rarrow\Escr_0 +24kl^2$, where $\Escr_0$ is the constant energy
defined in each of these problems.   (It is exactly this change which allows
one to easily insert a stiff perfect fluid in the Taub-Nut spacetime [41].)
Collisions with gravitational walls as
$\Omega\rarrow\infty$ decrease $|p_0|$  and therefore reduce the 
$\Omega$-velocity of the universe point between collisions.  If the universe
point does not escape from the gravitational walls first, then eventually
one has $d\beta/d\Omega< {1\over2}$ , after which the universe point can
no longer catch the faster moving gravitational walls and remains in this
free motion phase as $\Omega\rarrow\infty$. Thus such a stiff perfect fluid
(or scalar field [66]) leads to an end of an ``oscillatory approach to
the initial singularity".  (See below.) Similar calculations may be used to
study the effect of a nonzero cosmological constant which only directly
affects the equation of motion of $\beta^0$ and then indirectly affects the
remaining variables through the appearance of $\beta^0$ in their equations
of motion [67].     

What has been described so far are exact or qualitative solutions for the
case in which only one or two potentials in the supertime time gauge 
Hamiltonian play an important role at any given time.  The importance
of a given potential is described by a collision of the universe point
with the wall associated with that potential or with a corner
where two walls meet and a more careful description may be required. 
Each such phase of the motion is preceded and followed by asymptotic
Kasner phases which are related to each other by the exact or qualitative
solution for that collision.  Each collision is therefore interpreted as
a scattering problem.  As $\Omega\rarrow\infty$ one quickly reaches a stage
in which the description in terms of successive collisions with different
walls is valid and one may approximate the evolution by matching together
the Kasner asymptotes between collisions.

The class A diagonal case is easiest to describe in this manner.  As (4.36)
shows, the matter potential is exponentially cutoff as $\Omega\rarrow\infty$
and even when it dominates the spatial curvature, (4.37) shows that it has a
very shortlived effect as $\Omega\rarrow\infty$, i.e. ``the matter is 
dynamically unimportant" as $\Omega\rarrow\infty$.  In this limit the system
obeys the vacuum dynamics, leaving only the gravitational walls to determine
the qualitative solution.  Since these walls move with $\Omega$-velocity 
$1\over2$ between collisions while the universe point moves with unit
$\Omega$-velocity, in the semisimple case of Bianchi types VIII and IX where
these walls form closed trapping regions, the collisions never stop, leading
to an ``oscillatory approach to the initial singularity" in the words of
Lifshitz, Khalatnikov and Belinsky. 
The oscillations occur in the values of the variables
$e^{\beta^1}$, $e^{\beta^2}$ and $e^{\beta^3}$, a half period of an oscillation 
corresponding to two Kasner phases (``Kasner epochs" [24]) joined by a 
collision at 
the  maximum or minimum point.  A single excursion of the universe point 
into the vertex between two walls is called a Kasner era.  Except for 
collisions with
corners between two walls, these two types experience the same motion. However,
as $\Omega\rarrow\infty$ the size of the corner region where channel effects
are important becomes increasingly smaller in relation to the triangular
trapping region.  For types VI$_0$ and VII$_0$ the missing third wall provides
an escape route which leads to only a single Kasner era before the system
enters a final Kasner phase as $\Omega\rarrow\infty$, while for type II only
one collision with the single gravitational wall occurs between two asymptotic
Kasner epochs.  The Taublike cases for types VI$_0$, VIII and IX are special
cases characterized by channel trajectories which undergo a single collision
with one gravitational wall followed by an escape down an open channel in the
semisimple case, while the isotropic cases have no $\beta^{\pm}$ motion at all.
The type IX isotropic case is exceptional in that the matter is always
dynamically important since the positive spatial curvature makes the
vacuum super-Hamiltonian negative-definite.  The Kasner axes for the diagonal
case coincide with the body-fixed and space-fixed axes and are eigendirections
of the spatial Ricci curvature tensor.  

The symmetric cases (``nontumbling models" [68]) are much easier to 
describe qualitatively than the general case since the $\A G$-angular momentum 
$\s P_a$ is aligned with one of the diagonal axes which is in fact an 
eigendirection of the spatial Ricci curvature in all instances.
This axis is also an eigendirection of both {\bf g} and {\bf K}, so it
is a Kasner axis.  Only the 
1-parameter subgroup of $\A G$ which leaves this preferred axis invariant
is relevant to the dynamics; for this subgroup $\s P_a$ and $P_a$ coincide and
have a single nonvanishing component.  The same is true of $v_a^{\prime}$ 
and $v_a$ and of the components of the supermomentum $P^{\prime}(\bfdelta_a)$ 
and $P(\bfdelta_a)$ with respect to the primed and unprimed frames, all of 
which have a single (possibly) nonvanishing component along the preferred 
axis.  For the symmetric case $\Mscr_{S(a)}$, the index $a$ is associated 
with this preferred axis; as usual let $(a,b,c)$ be the associated cyclic
permutation of $(1,2,3)$. (Because of the block diagonal form of {\bf g} and 
{\bf K}, the remaining two Kasner axes are linear combinations of $e_b$ and
$e_c$.)  Then $v_a=v_a^{\prime}$ is a constant.  In  the
class A case excluding the degenerate types I and II, $a$ may assume any value
 and $l$ and $\s P_a=P_a$ are constants such that $e^{\alpha^a}{\s P}_a = 
-2klv^{\prime}_a$, leading to a single nonzero centrifugal potential which is
static in supertime time gauge.  In the class B case, only $a=3$ is allowed and 
$e^{\alpha^3}{\s P}_3 +ap_+=-2klv^{\prime}_3$, but neither $\s P_3$ nor $l$ is
a constant and the centrifugal potential is not static.

For the case $n^{(a)}\neq 0$ and $\hbox{rank {\bf n}}\,>1$, the walls associated
with the gravitational potentials and the single centrifugal potential define
closed trapping regions as is easily seen from Figure 9.  For the type $(ii)$
centrifugal potential, the pair of walls do not come into existence until 
$\zeta\leq1$ (see (4.3)) but since $|p_0|$ decreases towards zero as 
$\Omega\rarrow\infty$, this condition is eventually satisfied (unless $|p_0|$
has a limit away from zero as may occur in the case of a stiff perfect fluid).  
In the class A case one may  describe the $\Omega\rarrow\infty$ evolution in 
terms
of collisions with the centrifugal wall and the gravitational walls between
regimes of free motion.  Since these walls move with $\Omega$-velocity 0 and
$1\over2$ respectively between collisions, while the universe point moves with 
unit $\Omega$-velocity, in the case where closed trapping regions exist the 
collisions never stop, leading to an ``oscillatory approach to the initial
singularity".
In this case one must make a distinction between a Kasner epoch and an
ordinary epoch.  An ordinary epoch refers to a period of free motion in
the  $\beta^+\beta^-$ plane, while a Kasner epoch refers to 
a period of free motion with respect to the full rescaled DeWitt metric which 
includes
collisions with centrifugal walls.  In the latter case one can always
transform away the constant $\A G$-angular momentum $P_a$ by a constant linear
transformation from the space-fixed axes $e_b$ and $e_c$
to the Kasner axes. Since the Kasner
axes are time independent in a period of free motion, they can change only
during  collisions with gravitational walls.
 
As $|p_0|$ decreases due to collisions with gravitational walls, the single
centrifugal wall moves outward toward decreasing values of the centrifugal
potential where the type $(i)$ and $(ii)$ potentials rapidly approach the
asymptotic exponential potentials of type $(iii)$ and the distinction between
the three types of centrifugal potentials becomes unimportant.  Note also that
the value of $\overline{\Gscr}\null^{-1aa}$ at the universe point experiences
local maxima at each collision with the corresponding centrifugal wall
and these local maxima approach zero as the wall moves out.  Since
$\overline{\Gscr}\null^{-1aa}$ goes to zero and since $\s P_a$ is a constant
in the class A symmetric case $\Mscr_D$, the offdiagonal velocity
$\dot{W}{}\null^a=6\overline{\Gscr}\null^{-1aa}\s P_a$ also goes to zero.
The single nonvanishing automorphism variable $\theta$ is given by the
integral of $\dot{W}{}\null^a$.  Because $\overline{\Gscr}\null^{-1aa}$ is 
exponentially
cutoff, $\theta$ essentially changes only during a collision with the
centrifugal wall where the value of $\overline{\Gscr}\null^{-1aa}$ at the
universe point assumes a local maximum which decreases with each collision.
The discussion of (4.18) in fact shows that the total change $(\Delta\theta)
_{\sst bounce}$ of this variable during a single collision approaches zero
as $|p_0|\rarrow 0$.  An argument due to Lifshitz, Khalatnikov and 
Belinsky [26]
shows that the sum of these changes from any given time 
as $\Omega\rarrow\infty$ is finite, so $\theta$ approaches a limiting value
$\theta_0$ as $\Omega\rarrow\infty$.

  If $n^{(a)}=0$,
then the universe point bounces around between the centrifugal wall and a 
gravitational wall until it eventually enters a final Kasner phase which 
leaves both walls behind forever as $\Omega\rarrow\infty$.  In the type VIII
case with a type $(ii)$ centrifugal potential, the universe point  will 
collide with the gravitational walls until $|p_0|$ is lowered enough for the
pair of centrifugal walls to come into existence, after which the motion will
be confined to a smaller trapping region.  Similarly in the type VI$_0$ symmetric case with
a type $(ii)$ centrifugal potential (associated with $\Mscr_{S(3)}$ in the 
canonical case),
it is possible that a final Kasner phase is reached before the centrifugal
walls appear.  

Consider the analogy with the central force problem.  If one imagines a 
nonrelativistic particle in free motion on $R^3$ whose radial motion is 
continually reversed at large distances from the origin  by collisions with an
expanding spherical shell centered at the origin, the angular momentum will be
conserved  but the impact vector connecting the origin to the point of closest
approach of the particle to the origin will change at each collision.  A 
different translation of the origin will then be required in each free motion
phase between collisions to transform away the angular momentum.  The Kasner
axes are changed by collisions with the gravitational walls for a similar
reason.  Thus even after the motion in the offdiagonal variable $\theta$ 
essentially stops and one transforms to a diagonal gauge frame where its
limiting value $\theta_0$ is zero (so that $e_d=e_d^{\prime}$), two of the
Kasner axes do not coincide with the diagonal axes (otherwise the 
$\A G$-angular momentum would be zero) and continue to change at each collision
with a gravitational wall.
  
As long as the $a^2$-curvature potential is important in the class B case, this
description is not valid since the system will not experience free motion
between collisions with the walls and the analytic description of an
individual collision given above will not be valid.  However, eventually this
curvature potential becomes unimportant as $\Omega\rarrow\infty$ as indicated
by the above wall description and the evolution reduces to that of the 
corresponding class A model with one important exception.
Only the symmetric case $\Mscr_{S(3)}$ is allowed and it is not
the $\A G$-angular momentum $\s P_3=P_3$ which is a constant for the vacuum 
dynamics but rather the supermomentum $e^{\alpha^3}{\s P}_3 +ap_+$.  The 
equation of motion for $\s P_3$ in this case, namely $({\s P}_3)\dl{\ } =NQ_3
=48ae^{4(\beta^0+\beta^+)}e^{\alpha^3}\overline{\Gscr}_{33}$, is exactly 
$-ae^{-\alpha^3}$ times the equation of motion for $p_+$
(excluding the degenerate type V case), so even in the limit that the $a^2$
curvature potential is unimportant, $\s P_3$ still changes during collisions 
with gravitational walls to compensate for the change in $p_+$ which occurs
during these collisions, causing additional motion of the centrifugal wall.
This additional change in the $\A G$-angular momentum compared to the 
corresponding class A case then leads to an additional change in the Kasner
axes relative to that case.  Also since $n^{(3)}=0$ for the class B case,
closed trapping zones do not occur and a final Kasner phase exists as
$\Omega\rarrow\infty$ when the universe eventually leaves the walls behind.

This discussion has assumed that the matter is not dynamically important as in 
the diagonal case (when $\gamma\neq 2$).  The fluid variable $n$ may be 
expressed in the form $n= e^{3\Omega}l/u^{\sst \bot}$ where $u^{\sst\bot}\geq
1$.  In the class A case $l$ is a constant; for given values of $l$ and 
$\Omega$, $n$ can only be less than its corresponding value in the diagonal
case where $u^{\sst\bot}=1$ since $v_a\neq 0$ implies $u^{\sst\bot}>1$.
A similar statement holds for $\mu$ and $p$.  This means that the only
difference the nonzero value of $v_a$ can make that might lead to the
dynamical importance of matter in the symmetric case $\Mscr_{S(3)}$ would be
if $u^{\sst\bot}\gg1$.  In this case the matter super-Hamiltonian is 
approximated by the exponential tilt potential $U^{(a)}_{\sst tilt}$, but
since $v_a$ is a constant, the associated tilt wall moves out with unit
$\Omega$-velocity between collisions of the universe point with the
gravitational walls and becomes increasingly less important as $\Omega
\rarrow\infty$.  In this limit one is justified in ignoring the matter 
potentials.  However, the constant nonzero value of the fluid supermomentum
is required to allow a nonvanishing centrifugal potential (when the symmetric
case is nontrivial, i.e. not equivalent to a diagonal case).  

In the class B case, $l$ is not constant but satisfies the equation of motion
$(\ln l)\dl{\ }=24ae^{\beta^0+4\beta^+}v_3/v^{\sst\bot}$, so this argument does
not go through.  Because of the supermomentum constraint, $l$ must approach
a constant as $\Omega\rarrow\infty$ if the gravitational supermomentum is to
approach a constant as assumed above.  In fact since the equation of motion
for the supermomentum constraint is $(2klv_3)\dl{\ }=(e^{\alpha^3}+ap_+)\dl{\ }=
a\{ p_+,\Hscr_M^{\sst tilt}\}$,  $\dott{l}$ is negligible only
when the tilt potential has a negligible effect on the $\beta^+$ motion.
Apparently such a limiting stage is reached according to the work of
Peresetsky [35].

In the general case (``tumbling models" [68])
the angular momentum is not aligned with a diagonal axis
and the time dependence of $\s P_a$ complicates matters considerably by making 
the three centrifugal potentials implicitly time dependent.  Although one or
two components of $\s P_a$ may vanish at one instant of time, this can only
occur at isolated times without reducing the problem to a symmetric case, so
except for these exceptional moments of time, all three centrifugal potentials
are present and the varying values of the angular momenta induce time dependent
translations of each of these potentials.  For the class A case, $e^{\alpha^a+
\alpha^b}n^{ab}P_aP_b =e^{\alpha^a+\alpha^b} n^{ab}\s P_a \s P_b$ is a constant
of the motion [43], 
so only two degrees of freedom are involved in translating the centrifugal 
potentials.  Closed trapping regions exist for all Bianchi types but I and V
in the general case, leading to a never ending sequence of bounces from the 
various potentials, or an ``oscillatory approach to the initial singularity".

Examining Figure 9, one sees that the closed trapping regions are all bordered
by one gravitational wall and either two or three centrifugal potentials, 
depending on the relative magnitudes of the components of the offdiagonal
momenta.  Because the walls associated with $U^{(1)}_c$ and $U^{(2)}_c$ have
been drawn in symmetrical positions for the canonical types VI, VII and II, 
only two bounding centrifugal walls are shown for these types, but in general
the trapping region situation is described by the type IV figure with three
bounding centrifugal walls.  This is true even for the types VIII and IX where
a more detailed figure showing pairs of walls at varying distances from the
$\beta^+$-axes would make this possibility obvious.  However, since the 
gravitational walls move with $\Omega$-velocity $1\over2$ between collisions,
while the motion of the centrifugal walls is only due to the time dependence
of the offdiagonal momenta, the percentage of the trapping region that might
be cut off the corner opposite the gravitational wall by the third centrifugal
wall quickly goes to zero and the situation effectively looks like the one
described in the upper diagrams of Figure 9, unless something unusual 
happens in the evolution of the offdiagonal momenta.   

This is analogous to the argument which shows that collisions with the
tilt potential should be unimportant as $\Omega\rarrow\infty$. Neglecting the
velocity due to  the time dependence of $v_a^{\prime}$, these walls move
with unit $\Omega$-velocity between collisions so the percentage of the
trapping region cut off its corner by the tilt wall rapidly goes to zero. 
Put slightly differently, 
the truncated region quickly disappears on a 
scale which expands to keep up with the trapping region borders.  In 
the type IX case, compactness enables one to eliminate the qualifications
regarding the time dependence of $\s P_a$ and $v_a^{\prime}$ in these
arguments; the existence of the positive-definite quadratic constant of the
motion  
$|(2kl)^2n^{ab}v_a^{\prime}v_b^{\prime}| = |n^{ab}e^{\alpha^a+\alpha^b}
{\s P}_a{\s P}_b|$ [43]
provides an upper bound for the absolute value of each component.
One may compare the locations of the centrifugal and tilt walls with the 
locations they would have for each symmetric case characterized by the constant
upper bound of the corresponding component.  Due to the smaller value of 
${\s P}_a\null^2$ or $v_a^{\prime}\null^2$ compared to its upper bound, the 
actual wall locations must be outside the comparison walls relative to the
trapping region.  The tilt potentials can therefore have less effect than in 
the comparison symmetric cases where it has already been established that they
are unimportant in the limit $\Omega\rarrow\infty$.  Similarly the motion of
the centrifugal walls outward towards decreasing values of the associated 
potential is limited by the motion of the comparison wall which moves only due
to changes in $|p_0|$.  Thus the shape of a trapping region in the type IX
case is determined by one gravitational wall and two centrifugal walls in the
limit $\Omega\rarrow\infty$ and the arguments given for the symmetric case 
show that matter is not dynamically important in this limit.  This reasoning
does not extend to the remaining types but the work of Peresetsky [35]
seems to indicate that the results remain valid.  This question requires
further study in the present approach.

As $\Omega\rarrow\infty$, $\overline{\Gscr}\null^{-1aa}$ assumes it greatest
values during collisions of the universe point with the corresponding 
centrifugal wall, leading to significant values of $\dot{W}{}\null^a$ only during
such collisions according to (4.32).  This means that
the automorphism matrix {\bf S} essentially changes           
only during collisions with a centrifugal potential, during which it undergoes
a right translation by the matrix $e^{\theta\bfskappa\null_a}$ for a collision
with the potential $U^{(a)}_c$, where $\theta$ is given by the symmetric case
description of the collision already discussed.  The corresponding component
$\s P_a$ does not change much during this collision, justifying the 
approximation.  If the outward motion of the centrifugal walls towards 
decreasing values of their potentials due to decreasing $|p_0|$ is not 
compensated for by the
time dependence of the offdiagonal momenta, then $\overline{\Gscr}\null^{-1aa}$
approaches zero.  In the type IX case the bounds on the offdiagonal momenta lead
to the vanishing of the offdiagonal velocities and
an end to the offdiagonal motion.
However, this seems to be an effect which does not depend on the Bianchi type.
  
  Lifshitz, Khalatnikov and Belinsky [64]
claim that the unprimed and
primed gravitational and fluid supermomenta and the matrix {\bf S} approach
constants as $\Omega\rarrow\infty$. In the nonabelian class A case, all of the
centrifugal potentials would then become static since the $\A G$-angular
momenta $\s P_a$ would approach constants.  The evolution then would depend
only on the arrangement of the walls which enclose the trapping region.  All of
these walls would soon find themselves in the regime in which their potentials
are all exponential and the differences between Bianchi type are washed out.
For corresponding trapping regions in these nonabelian class A Bianchi type 
cases, the evolution becomes identical in this limit. In particular the change
in Kasner axes which occurs during a collision with the single gravitational
wall as computed by Lifshitz, Khalatnikov and Belinsky [64] 
does not
depend on Bianchi type.   In the class B case, excluding type V, $\s P_1$ and
$\s P_2$ would approach constants but $\s P_3$ would still change during
gravitational wall collisions as in the symmetric case, leading to an 
additional change in the Kasner axes relative to the class A models. 

The BLK discussion involves a trapping region formed by
the potentials $U^{(1)}_g$, $U^{(1)}_c$ and $U^{(3)}_c$; the outward motion of
the two centrifugal
walls as $\Omega\rarrow\infty$ leads to their condition $e^{\beta^1}\gg
e^{\beta^2}\gg e^{\beta^3}$ being satisfied by points in this trapping 
region, the inequality increasing with $\Omega$. The BLK limit for this
trapping region ($n^{(1)}\neq0$) is governed by the following supertime time
gauge Hamiltonian for the diagonal variables, with $\s P_1$, $\s P_2$, $l$ and
$v_a^{\prime}$ constants related by the supermomentum constraints, while
$\s P_3$ is determined by those constraints
$$\eqalign{ H_{BLK} &= \half\eta^{AB}p_Ap_B +6[e^{4\sqrt3\beta^-_1} 
(e^{-\alpha^1}n^{(2)})^2 {\s P}_1\null^2 + e^{4\sqrt3\beta^-}
(e^{-\alpha^3}n^{(1)})^2 {\s P}_3\null^2] \cr
& \qquad +\half (n^{(1)})^2 e^{4(\beta^0-2\beta^+)}\cr
& e^{\alpha^3}{\s P}_3 =-ap_+-2klv_3^{\prime} \ .\cr}\eqno(4.39)$$
Furthermore, as $|p_0|$ and therefore $|p_+|$ decrease towards zero with 
increasing $\Omega$, the term $ap_+$ becomes negligible in the
supermomentum constraint so that $\s P_3$ also approaches a constant.
In other words the BLK limit for the generic dynamics of all spatially 
homogeneous perfect fluid spacetimes ($\gamma\neq 2$)
allowing anisotropic spatial curvature (all but I and V) is eventually
described by a Bianchi type II model with frozen in values of the offdiagonal
variables, class B spacetimes passing through a type IV stage before reaching
this limit.  The relevant type II value of $\A g$ for this limit in the case
$n^{(1)}\neq0$ is the matrix Lie algebra of superdiagonal matrices, a
3-dimensional Lie algebra of Bianchi type II, for which the centrifugal
potentials are all of type $(iii)$. 

In order to understand the existence of this BLK limit using the present 
approach, one must 
examine in detail the equations of motion for the offdiagonal momenta and
fluid variables.  Although the BLK analysis may be correct, it would
certainly help to have a clearer discussion of these questions.  However,
this requires a more careful treatment than the limitations of time and
space permit in this article.

\section{Concluding Remarks} 

At one time or another spatially homogeneous cosmological models have captured
the imaginations of a significant fraction of the relativity community.  
Independent of the numerous reasons which have motivated their study, this
finite dimensional class of spacetimes within the context of a particular 
relativistic theory of gravitation represents an extremely rich 
mathematical system worthy of a careful and systematic treatment.
Its description brings together the fields of 
differential geometry, Lie group theory and classical mechanics in an elegant
example revealing many facets of each individual field and providing a 
finite dimensional setting within which certain questions about relativistic
theories of gravitation may be more easily explored.

The present article has endeavored to tie together numerous individual results 
concerning solutions of the Einstein equations of general relativity for
this class of cosmological models by constructing a single unifying framework
based on a few simple ideas.  Developing the consequences of these ideas
unavoidably leads to an unusual amount of detail.  These details have not been
sacrificed in the hope of establishing a clearer perspective of this  
area of cosmology.  

The key to the present discussion of spatially homogeneous dynamics is of
course diagonal gauge.  In the semisimple case of Bianchi types VIII and IX,
this gauge coincides with both the minimal strain and minimal distortion 
gauges introduced by Smarr and York for spatially compact and asymptotically
flat spacetimes [69]. 
For the remaining types where the correspondence
with the general theory breaks down [2], diagonal gauge generalizes the
attractive properties of the minimal strain minimal distortion gauge in the
semisimple case.  The decomposition of the usual synchronous gauge 
gravitational variables into diagonal and offdiagonal variables which 
accompanies the transformation to diagonal gauge in the semisimple case
is merely an application of the ideas put forth by Fischer, Marsden and 
York, among others (see ref.(62) for a summary and further references), in
describing the true degrees of freedom of the gravitational field and related
questions.   It is therefore clear that the power of the present approach
has come simply from taking seriously the ideas that have been developed in
the context of the three-plus-one approach to general relativity and
modifying them when necessary to fit the spatially homogeneous symmetry.

\appendix

\section{
The Adjoint Representation Trick}

If $e$ = $\{ e_{a}\}$ is a basis of the Lie algebra $g$
 of a 3-dimensional Lie group $G$, with structure constant tensor components
 $C^{a}{}_{bc}$, one
may introduce canonical coordinates $\{ x^{a}\}$ of the second
kind on the group [52]
in a local patch centered at the identity 
($x^{a}=0$) by the parametrization
$$x=\exp(x^{1}e_{1})\exp(x^{2}e_{2})\exp(x^{3}e_{3})\quad \in \quad G
 .\eqno(A.1)$$
The left invariant and right invariant frames $e$ and $\s e$ on $G$ are related
by the matrix adjoint representation of $G$
$$\eqalign{
        \s e_{a}&=R^{-1b}_{\hskip 13pt a}e_{b} \cr
        \hbox{\bf R}(x)&\equiv Ad_{e}(x) \equiv e^{x^{1}\hbox{\bf k}_{1}}
        e^{{x}^{2}
        \hbox{\bf k}_{2}}e^{{x}^{3}\hbox{\bf k}_{3}}
        \ \in \ Ad_{e}(G) \cr }\eqno(A.2)$$
where the expression for $\hbox{\bf R}(x)$ holds in the given coordinate patch
and $\hbox{\bf k}_{a} \equiv \hbox{ad}_{e}(e_{a}) \equiv (C^{b}{}_{ac})$ are
the adjoint
matrices for the given basis $e$, spanning the adjoint matrix Lie algebra
$\hbox{ad}_{e}(g)$ and satisfying $[\hbox{\bf k}_{a},\hbox{\bf k}_{b}] =
 C^{c}{}_{ab}\hbox{\bf k}_{c}$ which is the matrix form of the Jacobi identity.
The matrix 
{\bf R}  belongs to the linear adjoint matrix group $Ad_{e}(G)$ of $G$, which 
for a connected Lie group equals the inner automorphism matrix group 
$IAut_{e}(g)$ of the Lie algebra $g$. When $\{\hbox{\bf k}
_{a}\} $ are 
linearly  dependent matrices (i.e. the center of $g$
is trivial) and thus a basis of $\hbox{ad}_{e}(g)$ having the same 
structure
constant tensor components as $e$, then $\hbox{ad}_{e}(g)$ and $g$  are
isomorphic
and $Ad_{e}(G)$ is locally isomorphic to $G$, with $\{ x^{a}
\}$ serving as
the corresponding coordinates on the matrix group through (A.2).  This is
true for all Bianchi types except I, II and III.

When $C^{a}{}_{bc}$ belongs to $\Cscr_{D}$, the adjoint matrices are
$$\eqalign{
\hbox{\bf k}_{1}&=-n^{(2)}\hbox{\bf e}^{3}{}_{2}+n^{(3)}\hbox{\bf e}
   ^{2}{}_{3}-a\,\hbox{\bf e}^{3}{}_{1},\cr
\hbox{\bf k}_{2}&=-n^{(3)}\hbox{\bf e}^{1}{}_{3}+n^{(1)}\hbox{\bf e}
   ^{3}{}_{1}-a\,\hbox{\bf e}^{3}{}_{2},\cr
\hbox{\bf k}_{3}&=-n^{(1)}\hbox{\bf e}^{2}{}_{1}+n^{(2)}\hbox{\bf e}
   ^{1}{}_{2}+a(\hbox{\bf e}^{1}{}_{1}
+\hbox{\bf e}^{2}{}_{2}), \cr} \eqno(A.3)$$
where $\{ \hbox{\bf e}^{a}{}_{b}\}
 $ is the standard basis of $gl(3,R)$,enabling a 
matrix to  be expressed as $\hbox{\bf A}=A^{a}{}_{b}\hbox{\bf e}^{b}{}_{a}
$.  The
 matrix $\hbox{\bf R}(x)$ is then easily evaluated
$$ \hbox{\bf R}(x)=\left(\vcenter{\halign{$\ctr{#}$\quad &$\ctr{#}$\quad
   &$\ctr{#}$\cr 1&0&-ax^{1}\cr 0&c_{1}&-n^{(2)}s_{1}\cr
   0&n^{(3)}s_{1}&c_{1}\cr}}\right)
    \left(\vcenter{\halign{$\ctr{#}$\quad &$\ctr{#}$\quad &$\ctr{#}$\cr
   c_{2}&0&n^{(1)}s_{2}\cr 0&1&-ax^{2}\cr -n^{(3)}s_{2}&0&c_{2}\cr}}\right)
    \left(\vcenter{\halign{$\ctr{#}$\quad &$\ctr{#}$\quad &$\ctr{#}$\cr
   c_{3}e^{ax^{3}}&-n^{(1)}s_{3}e^{ax^{3}}&0 \cr
   n^{(2)}s_{3}e^{ax^{3}}&c_{3}e^{ax^{3}}&0 \cr 
   0&0&1\cr}}\right)\ . \eqno(A.4)$$

The following abbreviations and identities are useful, where $(a,b,c)$ is 
any cyclic permutation of $(1,2,3)$ and no indices are summed
$$
\eqalign{
& m^{(a)}=(-n^{(b)}n^{(c)})^{{1\over 2}}\,,\quad 
     \s m\null^{(a)}=(n^{(b)}n^{(c)})^{{1\over 2}}\,,\quad c_{a}=\hbox{cosh}\ 
      m^{(a)}x^{a}=\cos \s m\null^{(a)}x^a \,\cr
&s_{a}=(m^{(a)})^{-1} \hbox{sinh}\ m^{(a)}x^{a} =
       (\s m\null^{(a)})^{-1} \sin \s m\null^{(a)}x^{a}\,,\quad
       \lim_{m^{(a)}\rarrow 0} (c_{a},s_{a})=(1,x^{a})\,,\cr
& (c_{a})^{2}-(m^{(a)}s_{a})^{2}=(c_{a})^{2}+(\s m\null
       ^{(a)}s_{a})^{2}=1\,,\cr 
& d\, c_{a}=(m^{(a)})^{2}s_{a} dx^{a}\,,\quad 
       d\, s_{a}=c_{a} dx^{a}\,.\cr}
\eqno(A.5)$$

The canonical adjoint matrix groups (evaluating (A.4) at the canonical points
of $\Cscr _D$) are seen to be  IX: $SO(3,R)$ with $\hbox{\bf k}_a$ 
generating an active rotation about the $a^{th}$ axis, VIII: $SO(2,1)$ with
$\hbox{\bf k}_3$ generating an active rotation about the $3^{rd}$ axis and
$ \hbox{\bf k}_1$ and $\hbox{\bf k}_2$ passive and active boosts along the
$1^{st}$ and $2^{nd}$ axes,respectively,VII$_0$: $ISO(2,R)\ \simeq$ Euclidean 
group of the plane and VI$_0$: $ISO(1,1)\ \simeq$ Poincare group in 2 
dimensions.  The latter two matrix groups are better interpreted in terms
of their inhomogeneous action on $R^2$ by letting them act on the column
vector $(y_1,y_2,1)$, in which case $\hbox{\bf k}_3$ generates an active
rotation and a passive boost respectively of $R^2$ while $\{ 
\hbox{\bf k}_1,\hbox{\bf k}_2\}$ generate the translations.  
The canonical groups for Bianchi types VII$_h$, VI$_{h\neq -1}$, V and
IV differ in interpretation from types VII$_0$ and VI$_0$ only regarding 
the corresponding action of $\hbox{\bf k}_3$ on $R^2$.  In the first two 
cases the rotation or boost $e^{x^3\hbox{\bf k}_3}$ of Bianchi types VII$_0$
and VI$_0$ is accompanied by a dilation of $R^2$ by the factor $e^{a x^3}$
in types VII$_h$ and VI$_h$, while only the dilation is present for type 
V, and for type IV this dilation is instead accompanied by a null 
rotation of $R^2$ considered as a null 2-plane in Minkowski spacetime.  
In other words all of these matrix groups are isometric to 3-dimensional
subgroups of the conformal group of Minkowski spacetime. The canonical 
adjoint matrix groups for the remaining types I, II and III are 
respectively trivial(I) or abelian(II) or nonabelian(III) 
2-dimensional groups of $GL(3,R)$.

The Bianchi types fall naturally into three categories in this context:
(i) the types I, II and III, (ii) the remaining nonsemisimple types 
and (iii) the semisimple types VIII and IX.  Consider category (ii)
and assume  $n^{(3)}=0$ for simplicity.  The adjoint matrix group 
$T_2 \times _s Ad_3$ which through its action on the column vector 
$(y_1,y_2,1)$ is identified with the group of translations of $R^2$ 
together with a 1-dimensional subgroup of linear transformations of 
$R^2$ taken as a vector space.  Here $T_2 =\exp (\hbox{span} \{ 
\hbox{\bf e}^3{}_{1},\hbox{\bf e}^3{}_{2} \})$ corresponds 
to the translations and $Ad_3=\exp(\hbox{span} \{ \hbox{\bf k}_3 
\} )$ to the linear transformations.  Carrying out the 
matrix multiplication indicated in (A.4)   
$$\hbox{\bf R}(x)=\left(\vcenter{\halign{$\ctr{#}$\quad &$\ctr{#}$\quad
             &$\ctr{#}$\cr 
   c_{3}e^{ax^{3}}&-n^{(1)}s_{3}e^{ax^{3}}&-ax^{1}+n^{(1)}x^{2}\cr
   n^{(2)}s_{3}e^{ax^{3}}&c_{3}e^{ax^{3}}&-n^{(2)}x^{1}-ax^{2}\cr
   0&0&1 \cr}}\right)  \eqno(A.6)$$
and then considering the product $\hbox{\bf R}(x_3)=\hbox{\bf R}(x_1)
\hbox{\bf R}(x_2)$ of two such matrices, one may explicitly evaluate the
the multiplication function $x_3=\varphi (x_1,x_2)$ in these coordinates.  
Introducing the obvious notation $c_{3,a}= \hbox{cosh}\  m^{(3)}x^3_a$,
etc., one finds  
$$\eqalign{
   x^1_3&=e^{ax^3_1}(c_{3,1}x^1_2-n^{(1)}s_{3,1}x^2_2)+x^1_1 \cr
   x^2_3&=e^{ax^3_1}(n^{(2)}s_{3,1}x^1_2+c_{3,1}x^2_2)+x^2_1 \cr
   x^3_3&=x^3_1+x^3_2\ .\cr}\eqno(A.7)$$
Similarly by permutation of these formulas one can obtain the multiplication
law for all points of $\Cscr_D$ falling in category (ii).  With a little
effort one can write down a complicated generalization valid for case (iii)
as well, thus obtaining a local expression for the multiplication law on
$G$ which is in fact valid for all points of $\Cscr_D$.

For any Lie group the differential of the adjoint matrix satisfies the 
relations 
$$\hbox{\bf R}^{-1}d\hbox{\bf R}=\hbox{\bf k}_a\omega^a,\qquad
     d\hbox{\bf R}\ \hbox{\bf R}^{-1}= \hbox{\bf k}_a\s {\omega}\null^a\ ,
     \eqno(A.8)$$
where $\{ \omega^a \}$ and $\{ \s {\omega}\null^a 
\}$ are the invariant 1-form bases dual to $e$ and $\s e$
respectively.  When $\{ \hbox{\bf k}_a \}$ are
linearly  independent
matrices, these relations may be used to evaluate these 1-forms in the
local coordinates under consideration and then the corresponding bases 
$e$ and $\s e$ may be obtained from these expressions using duality.  
The result is
$$\vcenter{\halign{
        $\rt{#}$&$\ctr{#}$&$\lft{#}$\qquad
       &$\rt{#}$&$\ctr{#}$& $\lft{#}$\cr    
 \omega^{1} &\null=\null& e^{-ax^{3}}(c_{2}c_{3}dx^{1}+n^{(1)}s_{3}dx^{2}) &
 e_{1}& \null=\null&e^{ax^{3}}(n^{(2)}s_{3}\partial_{2}+c_{3}c_{2}^{\ \ -1}
 (\partial_{1}-n^{(3)}s_{2}\partial_{3}))\cr
 \omega^{2}& \null=\null& e^{-ax^{3}}(-n^{(2)}c_{2}s_{3}dx^{1}+c_{3}dx^{2}) &
 e_{2}& \null=\null& e^{ax^{3}}(c_{3}\partial_{2}-n^{(1)}s_{3}c_{2}^{\ \ -1}
 (\partial_{1}-n^{(3)}s_{2}\partial_{3}))\cr
 \omega^3& \null=\null & n^{(3)}s_2dx^1+dx^3 & 
 e_3 &\null=\null & \partial_3\cr
&&&&&\cr
\s  {\omega}\null^1 & \null=\null & dx^1+(n^{(1)}s_2-ax^1)dx^3&
 \s e_1 &\null=\null & \partial_1\cr
\s {\omega}\null^2 & \null=\null & c_1dx^2-(n^{(2)}s_1c_2+ax^2)dx^3&
 \s e_2 & \null=\null & c_1\partial_2-n^{(3)}s_1c_2^{\ \ -1}(\partial_3-n^{(1)}
  s_2\partial_1
      ))\cr
\s {\omega}\null^3 & \null=\null & n^{(3)}s_{1}dx^2 +c_1c_2dx^3 &
\s e_3 & \null=\null & n^{(2)}s_1\partial_2+c_1c_2^{\ \ -1}
    (\partial_3-n^{(1)}s_2\partial_1)\cr
    &&&&&\null+a(x^1\partial_1+x^2\partial_2)\ .\cr }}\eqno(A.9)$$  
These formulas also hold for all points of $\Cscr_D$, providing 
$\Cscr_D$-parametrized expressions for the collection of invariant fields
on the $\Cscr_D$-parametrized family of Lie groups $G$, with the 
$\Cscr_D$-parametrized local expression for the multiplication law given
by a suitable generalization of (A.7).  If $G$ is assumed to be simply
connected, then $\{ x^a \}$ are global coordinates on
$G\sim R^3$ for all but the type IX orbit where $G\sim S^3
\sim SU(2)$, 
and the group $G$ may simply be defined by these formulas, with 
restrictions on the ranges of the coordinates for points in the 
type IX orbit obtained by consideration of the basis $\A e=\{
-{1 \over 2}i(n^{(1)}n^{(2)})^{1 \over 2}\bfsigma_1,-{1 \over 2}i
(n^{(3)}n^{(1)})^{1 \over 2}\bfsigma_2,-{1 \over 2}i(n^{(1)}n^{(2)})
^{1 \over 2}\bfsigma_3 \}$ of the Lie algebra of $SU(2)$,where 
$\{ \bfsigma_a \}$ are the standard Pauli matrices.  In short,
the $\Cscr_D$-parametrized simply connected 3-dimensional Lie group 
$G_{\Cscr_D}$ has been constructed.  

By slightly modifying all of the above formulas, one may obtain expressions
for the quantities associated with the multivalued matrix group function 
$\A G$ on $\Cscr_D$ introduced in the third section.  Rather than giving 
formulas valid for all points of $\Cscr_D$, attention is restricted to those 
points for which $n^{(3)} =0$ for clarity.  Referring to (3.4), make the
following definitions
$$ \eqalign{ \{\bfkappa_a\} &=\{-b^{(2)}\bfe32 ,b^{(1)}
\bfe31 ,-a^{(1)}\bfe21 +a^{(2)}\bfe12\}\cr
\{{\A{\hbox{\bf k}}_a}\} &=\{-{\A n}\null^{(2)}\bfe32 ,
{\A n}\null^{(1)}\bfe31 ,-{\A n}\null^{(1)}\bfe21 +{\A n}\null^{(2)}\bfe12
\}\cr
C_3 &= \cos\ (a^{(1)}a^{(2)})^{1\over2}\theta^3\  \qquad 
S_3 = (a^{(1)}a^{(2)})^{-{1\over2}}\sin\ (a^{(1)}a^{(2)})^{1\over2}\theta^3
\ ,\cr}\eqno(A.10)$$
and let $({\A c}_3,{\A s}_3)$ be obtained from $(c_3,s_3)$ by the 
replacement $(n^{(a)},x^a) \rarrow ({\A n}\null^{(a)},\theta^a)$.
Then one can write down the following formulas by inspection
$$ \vbox{\hbox{$\hbox{\bf S}(\theta) =\left(\vcenter{\halign{$\ctr{#}$\quad&
$\ctr{#}$\quad&$\ctr{#}$\cr C_3&-a^{(1)}S_3&b^{(1)}\theta^2\cr
a^{(2)}S_3&C_3&-b^{(2)}\theta^1\cr 0&0&1\cr}}\right)\qquad
{\A{\hbox{\bf R}}}(\theta) =
 \left(\vcenter{\halign{$\ctr{#}$\quad&$\ctr{#}$\quad&$\ctr{#}$\cr
{\A c}_3&-{\A n}\null^{(1)}{\A s}_3&{\A n}\null^{(1)}\theta^2\cr
{\A n}\null^{(2)}{\A s}_3&{\A c}_3&-{\A n}\null^{(2)}\theta^1\cr
0&0&1\cr}}\right) $} \hbox{$\null$ }\hbox{ } \hbox{$\vcenter{
\halign{$\qquad\rt{#}\null=\null$&$\lft{#}\qquad$&$\rt{#}\null=\null$&
$\lft{#}$\cr
\du W\null^1&{\A c}_3\du\theta\null^1 +{\A n}\null^{(1)}{\A s}_3\du\theta
\null^2&P_1&{\A n}\null^{(2)}{\A s}_3p_1 +{\A c}_3p_2\cr
\du W\null^2&-{\A n}\null^{(2)}{\A s}_3\du\theta\null^1+{\A c}_3\du\theta
\null^2&P_2&{\A c}_3p_1-{\A n}\null^{(1)}{\A s}_3p_2\cr
\du W\null^3& \du\theta\null^3&P_3&p_3\cr
&&&\cr
\dot{W}{}\null^1& \du\theta\null^1 +{\A n}\null^{(1)}\theta^2\du\theta\null^3&\s P_1&p_1
\cr  \dot{W}{}\null^2&\du\theta\null^2- {\A n}\null^{(2)}\theta^1\du\theta\null^3&
\s P_2&p_2\cr \dot{W}{}\null^3&\du\theta\null^3&\s P_3&p_3-{\A n}\null^{(1)}\theta^2
p_1+{\A n}\null^{(2)}\theta^1p_2\ \ .\cr}}$}}\eqno(A.11)$$
It is worth noting that the only values of $\A G$ which are not adjoint matrix
groups of some group are those for which $\hbox{rank} \ \A {\hbox{\bf n}} =1$.
These are matrix groups of Bianchi type II.  
The centrifugal potentials are all
exponential (type $(iii)$) for such values of $\A G$.  
For example, in the case $n^{(3)}=0$, 
setting $n^{(2)}=0$ makes $a^{(2)}=0$ and gives such a value of $\A G$; its Lie
algebra consists of superdiagonal matrices.  

The invariance of the structure constant tensor under the action of the
automorphism group
$$j_{\hbox{\bf A}}(C^a{}_{bc}) = C^a{}_{bc}\ , \qquad \hbox{\bf A}\in Aut_e(g)
\eqno(A.12)$$
may be written in matrix form using the adjoint matrices
$$  \hbox{\bf Ak}_a\hbox{\bf A}^{-1} = \hbox{\bf k}_bA^b{}_{a}\ .\eqno(A.13)
$$
Similarly the tensor $\delta^{\hskip 4pt b}_{a\hskip 4pt c} \equiv
C^b{}_{ac}-2\delta^b{}_{a}a_c$ which is related to the matrices $\bfdelta_a$
of (2.49) in the same way $C^b{}_{ac}$ is related to $\hbox{\bf k}_a$ is also
invariant under the automorphism group so the matrices $\bfdelta_a$ also 
satisfy (A.13).  This fact was used in (3.37).

\section{
Automorphism Matrix Groups}

Discussion of the matrix group of the $\Cscr_D$-parametrized Lie algebra
$g_{{\cal C}_D}$ leads one to consider the four categories of table II labeled by the integers
0, 1, 2, 3 which represent the orbit dimensions within $\Cscr_D$. Discrete automorphisms will be ignored here, being discussed at length in ref.~[43], so only the identity component 
$Aut_e(g)^+ \subset Aut_e(g)$ of the matrix automorphism group and the identity component $SAut_e(g) = SL(3,R) \cap Aut_e(g)^+$ of the special automorphism matrix group will be considered. 
Deqignate their matrix Lie algebras by $aut_e(g)$ and $saut_e(g)$. The adjoint-matrix Lie algebra ${\rm ad}_e(g)$ is a Lie subalgebra of 
$aut_e(g) = der_e(g)$ which consists of the matrices of 
derivations of the Lie algebra $g$.

For the first (Abelian) category, the full matrix groups are $GL(3,R)$ and
$SL(3,R)$ respectively, while 
for the last (semi-simple category) one has $Aut_e(g)
= SAut_e(g)^+ = Ad_e(G)$ which has been discussed in appendix A. For the third category, assume $n^{(3)} = 0$ for uniformity of discussion. 
Here $Aut_e(g)$ is 4-dimensional, the extra 
dimension relative to the adjoint subgroup $Ad_e(G)$ arising from the addition of the matrix 
$\hbox{\bf I}_3 = \diag(1, 1, 0)$ to the basis of ${\rm ad}_e(g)$, generating the scaling matrix $e^{\theta {\bf I}_3} = \diag(e^\theta, 
e^\theta, 1) \in Diag(3,R)^+$. For the class A types of this category $SAut_e(g) = Ad_e(G)$, while, 
in the class B case, replacing the generator ${\bf k}_3$ of ${\rm ad}_e(g)$ by 
${\bf k}_3^0$ leads to a basis of $aut_e(g)$,                       
where $\{ {\bf k}_a^0 \}$ are the matrices obtained from (A.3) by setting the structure constant $a$ to zero. 
Thus for this category $SAut_e(g)^+ = T_2 \times_s Ad_3^0$, where 
$Ad_3^0 \equiv \exp [{\rm span}\{{\bf k}_3\}]$ and
the group $T_2 \equiv \exp ({\rm span}\{\bfe^3{}_1, \bfe^3{}_2 \})$ is generated by ${\rm span}\{ {\bf k}_1, {\bf k}_2 \}$ (except for
Bianchi type VI$_{-1} \equiv$ III, where
${\bf k}_1$ and ${\bf k}_2$ are linearly dependent, satisfying
$a{\bf k}_1 + n^{(1)} {\bf k}_2 = 0$, so only a 1-dimensional subgroup of 
$T_2$ corresponds to an adjoint transformation).

For the remaining category containing Bianchi types II and V, some notation is required. 
Let $GL_{2,3}$ be the subgroup of $GL(3,R)$ isomorphic to $GL_{2,R}$ which leaves the 3rd axis of 
$R^3$ fixed, let $SL_{2,3} = GL_{2,3} \cup SL(3,R)$ and let $T_2^T$
be the transpose of the matrix group $T_2$; let $ $ and $ $ be the corresponding Lie algebras. 
Again for uniformity of discussion assume $n = \diag(0, 0, n^{(3)})$ for type II. For Bianchi type V, one has $Aut_e(g) = T_2 \times_s GL_{2,3}$ and 
$S (g) = T_2 \times_s SL_2$ while for type II $SAut_e(g) = T_2^T\times_s SL_{2,3}$; 
$Aut_e(g)^+$ is obtained
by adding the diagonal automorphism generator $\diag(1, 1, 2)$ to a basis of the Lie algebra of 
$SAut_e(g)$, generating the scaling $\diag(e^\theta, e^\theta, e^{2\theta}) \in Diag(3,R)^+$.

Introduce the map (homomorphism) ${\rm ad}: aut(G) \to aut(g)$ 
by ${\rm ad}(\xi) X =[\xi,X]$ for $\xi \in aut(G)$ and $X \in g$, and let ${\rm ad}_e(\xi)$ be the matrix of ${\rm ad}(\xi)$ with respect to the basis $e$ of $g$; 
restricting the domain of {\rm ad} to $g$ gives the adjoint representation of $g$. For a simply connected Lie group $G$ this map {\rm ad} is a Lie algebra isomorphism; one can therefore, 
consider the inverse map ${\rm ad}_e^{-1}: {\rm ad}(g) \to aut(G)$ which associates a generating vector 
field of the automorphism group of $G$ with each matrix of the matrix Lie algebra of the 
automorphism group of $g$. For example, the adjoint matrices map onto elements of ${\rm ad}(G)$ 
with ${\rm ad}_e^{-1}({\bf k}_a) = e_a - \tilde e_a$, 
the coordinate representation of which may be read off from (A.9).
For the nonsemisimple Bianchi types where $aut(g)$ is larger than ${\rm ad}(G)$, the coordinate representation of the remaining automorphism generators is also easily obtained. For the third category only one 
additional linearly independent such generator exists given by 
${\rm ad}_e^{-1}({\bf I}_3) = - (x^1 \partial_1 + x^2 \partial_2)$ for those points of ${\cal C}_D$ for 
which $n^{(3)} = 0$. For the first (Abelian) category where 
$aut_e(g)= gl(3,R) \ni {\bf B}$, one has 
${\rm ad}_e({\bf B}) = -B^a{}_b x^b \partial_a$. The same 
formula holds for Bianchi type V if ${\bf B}$ belongs to the Lie algebra $gl_{2,3}$, leaving only the remaining Bianchi type II. 
Here only the Lie subalgebra 
$sl_{2,3} \oplus span\{ \diag (1, 1, 2)\} \subset aut_e(g)$ need be considered, assuming $n = \diag(0, 0, n^{(3)})$ for simplicity. Again the Abelian formula holds for all diagonal elements of this Lie subalgebra,
leaving only off-diagonal elements to be considered. 
Here the formula
$$
	{\rm ad}_e^{-1}(-q^{(1)}\bfe^2{}_1 + q^{(2)}\bfe^1{}_2) 
 = q^{(l)} x^2 \partial_1- q^{(2)} x^1 \partial_2 
  + \fraction12 n^{(3)} [ q^{(2)} (x^l)^2 - q^{(l)}(x^2)^2] \partial_3
\eqno(B.1)$$
is essentially due to Bianchi [10].
An important thing to understand about the automorphism group is how it acts on the dual space $g^\ast_{{\cal C}_D}$ of left 
invariant 1-forms on ${\cal C}_D$. This is important since both the supermomentum ${\cal H}^G_a \omega^a$ and fluid current 1-form 
$v_a \omega^a $ transform under this action, namely
$$
	s_a \omega^a \in g^\ast \to z_b A^{-1}{}^b{}_a \omega^a\ ,\qquad
   {\bf A} \in  Aut_e(g)\ .
\eqno(B.2)$$
The orbit space then gives one information about inequivalent initial data, while the orbits themselves give 
information about constants of the motion if the action is intransitive [43]. This automorphism group action is 
also important in establishing a canonical form for the Kasner axes in the BLK limit [35, 64].

Define $z\equiv |n^{ab} z_a z_b|^{1/2}$  
and $\gamma^{ab} \equiv  {\rm sgn}(\det {\bf n}) n^{ab}$. 
For the semi-simple types IX and VIII, $\gamma^{ab}$ 
are the components of an 
inner product of signature $(+++)$ and $(-++)$, respectively, 
and the norm $z$ associated with this inner 
product is an invariant of the automorphism action. The orbits are just surfaces of constant $z$. In the canonical 
case these are just the origin-centered spheres of radius $r$ for type IX and pseudospheres of radius $z$ for type 
VIII. For the latter type, $\varepsilon = sgn(\gamma^{ab} z_a z_b)$ may assume the invariant values 1, 0 and -1 for the two simply 
connected timelike hyperboloids, the null cone and the nonsimply connected spacelike hyperboloids, 
respectively. (The timelike directions are associated with the rotations and the spacelike directions with the 
boosts of $SO_{2,1}$, while the null directions are associated with the null rotations.) For the nonsemi-simple types; 
$n^{ab}$ are the components of a degenerate quadratic form. The norm $z$ is a constant only for the special 
automorphism action. For the canonical type-VII point, the orbits of $SAut_e(g)$ are cylinders of radius $z$ about the 
$z_3$-axis which consists of fixed points, while for the canonical type-VI point each value of $z\neq 0$ consists of 4 
disconnected orbits (`hyperbolic cylinders') equivalent under the action of discrete automorphisms, with $z = 0$ 
consisting of 4 disconnected half-plane orbits meeting at the $z^3$ axis of fixed points. Under the action of $Aut_e(g)$ 
all points with $z \neq 0$ are equivalent. For type-IV points with $n^{(l)} \neq 0$, the orbits of $SAut_e(g)$ are the planes of 
constant $|z_1|$ for $z \neq 0$, but the lines parallel to the $z_3$-axis in the plane $z_1 = 0$ for $z = 0$. For type-II points with 
$n^{(3)}\neq 0$, $SAut_e(g)$ acts on each plane of constant $|z_3|$ as the inhomogeneous special linear group of the plane, 
with all points Of $z_3 \neq 0$ being equivalent under $Aut_e(g)$, but only as the special linear group on the plane $z_3 = 0$ 
with all points of this plane except the origin equivalent under $Aut_e(g)$. For type V the $z_3$-axis consists of fixed 
points while all other points belong to the same orbit for both $Aut_e(g)$ and $SAut_ e(g)$. For type I only two orbits 
exist for both groups, the origin and all other points.

\section{
A Spatially Homogeneous Perfect Fluid}

Following the notation of chapter~23 of [18], consider a spatially homogeneous perfect fluid with energy density 
$\rho$, pressure $p$, 4-velocity field $u$; baryon number
density $n$, chemical potential $\mu = (\rho + p) /n$ and 
synchronous-gauge energy-momentum tensor
$$
	T^\alpha{}_\beta 
  = (\rho+p) u^\alpha  u_\beta +p \delta^\alpha{}_\beta \ .
\eqno(C.1)
$$
For the equation of state $p = (\gamma -1) \rho$, $n$ may be taken as the single independent thermodynamic variable, in terms of which the 
others may be expressed as follows
$$
  \rho = n^\gamma\ ,\	
  p = (\gamma - 1) n^\gamma\ ,\
  \mu = \gamma n^{\gamma-1}\ ,
\eqno(C.2)$$
where a constant of integration has been eliminated by a redefinition of $n$. Following Taub [23,70], introduce the spatial 
circulation 1-form $v$ with components $v_a = \mu u_a$ and the spatial scalar density
$$
\ell = n g^{1/2} u^\bot = n \mu^{-1} g^{1/2} v^\bot\ ,\  
\hbox{where } u^\bot = (1+ u_au^a)^{1/2}\ .
$$
Under the change of basis $\overline{e}_a = A^{-1}{}^b{}_a e_b$, the fluid variables $(\ell, v_a) $ transform in the following way
$$
  (\ell, v_a)	= f_{\bf A}(\ell, v_a) 
    = ( |\det({\bf A})|^{-1} \ell, v_b A^{-1}{}^b_a) \ .
\eqno(C.3)$$
The matter super-Hamiltonian and supermomentum may be expressed in terms of these variables as follows
$$
 {\cal H}^M = 2k \ell v^\bot - 2kpg^{1/2}\ , \
        {\cal H}_a^M = 2k \ell v_a \ .
\eqno(C.4)$$
The matter super-Hamiltonian, considered as an independent function of the spatial metric and the matter variables $(n,\ell, v_a)$, 
satisfies (2.50) and, therefore, acts as a potential for the matter driving force $T^\ast$.

The equations of motion for the fluid variables $(\ell, v_a)$ in almost synchronous gauge are [43]
$$
 \ell = N \ell (v^\bot)^{-1} 2 a^c v_c\ ,\ 	
   v_a =N (v^\bot)^{-1}C^b{}_{ba} v^c\ ,
\eqno(C.5)$$
while the defining relation for $\ell$ may be used as an equation implicitly defining $n$ in terms of $(gab, \ell, v_a)$. The fluid constants of 
motion in almost synchronous gauge are described elsewhere [43]. In the class A case, for example, $l$ and 
$V \equiv |n^{ab}v_av_b|^{1/2}$ are constants of the motion, while 
$\ell^{-1/2} V$ is a class B constant of the motion; others exist, however, with at least two 
nontrivial constants of the motion of the motion in all cases.

\section{
The Kantowski-Sachs models.}

The only spatially homogeneous space-times not described by a Bianchi type model are the Kantowski-Sachs space-times [71,72] which have a 4-dimensional isometry group acting transitively on the hypersurfaces of homogeneity (implying local rotational symmetry) but no 3-dimensional subgroup which acts simply transitively on these hypersurfaces. The spatial metrics 
of such space-times were also classified by Bianchi [10] in his categorization of all Riemannian 3-manifolds which 
admit a Lie group of isometries. Their isometry group is the direct product group $R \times SO(3,R)$ acting on the 
3-manifold $R \times S_2$ (in the simply connected case), where the additive group of real numbers acts on $R$ by translation 
and the special orthogonal group acts isometrically on $S_2$ with its standard metric.

To appreciate the relationship of these models to the Bianchi type models, consider the following choice of Euler 
angle coordinates for the class A submanifold of ${\cal C}_D$ for which $n^{(1)} \neq 0$:
$$
 	x = \exp(y^2 e_3) \exp(y^l e_2) \exp(y^3 e_3) \in G \ .
\eqno(D.1)$$
Using the trick of appendix A, one finds the coordinate expressions for the invariant fields to be
$$
\meqalign{
\omega^1 &= n^{(1)} (s_3 dy^1 - c_3 s_{2,1} dy^2)\ ,
&e_1 = (n^{(1)})^{-1}[ (m^{(3)})^2 s_3 \partial_1 
\cr & &\qquad\qquad
             -(s_{2,1})^{-1}(\partial_2 - c_{2,1} \partial_3)] \ ,\cr
\omega^2 &= c_3 dy^1 + (m^{(3)})^2 s_3 s_{2,1} dy^2\ ,
&e_2 = c_3\partial_1 
             +s_3(s_{2,1})^{-1}(\partial_2 - c_{2,1} \partial_3) \ ,\cr
\omega^3 &= c_{2,1} dy^2 + dy^3\ ,
&e_3 = \partial_3 \ ,\cr
& & \cr
\tilde\omega^1 &= n^{(1)} (-s_{3,2} dy^1 + c_{3,2} s_{2,1} dy^3)\ ,
&\tilde e_1 = (n^{(1)}{}^{-1}[ -(m^{(3)})^2 s_{3,2} \partial_1 \cr
& & \qquad\qquad
         +c_{3,2}(s_{2,1})^{-1}(\partial_3 - c_{2,1} \partial_2)] \ ,\cr
\tilde\omega^2 &= c_{3,2} dy^1 + (m^{(3)})^2 s_{3,2} s_{2,1} dy^3 \ ,\qquad
&\tilde e_2 = c_{3,2}\partial_1 
             +s_{3,2}(s_{2,1})^{-1}(\partial_3 - c_{2,1} \partial_2) \ ,\cr
\tilde\omega^3 &=  dy^2 + c_{2,1} dy^3\ ,
&\tilde e_3 = \partial_2 \ ,\cr
& & \cr}
\eqno(D.2)$$
where the notation of (A.5) and (A.7) is used with the replacement of $x$ by $y$. Note that these coordinates are 
singular at the identity ($y^l = 0$).

The left coset space $X= G/\exp({\rm span} \{e_3\})$ is obtained by identifying points of $G$ along integral curves of $e_3 = \partial_3$ (the 
orbits of right translation by the subgroup $\exp({\rm span}\{e_3\})$, namely the $y^3$-coordinate lines of these local 
coordinates which are comoving with respect to $e_3$). $\{y^l, y^2\}$ are local coordinates on $X$ which reduce to standard 
spherical coordinates $\{\theta,\phi\}$ on $S_2$ at the canonical type-IX point of ${\cal C}_D$. The right invariant vector fields $\tilde e_a$, since 
they are invariant along $e_3$, project to fields $\xi_a\equiv\xi(e_a)$ on the quotient space $X$ obtained by ignoring their third 
components in these local coordinates. $\{\xi_a\}$ is the image basis of generators of the natural left translation action of $G$
on $X$
$$
\eqalign{
& \xi_1 =(n^{(1)})^{-1}[ -(m^{(3)})^2 s_{3,2} \partial_1 
             -s_{3,2}c_{2,1}(s_{3,1})^{-1}\partial_2 \ ,\cr
& \xi_2 = c_{3,2} \partial_1 - s_{3,2}c_{2,1}\partial_2 \ ,\cr
& \xi_3 =\partial_2 \ ,\cr
& [\xi_a,\xi_b] = - C^c{}_{ab} \xi_c\ ,\cr}	
\eqno(D.3)$$

The isotropy group at the identity coset $\exp({\rm span}\{e_3\}) \in X$ of this left translation action of $G$ is just $\exp({\rm span} \{e_3\})$.
Now consider the following left invariant second-rank symmetric covariant tensor $^2\hbox{\sl g}_\kappa$ on $G$, with $\kappa \equiv m^{(2)}$:
$$
\eqalign{(m^{(2)})^2\, ^2\hbox{\sl g}_\kappa 
  &= (m^{(2)})^2 \omega^2\otimes\omega^2
  + (m^{(1)})^2 \omega^1\otimes\omega^1 \cr
  &= (m^{(2)})^2 [ dy^1\otimes dy^1
  + (m^{(3)} s_{2,1})^2 dy^2\otimes dy^2]\ . \cr
}
\eqno(D.4)$$
This is also invariant under right translation by the subgroup 
$\exp({\rm span} \{e_3\})$ and so projects to a left invariant tensor on $X$. For 
$(m^{(3)})^2 > 0$ this is a Riemannian metric on $X$ of constant Gaussian curvature $R^{12}{}_{12} =-\kappa^2$. (For $(m^{(3)})^2 < 0$ it is a 
pseudo-Riemannian metric of constant curvature, but at $m^{(3)} = 0$ it is degenerate.

If Riemannian metrics are of interest, one might as well consider only the line segment
$$
	\Gamma \equiv \{(a, {\bf n}\} \in {\cal C}_D | a=0;\ 
        {\bf n} = \diag (1, 1, n^{(3)}),\ n^{(3)}\in [-1,1]\}
\eqno(D.5)$$
connecting the canonical type-VIII,-VII$_0$, and -IX points of ${\cal C}_D$. The most general Riemannian metric on the product manifold 
$R \times X$ invariant under the natural left action of the direct-product group $R \times G$ (where $R$ is the additive group of real numbers, 
with coordinate $u$, and only the groups $G$ parametrized by the line segment $\Gamma$ are considered) is
$$
 	^3\hbox{\sl g}_\kappa = \overline{g}_{22} \, du \otimes du 
         + \overline{g}_{22} \,^2\hbox{\sl g}_\kappa \ ,
\eqno(D.6)$$
where $\overline{g}_{22}$ and $\overline{g}_{33}$ are constants and now 
$\kappa^2 = - n^{(3)}$. This is in fact the class of metrics studied by Kantowski and Sachs. 
However, when $n^{(3)} < 0$, these metrics are locally rotationally symmetric Bianchi type metrics, as shown by Bianchi [10]. The 
case $n^{(3)} = 0$ is obviously just the locally rotationally symmetric type-I or type-VII$_0$ metric expressed in cylindrical 
coordinates.
For $n^{(3)} < 0$, three inequivalent classes of standard coordinate systems exist for the constant-negative-curvature 2-manifold $X$ 
[73,74], namely coordinates chosen to be comoving with respect to 
$\xi_1,\xi_3$ or $\xi_1+\xi_3$ which correspond, respectively, to 
boosts, rotations and null rotations in the simply-connected covering group $\overline{SO_{2,1}}$ of $SO_{2,1}$ which is the value $G$ assumes at 
$n^{(3)} = -1$. The coordinates 
$\{y^l, y^2\} \equiv \{\nu_2,\eta_2\}$ are 
comoving with respect to $\xi_3$. For the canonical type-VIII case 
$n^{(3)} =-1$, choosing new coordinates $\{\overline{x}{}^l, \overline{x}{}^3\} \equiv \{\nu_3,\eta_3\}$
as in (2.34) of ref.~[73] which are comoving with respect to
$\xi_1+\xi_3$ and defining $\overline{x}{}^2\equiv u$, one has
$$
 	^3\hbox{\sl g}_1 
= \overline{g}_{22} d\overline{x}{}^2 \otimes d\overline{x}{}^2
 + \overline{g}_{33} (d\overline{x}{}^3 \otimes d\overline{x}{}^3
+ e^{-2\overline{x}{}^3} d\overline{x}{}^1 \otimes d\overline{x}{}^1) \ ,
\eqno(D.7)$$
which has the component matrix ${\bf g} 
= \diag (\overline{g}_{33}, \overline{g}_{22},\overline{g}_{33})
\in {\cal M}_{T(2)}$ with respect to the type-III $\equiv$ VI$_1$ 
frame $\overline{e}$  of (3.52) evaluated 
at $q = a = 1/2$, i.e. this is just the locally rotationally symmetric Bianchi type-III metric, whose spatial curvature matrix (in the 
same frame) is
$$
	\overline{g}_{33} \overline{{\bf R}} 
  = - \kappa^2 \diag (1, 0, 1)\ ,\qquad
 \kappa = 1 \ .
\eqno(D.8)$$

The Einstein equations for a spatially homogeneous perfect-fluid spacetime with metric
$$
 	^4\hbox{\sl g} 
   = N^2 dt \otimes dt +	^3\hbox{\sl g}_\kappa\ ,\qquad 
  \kappa^2\in[-1,1]\ ,
\eqno(D.9)$$
differ from this case only in that one retains the factor $\kappa^2$ in the spatial curvature. The value $\kappa = 0$ corresponds 
to the locally rotationally symmetric type-I or type-VII$_0$, case, while the complex rotation $\kappa = 1 \to \kappa = i$ takes 
one to the Kantowski-Sachs case. The equations for 
$(N, \overline{g}_{22}, \overline{g}_{33})$ 
and the fluid variables in the Kantowski-Sachs 
case $\kappa= i$ are identical with the $\kappa = 1$ case with the exception that the sign of the spatial curvature changes. In 
other words, if one solves the locally rotationally symmetric type-III equations with $\kappa$ left as an arbitrary 
parameter, one may obtain the Kantowski-Sachs solutions by the analytic continuation $\kappa = 1 \to \kappa = i$ [74]. The 
Kantowski-Sachs case could, therefore, be included in table II connected by horizontal dots to Bianchi type III.
The Kantowski-Sachs metric is also related to the locally rotationally sym-
metric Bianchi type-IX metric by a contraction of the group action in which the length of the subgroup 
$\exp({\rm span} \{e_3\}) \sim S_1$ becomes infinite; a similar contraction leads from the locally rotationally symmetric Bianchi 
type-VIII metric to the locally rotationally symmetric type-III metric. The singular transformation of the 
vacuum solutions induced by these group contractions is described explicitly in ref.~[74].

\end{document}